\documentclass[12pt,twoside]{report}
\normalsize
\usepackage{xcolor}
\usepackage{datetime2}
\usepackage[numbers]{natbib}

\newcommand{\reporttitle}{Varying Constants and the Brans-Dicke Theory: a new Landscape in Cosmological Energy Conservation}
\newcommand{\reportauthor}{Paolo Massimo Bassani}
\newcommand{\supervisor}{Prof João Magueijo}
\usepackage[a4paper,hmargin=2.8cm,vmargin=2.0cm,includeheadfoot]{geometry}
\usepackage{textpos}
\usepackage{natbib} % for bibliography
\usepackage{tabularx,longtable,multirow,subfigure,caption}%hangcaption
\usepackage{fncylab} %formatting of labels
\usepackage{fancyhdr} % page layout
\usepackage{url} % URLs
\usepackage[english]{babel}
\usepackage{amsmath}
\usepackage{graphicx}
\usepackage{dsfont}
\usepackage{epstopdf} % automatically replace .eps with .pdf in graphics
\usepackage{backref} % needed for citations
\usepackage{array}
\usepackage{latexsym}
\usepackage[pdftex,pagebackref,hypertexnames=false,colorlinks]{hyperref} % provide links in pdf

\hypersetup{pdftitle={},
  pdfsubject={}, 
  pdfauthor={},
  pdfkeywords={}, 
  pdfstartview=FitH,
  pdfpagemode={UseOutlines},% None, FullScreen, UseOutlines
  bookmarksnumbered=true, bookmarksopen=true, colorlinks,
    citecolor=black,%
    filecolor=black,%
    linkcolor=black,%
    urlcolor=black}

\usepackage[all]{hypcap}

\usepackage{color}
\usepackage{graphicx}% Include figure files
\usepackage{dcolumn}% Align table columns on decimal point
\usepackage{bm}% bold math
\usepackage{hyperref}% add hypertext capabilities
\usepackage[mathlines]{lineno}% Enable numbering of text and display math
\usepackage{amsmath, bm}
\usepackage{amssymb}
\usepackage{graphicx}
\usepackage{physics}
\usepackage{slashed}
\usepackage{systeme}

\usepackage{scalerel,stackengine}
\stackMath
\newcommand\reallywidehat[1]{\savestack{\tmpbox}{\stretchto{\scaleto{\scalerel*[\widthof{\ensuremath{#1}}]{\kern-.6pt\bigwedge\kern-.6pt}{\rule[-\textheight/2]{1ex}{\textheight}}}{\textheight}}{0.5ex}}\stackon[1pt]{#1}{\tmpbox}}
\usepackage[autostyle]{csquotes}
\allowdisplaybreaks

\newcommand{\N}[0]{\mathds{N}} % natural numbers
\date{\today}

\begin{document}

% load title page
% Last modification: 2015-08-17 (Marc Deisenroth)
\begin{titlepage}

\newcommand{\HRule}{\rule{\linewidth}{0.5mm}} % Defines a new command for the horizontal lines, change thickness here

%----------------------------------------------------------------------------------------
%	LOGO SECTION
%----------------------------------------------------------------------------------------

\center % Center remainder of the page

%----------------------------------------------------------------------------------------
%	HEADING SECTIONS
%----------------------------------------------------------------------------------------

\textsc{\Large Imperial College London}\\[0.5cm] 
\textsc{\large Department of Physics}\\[0.5cm] 

%----------------------------------------------------------------------------------------
%	TITLE SECTION
%----------------------------------------------------------------------------------------

\HRule \\[0.4cm]
{ \huge \bfseries \reporttitle}\\ % Title of your document
\HRule \\[1.5cm]
 
%----------------------------------------------------------------------------------------
%	AUTHOR SECTION
%----------------------------------------------------------------------------------------

\begin{minipage}{0.4\textwidth}
\begin{flushleft} \large
\emph{Author:}\\
\reportauthor % Your name
\end{flushleft}
\end{minipage}
~
\begin{minipage}{0.4\textwidth}
\begin{flushright} \large
\emph{Supervisor:} \\
\supervisor % Supervisor's Name
\end{flushright}
\end{minipage}\\[4cm]

%----------------------------------------------------------------------------------------
%	FOOTER & DATE SECTION
%----------------------------------------------------------------------------------------
\vfill % Fill the rest of the page with whitespace
A thesis submitted in partial fulfilment of the requirements
for the degree of Master of Science in Quantum Fields and Fundamental Forces\\
in Theoretical Physics at Imperial College London\\[0.5cm]

\makeatletter
\@date 
\makeatother

\end{titlepage}

% page numbering etc.
\pagenumbering{roman}
\clearpage{\pagestyle{empty}\cleardoublepage}
\setcounter{page}{1}
\pagestyle{fancy}

%%%%%%%%%%%%%%%%%%%%%%%%%%%%%%%%%%%%
%%%%%%%%%%%%%%%%%%%%%%%%%%%%%%%%%%%%
\begin{abstract}
We develop the Brans-Dicke theory of gravity in the context of varying constants of Nature. Using the unimodular formalism of General Relativity, we create a platform to provide physical relational times giving the evolution of physical constants. We therefore review the ideas and experiments behind varying constants, mostly focusing on the speed of light and the gravitational constant. Then, we apply this idea to the energy conservation in cosmology, illustrating the arising patterns.
Motivated by a varying gravitational constant resulting from Mach's principle, we develop the unimodular formalism of varying constants in the Brans-Dicke theory. Doing so, we obtain several original results, some of which can be compared with phenomenological observation. Finally, we suggest how a varying Brans-Dicke parameter could be linked to the Cosmological Constant problem.

\end{abstract}

\cleardoublepage
%%%%%%%%%%%%%%%%%%%%%%%%%%%%%%%%%%%%
%%%%%%%%%%%%%%%%%%%%%%%%%%%%%%%%%%%%
\section*{Acknowledgments}
Firstly, I would like to thank my supervisor, Professor João Magueijo. His support, guidance, insights and creativity have been invaluable towards the success of this dissertation. I am greatly beholden for the opportunity of having interacted and exchanged ideas with a person of his human and intellectual stature. His example will be an important part of my future life.
\\
\\
Secondly, I would like to express my endearment to all the friends I have shared this experience with: Giacomo, Farbod, Greg, Frank, Coen, David, Abdulaziz and Lorenzo. The long discussions on crazy ideas, the support during this Master and the happy moments together have enriched my life with the gift of friendship, which I will always carry with me. Also, I would like to thank my friend Marcel for the great help provided. Honorable mention goes to Ray, for the important discussions and insights. A special thought goes to Altay, whose unstoppable enthusiastic optimism and faith in physics have motivated me to undertake this Master until its completion.
\\
\\
I am grateful to the unique Flower of the Desert that accompanies me through life: your delicacy, kindness and loving care have been the strength and inspiration throughout this time. Above all, your hopefulness and support haven been the emotional pillar of this accomplishment.  
\\
\\
Infine, la mia più sconfinata riconoscenza e infinita gratitudine vanno alle persone più importanti: la mia famiglia. Grazie Druso, per essere sempre stato giocoso, allegro e sempre disponibile a passare del tempo con me. Grazie Costanza, per avermi sempre sopportato, ascoltandomi con pazienza nei momenti più importanti, mostrandomi il giusto valore delle cose. Sopratutto, grazie per essere sempre stata mia sorella: spero, da grande, di poter essere tuo fratello.\\
Grazie Papà, per le importanti riflessioni e lo sprone necessario per completare questo Master. Il tuo incoraggiamento e il tuo animo originale e indipendente mi hanno spinto a continuare il mio percorso con spirito critico e determinazione.\\
Grazie Mamma, per il tuo sconfinato affetto e la tua celestiale amorevolezza che mi hanno sempre accopagnato, senza interruzioni, dalla nascita fino ad ora. Grazie per essere sempre stata al mio fianco, parlandomi, ascoltandomi nei momenti più difficili. La tua frizzante curiosità e il tuo amore per la Natura hanno sempre affascinato il mio intelletto, ispirandolo a ricercare la conoscenza con gioia e purezza.\\
A tutti voi, per tutto quello che mi avete donato, ma sopratutto, per essere stati la mia famiglia, esprimo il mio più sincero affetto: vi voglio bene.

\newlength\longest

\clearpage

\thispagestyle{empty}

\vspace*{\fill}

\begin{quote}
\begin{center}
    \textit{``Humana ante oculos foede cum vita iaceret
in terris oppressa gravi sub religione
quae caput a caeli regionibus ostendebat
horribili super aspectu mortalibus instans,
primum Graius homo mortalis tollere contra
est oculos ausus primusque obsistere contra,
quem neque fama deum nec fulmina nec minitanti
murmure compressit caelum, sed eo magis acrem
irritat animi virtutem, effringere ut arta
naturae primus portarum claustra cupiret.
Ergo vivida vis animi pervicit, et extra
processit longe flammantia moenia mundi
atque omne immensum peragravit mente animoque,
unde refert nobis victor quid possit oriri,
quid nequeat, finita potestas denique cuique
quanam sit ratione atque alte terminus haerens.
Quare religio pedibus subiecta vicissim
obteritur, nos exaequat victoria caelo.''}
\end{center}
\end{quote} 
\begin{center}
    \textit{De Rerum Natura}\:-\:{\textit{Titus Lucretius Carus}}
\end{center}
\vspace*{\fill}

\clearpage

\clearpage{\pagestyle{empty}\cleardoublepage}
%%%%%%%%%%%%%%%%%%%%%%%%%%%%%%%%%%%%
%%%%%%%%%%%%%%%%%%%%%%%%%%%%%%%%%%%%
%--- table of contents

\fancyhead[RE,LO]{\sffamily {Table of Contents}}
\tableofcontents 

\clearpage{\pagestyle{empty}\cleardoublepage}
\pagenumbering{arabic}
\setcounter{page}{1}
\fancyhead[LE,RO]{\fontsize{10}{12} \slshape \rightmark}
\fancyhead[LO,RE]{\fontsize{10}{12} \slshape \leftmark}

%%%%%%%%%%%%%%%%%%%%%%%%%%%%%%%%%%%%
%%%%%%%%%%%%%%%%%%%%%%%%%%%%%%%%%%%%
\chapter{Introduction}

In this chapter, we present the three main background theories -- Unimodular Gravity, the Constants of Nature and the Brans-Dicke theories -- to propel the investigation in the next chapters. The first one provides the mathematical framework to rigorously implement the idea of varying constants in General Relativity and in Brans-Dicke theory. The second theory explores the philosophical and physical depths behind constants and motivates why they could, paradoxically after all, be variable. Furthermore, it provides an insight into the broad landscape of roles the speed of light plays in physical theories. Finally, the third section introduces the Brans-Dicke theory of gravity, from its connection with the idea of varying $G$ to its fields equations, employing Mach's principle. The Brans-Dicke theory will be the basis for the original research presented in Chapter 3, so it is particularly important towards the results obtained and their interpretation. To conclude, section \eqref{overview_section} summarises how the ideas in these three sections empower Varying Constant Theories (VCT). We also explain the aims of this research which will be reviewed and assessed in Chapter 4 in light of the obtained results.

\section{Unimodular Gravity} \label{UniIntro}

In this section, we will explore the implementation of unimodular gravity in General Relativity. In doing so, we will distinguish between two main formalisms: Einstein's pure unimodular gravity and the Henneaux-Teitelboim theory. The former is called so because it was originally suggested by Einstein as a way to simplify calculation in General Relativity by requiring that the determinant of the metric $\sqrt{-g} = 1$. In 1989, the latter was proposed to generalise Einstein's idea, allowing for a less restrictive condition, i.e., $\sqrt{-g} = {\partial_\mu  \mathcal{T}^\mu}$, where $\mathcal{T}^\mu$ is a generic vector density. As we will see, this extension allows to associate different constants with their respective conjugate relational times, a central property of theory upon which most of the results in this work are based.

\subsection{Einstein's pure unimodular gravity}

Since its inception, Einstein's Theory of General Relativity (EFE) has been the most experimentally successful attempt at describing the phenomenon known as gravitation \cite{berti2015testing}. Einstein's theory victoriously survived experimental tests ranging from the perihelion procession of Mercury to light deflection due to the Sun's gravitational field \cite{berti2015testing}. More recently, predictions from General Relativity have been tested in more exotic phenomena such as gravitational lensing and gravitational waves \cite{abbott2016observation}, once again confirming this theory. \\However, one observational evidence controverts the Einstein Field Equations as they were presented originally in the famous 1915 paper titled \emph{The Field Equations of Gravitation} \cite{einstein1915field}. In fact, the first field equations for a vacuum solution,
\begin{equation}
    R_{\mu \nu} - \frac{1}{2}R g_{\mu \nu} = 0
\end{equation}
predicted a dynamical universe with space evolving in time. Rejecting this prediction, in 1917 Einstein introduced a parameter, $\Lambda$, called the cosmological constant. It allowed the community to believe that the universe could remain static and eternal, completely deprived of any evolving dynamic (the so-called Einstein static universe solution). It was not long after his addition of $\Lambda$ that Hubble discovered the relationship between velocities and redshift of distant galaxies and concluded the opposite: the universe was expanding. This experimental evidence combined with theoretical works by Friedmann (the universe is expanding regardless of the value of $\Lambda$) led Einstein to call $\Lambda$ his \emph{biggest blunder}, thus changing his views on the nature of the Universe.\\
Since then, the necessity for the cosmological constant has been a matter of debate until 1998, when it was proven that the expansion of the universe is accelerating, requiring a positive cosmological constant \cite{riess1998observational}. Interestingly, Einstein's \emph{biggest blunder} turned out to be a necessary addition to the field equations describing a dynamical universe subject to an expansion possibly driven by $\Lambda$.\\
\\
The appearance of the cosmological constant in the field equations has been since then a matter of active debate from a theoretical point of view. In fact, it is not clear, \emph{a priori}, why $\Lambda$ should be constant and how it actually enters the field equations. Einstein's unimodular formulation of General Relativity has been an attempt to better justify its appearance. Frivolously, before its inception, $\Lambda$ would simply be  inserted in the Einstein-Hilbert action with matter as follows \cite{d2022introducing}:\\
\begin{equation}
    S_{EH} = \frac{c^4}{16 \pi G} \int{d^4x \: \sqrt{-g} \: R} \: + S_m \rightarrow \: S_{EH} = \frac{c^4}{16 \pi G} \int{d^4x \: \sqrt{-g} \: (R - 2\Lambda)} + S_m \label{EHstart}
\end{equation}\\
where $S_m$ is the matter action. It is clear that using the full action above, when deriving the EFEs, we arrive at a set of field equations where $\Lambda$, from a purely mathematical point of view, does not need to be constant. Varying action (\ref{EHstart}), we obtain the field equations:\\
\begin{equation}
    \frac{\delta S_{EH}}{\delta g^{\mu \nu}} = 0 \Leftrightarrow R_{\mu \nu} - \frac{1}{2}R g_{\mu \nu} + \Lambda g_{\mu \nu} = \frac{8 \pi G}{c^4}T_{\mu \nu} \label{EFE_1}
\end{equation}\\
where we again stress that $\Lambda$ is taken to be constant purely for the observational reasons \cite{riess1998observational}, but it does not need to be. The idea of unimodular gravity, first formulated by Einstein himself as a mean to unify gravity and matter \cite{einstein1923electrodynamics}, starts from imposing the constraint $\sqrt{-g} = 1$ to action (\ref{EHstart}), a condition that leads to preserving the determinant of the metric when varying the action,
\begin{equation}
    \frac{\delta \sqrt{-g}}{\delta g^{\mu \nu}} = 0 \label{Uni_22}
\end{equation}
Furthermore, consider the infinitesimal variation of the metric
\begin{equation}
    \delta g_{\mu \nu}(x) = \nabla_\mu k_\nu + \nabla_\nu k_\mu \label{killing}
\end{equation}
where $k^\mu$ is the transformation's gauge vector and $x$ are the general coordinates, under the infinitesimal coordinate transformation
\begin{equation}
    x^\mu \mapsto x'^\mu = x^\mu + k^\mu 
\end{equation}
Using the chain rule on (\ref{Uni_22}), we can get a full expression for the variation of $\sqrt{-g}$ the same way it is done in \cite{bufalo2015unimodular}:
\begin{equation}
    \delta \sqrt{-g} = \frac{1}{2} \sqrt{-g} g^{\mu \nu} \delta g_{\mu \nu} = 0 \implies g^{\mu \nu} \delta g_{\mu \nu} = 0 \label{chain}
\end{equation}
where $g = \det g_{\mu \nu}$. Finally, we can plug (\ref{killing}) into the last expression of (\ref{chain}), contract the indices with $g^{\mu \nu}$, and arrive at a condition
\begin{equation}
    \nabla_\mu k^\mu = 0
\end{equation} 
for the gauge vector $k_\mu$.\\
The transformations on $k^\mu$ are called transverse diffeomorphisms. They can be interpreted as symmetric properties of the metric and are naturally arising from the unimodular constraint imposed on $\sqrt{-g}$. It is now possible to introduce the unimodular constraint by modifying the Einstein-Hilbert action with the addition of a Lagrange multiplayer $\lambda$
\begin{equation}
    S_{EH} = \frac{c^4}{16 \pi G} \int{d^4 x \: \sqrt{-g}R} - \int{d^4 x \: \lambda(x) (\sqrt{-g} -1)} + S_m
\end{equation}
Varying this action with respect to the Lagrange multiplier scalar $\lambda(x)$ gives the unimodular constraint $\sqrt{-g} = 1$. On the other hand, performing the usual variation with respect to $g^{\mu \nu}$ gives the field equations 
\begin{equation}
    R_{\mu \nu} - \frac{1}{2}R g_{\mu \nu} + \frac{8 \pi G}{c^4} \lambda(x) g_{\mu \nu} = \frac{8 \pi G}{c^4}T_{\mu \nu} \label{EFE_uni_22}
\end{equation}
Finally, by expressing (\ref{EFE_uni_22}) for $\lambda$ and then taking its divergence we get
\begin{equation}
    \nabla_\mu \lambda = \nabla^\nu T_{\mu \nu} - \frac{c^4}{8 \pi G} \biggl [\nabla^\nu R_{\mu \nu} - \frac{1}{2} \nabla_\mu R \biggl] = 0
\end{equation}
which is equal to zero because of the energy-momentum conservation, i.e., $\nabla^\nu T_{\mu \nu} = 0$ and because of the contracted Bianchi identities $\nabla^\nu R_{\mu \nu} - \frac{1}{2} \nabla_\mu R = 0$. Given that and that $\lambda$ is a scalar, we can write 
\begin{equation}
    \partial_\mu \lambda = 0
\end{equation}
which tells us that $\lambda$ is indeed an integration constant as we wanted. Therefore, comparing (\ref{EFE_uni_22}) with (\ref{EFE_1}), we see that the dynamics of both theories are the same with the cosmological constant of 
\begin{equation}
    \Lambda = \frac{8 \pi G}{c^4} \lambda
\end{equation}
This result confirms once again that the cosmological constant can be derived in the field equations by imposing the unimodular condition on the determinant of the metric. However, as we will see in the next section, it is possible to go beyond this result, introducing a formalism that will turn out essential for the implementation of varying constants in cosmology
\cite{einstein1952gravitational}.

\subsection{Henneaux-Teitelboim theory}

Generalising Einstein's idea, Henneaux and Teitelboim formulated a fully diffeomorphism-invariant extension of the unimodular condition by introducing the divergence of a vector field density as follows \cite{henneaux1989cosmological}:
\begin{equation}
    \sqrt{-g} = {\partial_\mu  \mathcal{T}^\mu_\Lambda} \label{uni}
\end{equation}
where the range of the index is $\mu = 0...3$.
\\We will focus now on condition (\ref{uni}) as this will become central to our developments of unimodular time and other times. Therefore, following \cite{henneaux1989cosmological}, it is possible to write action (\ref{EHstart}) as the sum of two actions:\\
\begin{equation}
    S = S_0 + S_U 
\end{equation}\\
where $S_0$ is, in our case, the EH action, but it could be any base action, whereas $S_U$ is the unimodular term defined as:\\
\begin{equation}
    S_U = \int{d^4 x \: \Lambda \: \partial_\mu  \mathcal{T}^\mu_\Lambda} \label{UNI}
\end{equation}\\
which, combined with $S_0$, and integrated by parts, leads to the final result\\
\begin{equation}
    S_0 \rightarrow S = S_0 + \int{d^4 x \: \Lambda \: \partial_\mu  \mathcal{T}^\mu_\Lambda} = S_0 - \int{ d^4 x \:  (\partial_\mu \Lambda) \mathcal{T}^\mu_\Lambda} \label{HT}
\end{equation}\\
where $S_0$ is a base action which, for the moment, is the $EH$ action as defined in (\ref{EHstart}), but will be modified to Brans-Dicke in the following sections. It is also important to note that in (\ref{HT}) $\Lambda$ is not anymore, \emph{a priori}, the cosmological constant, but just a scalar, which happens to appear in the EFEs as a result of being an integration constant. This allows us to transform the cosmological constant from a fixed parameter to an integration constant, which will be such only in the equations of motion.
In fact, calculating the equations of motion from action (\ref{HT}) we arrive at\\
\begin{equation}
    \frac{\delta S}{\delta \Lambda} = \frac{\delta S_0}{\delta \Lambda} + \frac{\delta S_U}{\delta \Lambda} = 0 \Leftrightarrow \sqrt{-g} = \partial_\mu \mathcal{T}^\mu_\Lambda \label{HTcond}
\end{equation}\\
\begin{equation}
    \frac{\delta S}{\delta \mathcal{T}^\mu_\Lambda} = \frac{\delta S_U}{\delta \mathcal{T}^\mu_\Lambda} = 0 \Leftrightarrow \partial_\mu \Lambda \approx 0 \label{Lambd}
\end{equation}\\
Firstly, equation (\ref{HTcond}) is the Henneaux-Teitelboim condition introduced in (\ref{uni}). This proves that the condition holds on the level of the equations of motion. On the other hand,  (\ref{Lambd}) proves that now $\Lambda$ is a constant only on the shell, i.e., when it obeys the equation of motion.\\
\\For now, we wish to understand the relation between $\Lambda$ and the vector density $\mathcal{T}^\mu_\Lambda$, as well as what the interpretation of this density could be.\\
An important insight into this question is achieved by considering the $3 +1$ foliation of space-time induced by the ADM formalism in $S_U$. Upon the $3+1$ splitting of the time and space components of $\mathcal{T}^\mu_\Lambda$ as $\mathcal{T}^\mu_\Lambda$ = $\mathcal{T}^0 + \mathcal{T}^i$, where $ i = 1, 2, 3 $, the unimodular action (\ref{UNI}) becomes
\begin{equation}
    S_U = \int {dt \: d^3 x \: (\Lambda \: \dot{\mathcal{T}^0_\Lambda} + \Lambda \: \partial_i \mathcal{T}^i_\Lambda)} \label{UU}
\end{equation}\\
This shows that $\mathcal{T}^0_\Lambda$ is the canonical conjugate of $\Lambda$ as they form a canonical pair of variables in the Hamiltonian formalism. Furthermore, following \cite{henneaux1989cosmological}, the only propagating mode of $\mathcal{T}^\mu_\Lambda$ is the zero-mode of its time component as defined below:\\
\begin{equation}
    T_\Lambda = \int{d^3x \: \mathcal{T}^0_\Lambda}
\end{equation}
\\
This zero-mode $T_\Lambda$ defines a time \cite{bombelli1991time} that is naturally canonical to the cosmological constant. Importantly, the notion of a canonical time to a constant is intimately related with relational quantum mechanics \cite{rovelli1996relational} and, more generally, the problem of time in quantum mechanics. This allows to create relational times which depend on their canonical pair (the constant) instead of an absolute parameter time $t$. As we will see, this formalism provides a powerful implementation of Leibniz's idea of relational events, and, therefore, times \cite{arthur2021leibniz}. \\
Having established this framework where the cosmological constant forms a conjugate pair with the identified unimodular time $T_\Lambda$, we can generalise this to any constant of nature appearing in our base action $S_0$. In fact, instead of having only one constant in the unimodular term, we could have a full vector of constants called $\bm{\alpha}$ conjugate to a set of different corresponding relational times $\bm{T_\alpha}$. Since any $\bm{\alpha}$ is conjugated with its $\bm{T_\alpha}$, we start to see here how some physical constants will become relational clocks, determining the evolution of other constants. Using this generalisation in action (\ref{UNI}), we arrive at an additional unimodular-like term in the action,\\
\begin{equation}
    S_{\bm{\alpha}} = \int \:{ d^4 x \: \bm{\alpha} \: \partial_\mu  {\mathcal{T}^\mu_\alpha}}
\end{equation}\\
which, to conclude, will give us the full action\\
\begin{equation}
    S = S_0 - \int{ d^4 x \: \bm{\alpha} \: \partial_\mu \mathcal{T}^\mu_{\alpha}} \label{starting}
\end{equation}\\
From this point on, we will be able to use the formalism introduced in action (\ref{starting}) to derive a set of time evolution equations for different constants. Following this procedure, we will manage to demote any constant appearing in the base action to mere constants of motion, thus obtaining the set of relational times canonical to them. The constant and the associated time will form a canonical conjugate pair as prescribed by the Hamiltonian formalism, as we will see in the next chapter. These times will then be used to provide evolution parameters for other constants, as we will see. To conclude, we will mostly focus our investigation on using the Brans-Dicke action as a base theory for reasons that will become clearer later. This will be the core of Chapter 3 and its results.

\section{The Constants of Nature} \label{constant_section}

\begin{quote}
    ``\:\textit{Tempora Mutantur et Nos Mutamur in Illis}\:''
\end{quote}
What makes the Reality we perceive unique? What is the property, the characteristic so deeply connected to our Universe to make it one of a kind, and therefore differentiable from anything else? Something of this sort should certainly be the pillar of the world, a unique feature, so exclusive to the Universe to actually make it The Universe. As such, it most definitely would need to be a fixed property, eternal and equal everywhere and at the heart of anything else populating the Universe: it should be constant, a constant.\\
\\
In fact, as the word suggests, constants are values that not only remain the same everywhere and at any time (in the past and in the future), but also, and in parts \textit{because} of their universality, they are the fundamental constituents of every physical theory. This is because their spacetime invariance characterises uniquely a set of phenomena (which is the case for some constants) or even the essence of the Universe (which is the case for three constants). Bringing this concept to an extreme, imagine a Universe where the very constant that make it such vary, not being constant anymore. Since these constant are the essence, the origin, the $A \rho \chi \acute{\eta}$ of the Universe, changing them would result in changing the Universe. In other words, the Universe would not be the same one as before, it would become another Universe, since its most fundamental identifier are subject to change. This inevitably leads to think that, since the constant are not invariant, then some other property of the Universe must be the truly fundamental one, and what is better then the quantity these constant are changing with respect to for this role. If we assume time-dependent constants (it will turn out that this is not the only possibility), it is then time the next subject of our interest.\\
\\
As we have seen before, the generalised unimodular formalism provides excellent definitions of times for the evolution of the constants, but it also does much more then that. It enables us to implement Leibniz's  idea of relational times, so intimately connected with the problem of time in quantum mechanics \cite{fortin2022relational}. Finally, it shows us Nature's subtle irony: certain constants vary while other provide definition of the relational times, but this situation can be reversed. This leads to a racy ambiguity, where the role of a constant and that one of a time can be interchanged and mixed: where the ultimate truth lies is a question for the future.\\
\\
Therefore, in this section, we explore the heuristic and the ideas behind the concept of a constants and their potential variability in time. Firstly, we investigate the nature of constants, understanding why they arise, what property makes them similar or differentiate them into categories and how they relate to the physical laws. Secondly, subsection \eqref{vary_vary} presents the experimental hints and evidences for varying constants. Furthermore, it also reviews the early theoretical works on the subject, developing an intuition for the next chapters. Finally, subsection \eqref{different_c} provides a crucial specification to the ideas illustrated before: the ``different'' speeds of light. It does so by showing how a single constant like $c$ could have different roles, depending on the phenomenon it describes. This differentiation will turn out to be very important in Chapter 3, where ``different'' speeds of light can lead to different scenarios in the context of energy conservation.

\subsection{The Essence of a Constant}

\begin{quote}
    ``\:\textit{Multa sunt quae esse concedimus; qualia sunt? Ignoramus.}\:''
\end{quote}
What is a constant? It is a simple question at the heart of our understanding of Reality, the very basis of the fundamental phenomena constructing this Universe, that will prove rather deceptive and elusive to grasp. We will, nevertheless, try our best to succeed.
To shed some light on this topic, it is enlightening to distinguish the domains of constants: mathematical and physical ones. Our main interest will pivot around the latter, but we believe it will be useful to briefly mention mathematical constants and their relevance to this investigation. Mathematical constants are fundamentally different from physical ones in one property: they cannot be measured, and therefore, they are all dimensionless numbers without any uncertainty. In this regard, they are certainly more abstract than physical ones, as they appear as recurring numbers across mathematics, almost being structures of the model we use to describe Nature. Another interesting fact about mathematical constants is that some of them appear in physical laws, but no measured physical constants appear in mathematics.
The most evocative example is, by far, the most ancient constant known: $\pi$. It is defined as \cite{acheson2020wonder}
\begin{equation}
    \pi := \frac{C}{d}
\end{equation}
i.e., the ratio of the circumference of a circle (a 2D object, more generally, a 1-sphere $\bm{\mathcal{S}^1}$) to its diameter. It appears in any classical physical law possessing some periodic property \cite{halliday2013fundamentals}; it  plays a role in General Relativity, being in the EFE constant $\kappa = \frac{8 \pi G}{c^4}$ \cite{d2022introducing}, to Quantum Mechanics, e.g., in the Uncertainty Principle \cite{griffiths2018introduction}, defining reduced Planck's constant as $\hbar = \frac{h}{2 \pi}$. The proliferating appearance of $\pi$ in so many different fields is incredible, but its most striking property is indeed its constancy: the value of $\pi$ is invariant in space, time and other dimensions ($\pi$ is the same no matter the dimension in which a $n$-sphere is considered \cite{finch2003mathematical}). This happens because $\pi$ is actually the ratio of two geometrical quantities that, by the construction of space, are related exactly by it: no matter what, when one is changed, the other one will change proportionally, producing always the same value of $\pi$. This suggests the first intuition on what a constant is: a fixed value linking two concepts or quantities (which often might not be evidently related, see subsection \eqref{different_c}) so defining their absolute and universal nature. From this definition, we can easily understand why such a profound quantity like a constant must be invariant in space, time and dimensions: if it was not, it would not describe the $n$-sphere anymore, but something else. The relationships between abstract elements of an $n$-sphere define $\pi$ and $\pi$ defines those relationships.
\\
\\
As we will see for physical constants, equations, numbers and functions describe $\textit{how}$ a system evolves and interacts with other systems, but constants provide the essential information of $\textit{what}$ that system is, its fundamental nature that makes it different from anything else and equal to anything sharing the same constant. This is due to the principle of identity of indiscernible \cite{hacking1975identity}. This powerful idea, expressed in its more concrete as well as abstract realisations, has been at the heart of great discoveries such as the equivalence of mass and energy (where the constant $c$ is really the link between the two) or the equivalence principle
\cite{einstein1908relativitatsprinzip}. Now that we have established the starting point of our constants discussion, we can explore the more natural topic of physical constants.
\\
\\
What is a constant? A physical one, this time. Expectedly, a physical constant usually is a dimensional numerical value arising in the description of a natural phenomenon. As such, it arises in the same way whenever that phenomenon, its outcomes or other qualitatively relatable phenomena, are experienced, making it the characteristic and often defining property of the phenomenon. The crucial point is that physical constants are the result of measurements and dimensional, meaning they carry information about the measurement convention. Units intrinsically distinguish these quantities. Some constants measure velocity (the state of motion of an object), others measure charge, mass, other temperature and more. We could, more abstractly, define physical constants, following \cite{uzan2011varying}, as \emph{any parameter that cannot be explained by the theory it describes}. This highlights the first fork in our analysis: fundamental and non-fundamental constants. In fact, while the first ones cannot possibly be determined in terms of any other parameter of the theory, the second ones are often made up of fundamental constants or are kept because of historical uses.  More specifically, fundamental constants are impossible to compute from the theoretical framework as this would mean they obey some dynamical equation; resultantly, they would not be constant anymore. Conversely, from the experimental point of view, the fundamental constants must and have been measured, thus confirming the validity of the theory and fixing the value of the constant up to the experiment's accuracy.\\
\\
Following the idea behind fundamental constants, Weinberg defined them as ``\emph{constants that appear in the laws of nature at the deepest level that we yet understand, constants whose value we cannot calculate with precision in terms of more fundamental constants, not just because the calculation is too complicated (as for the viscosity of water or the mass of the proton) but because we do not know of anything more fundamental}"\cite{weinberg1983overview}. This definition remarks the nature of fundamental constants: they create the basis upon which our physics, our understanding of Reality is constructed. They quite literally define the Universe, giving the \emph{what} property of objects as described for mathematical constants. Another crucial point is that fundamental constants retain their title until another more general theory can explain their origin. Therefore, when we discuss variable fundamental constants, whether they are varying with time \cite{magueijo2023evolving} (as we will consider here), with respect to space \cite{levshakov2009spatial} or something else, or they are just dynamical quantities explained by a more general theory, the constants will lose their fundamental character. This will expectedly and  drastically change their nature, leading to, as an example above all, a speed of light that is no longer observer-invariant, that breaks Lorentz invariance \cite{albrecht1999time} and energy conservation.\\
\\
We conclude by analysing the current set of fundamental physical constants and by providing a classification in which three constants stand out above all the others. As aforementioned, physical constants could be divided into fundamental and non-fundamental ones; the former cannot be expressed in terms of anything more fundamental but can be combined to give the latter. These are the constants describing the four fundamental interactions known thus far: electromagnetism, weak interaction, strong interaction and gravity. Therefore, considering these theories, we have 22 unknown constants in total\cite{uzan2011varying}: six Yukawa couplings for the quarks, three Yukawa couplings for the leptons, two parameters for the Higgs field, four parameters for the CKM matrix, three couplings for the symmetry group of the Standard Model, a phase for the QCD vacuum \cite{robinson2011symmetry} and, finally, the speed of light, the gravitational constant and Planck's constant.\\ Interestingly, most of these fundamental constants come from the Standard Model and are related to the individual particles, fields, and their symmetries. They are, moreover, different in nature from $c$, $G$ and $\hbar$, as these appear to be more universal and less specific to a particular physical phenomenon as the Standard Model ones. This is hardly a coincidence, as fundamental constants can be divided into three different subcategories: constants characteristic of a particular system, constants characteristic of a class of physical phenomena, and truly universal constants \cite{levy1979importance}. It is clear how the Standard Model constants belong predominantly to the second category, while $c$, $G$ and $h$ are part of the universal constants, as they describe the very structure of Reality. Furthermore, as observed in \cite{levy1979importance}, \cite{uzan2008natural} and \cite{ellis2005c}, they have the property of being concept synthesisers and reference scales for physical theories. \\
Firstly, equations like $E = c^2 m$, $E = \hbar \omega$ and $G_{\mu \nu} = \frac{8 \pi G}{c^4}T_{\mu \nu}$ show how each of these three constants allows linking concepts that are apparently unrelated. They are the conversion, the bridge between different aspects of the same Reality. Secondly, $c$ and $\hbar$ set the physical scale of applicability of theories: the speed of light determines an upper bound for causality, establishing when systems become relativistic, while Planck's constant defines the scale of quantum phenomena. The same role is not clearly defined for $G$, even though it could be interpreted as setting the scale at which gravitational phenomena become relevant \cite{uzan2008natural}. Nevertheless, we clearly see how these three constants play a more fundamental and differentiated role than the others, to the point that we might be tempted to call them, in an ecclesiastic spirit, the Holy Trinity.

\subsection{Constants that Vary} \label{vary_vary}
\begin{quote}
    ``\:$\Pi \acute{\alpha} \nu \tau \alpha \: \rho \epsilon \tilde{\iota}$\:''
\end{quote}
In the previous subsection, we have explored the properties that substantiate and differentiate constants. As exemplified, an essential property of a constant is, tautologically, its constancy in time, space, dimensions, etc. More generally, a fundamental constant, to be such, should not have any dependency whatsoever, whether it is from other constants or from the structure of Reality. It is an absolute, a quantity only measurable and, if we wish, \textit{a Deo datum}. However, the idea at the heart of this work challenges this view by asking: what if the fundamental constants of Nature are not actually constant? That means that among the many mechanisms that could result in varying constants models, we assume fundamental constants to be varying in the sense of being subject to evolution, resulting in their dynamics. This postulate inevitably leads to the second question of evolution with respect to what. In this work, we will use the idea of relational times as parameters determining the evolution of the constant, as provided by the unimodular formalism. Lastly, we will only consider physical constants appearing in the Einstein-Cartan, and Brans-Dicke actions provided with a perfect fluid matter content. This is the case because the unimodular formalism presented in section \eqref{UniIntro} was originally designed to be applied to gravitational types action. Resultantly, from the 22 fundamental constants mentioned earlier, the ones associated with the Standard Model will be neglected, even though they could become the object of future studies.
\\
\\
Firstly, we would like to explore the heuristic behind varying constants, so as to understand why such a hypothesis could, after all, not be too insane. Everything started with the possibility that $G$, the gravitational constant, could vary depending on cosmic time. In 1938, Dirac formulated the Large Number Hypothesis \cite{dirac1938new}, creating the basis for varying constants. He noticed a remarkably close coincidence between two ratios, namely that \cite{chiba2011constancy}
\begin{equation}
    N_1 = \frac{e^2}{G m_p m_e} \simeq 2 \cross 10^{39}
\end{equation}
\begin{equation}
    N_2 = \frac{c H_0^{-1}}{e^2 m_e^{-1} c^{-2}} \simeq 2 \cross 10^{40}
\end{equation}
where $N_1$ is the ratio of electrostatic to gravitational force between the proton and the electron, while $N_2$ is the ratio of the Hubble horizon radius today, $H_0^{-1}$, to the classical electron radius. Intrigued by their closeness, he thought that the two numbers should be the same up to a factor of order unity such that 
\begin{equation}
    N_{1} = N_{2} \cross \mathcal{O}(1) \label{dir_ac}
\end{equation}
Suppose that this relationship was valid, then we expect it to hold at all cosmological times, from the beginning of the Universe till our time. However, $N_{2}$ is not a constant ratio, as it depends on $H_0^{-1}$, the current Hubble radius. Since the Universe is expanding \cite{riess1998observational}, $H_0^{-1}$ varies in time accordingly, making $N_2$ varying. But a non-constant $N_2$, by virtue of relation \eqref{dir_ac}, requires also $N_1$ to vary over time. Dirac interpreted this fact as a varying $G$ over cosmological time as $G \propto t^{-1}$. Alternatively, one could also explain a varying $N_1$ as a constant $G$ but a varying fine structure constant according to $\alpha \propto t^{\frac{1}{2}}$. Interestingly, a third possibility would be that the Hubble radius in $N_2$ is constant but the speed of light is varying similarly to VSL theories \cite{albrecht1999time}; though fascinating, this scenario would require a whole discussion itself tangential to this work.\\
Several experimental tests have been conducted to check the variability of $G$ \cite{hellings1989experimental} \cite{damour1991orbital} \cite{teller1948change} \cite{degl1995time}, from Solar System constraints to binary pulsars to stellar constraints all the way up to cosmological one \cite{uzan2011varying}, converging on similar variations for $G$. For example, the change in $G$ over time was estimated using binary pulsars to be \cite{damour1988limits}
\begin{equation}
    \frac{\dot{G}}{G} = (1.0 \pm 2.3) \cross 10^{-11} \: yr^{-1}
\end{equation}
suggesting a minimal, but measurable change in $G$ over time.
\\
\\
On the other hand, we might interpret \eqref{dir_ac} as an equation where $G$ is constant but $\alpha$ is varying. We now follow this line of reasoning. This possibility is particularly interesting as it will closely relate to the variability of $c$, as we will see. Plenty of different experiments have been performed to test the variability of $\alpha$, ranging from quasar spectra \cite{webb1999search} to atomic clocks \cite{blatt2008new} and absorption lines spectra \cite{kanekar2004conjugate}, including measurements of CMB in the WMAP mission \cite{martins2004wmap}. However, among all these different measurements, the most interesting, constraining, and sensitive one is the Oklo event \cite{kanekar2004conjugate}. Approximately $1.8 \cross 10^9 \: yr$ in Gabon a Uranium deposit reached the critical phase of fission reaction due to the extreme coincidence of favourable conditions (abundance of naturally enriched ${}^{235}U$, low concentration of neutron absorbing materials and water presence that acted as moderator), turning into a natural nuclear reactor \cite{uzan2011varying}. The fission continued for a few million years consuming part of its fuel. Imagine what a strange and breathtaking experience it would be to witness the spontaneous beginning of fission in a primordial Earth, millions of years before Fermi managed to start the first man-made nuclear reaction!
The reason why the Oklo phenomenon quickly became relevant for testing the variability of $\alpha$ is the ratio of two isotopes of samarium found in the mine. While the ratio between ${}^{149}Sm$ and ${}^{147}Sm$ is usually $0.9$, the same ratio in the Oklo mine was measured to be around $0.02$ \cite{chiba2011constancy}. This discrepancy in the measurements could be attributed to a change in $\alpha$ \cite{shlyakhter1976direct}, as the Oklo reaction happened in a past enough time. Several studies were conducted on the constraints put on the variability of $\alpha$ by the Oklo phenomenon, leading to the most stringent one of \cite{fujii2000nuclear}:
\begin{equation}
    \frac{\dot{\alpha}}{\alpha} = (0.2 \pm 0.8) \cross 10^{-17} \: yr^{-1}
\end{equation}
It shows that the fine structure constant might indeed vary with time. Together with Dirac's Large Number Hypothesis and the observations made regarding a varying $G$, this result motivates our work on the possibility that further physical constants might vary. However, before mentioning the last notable motivation for varying constants, we will show how a varying $\alpha$ links particularly well with a varying speed of light.
\\
\\
The fine structure constant is defined as \cite{griffiths2005introduction}
\begin{equation}
    \alpha := \frac{e^2}{\hbar c}
\end{equation}
where $e$ is the electron's charge, $\hbar$ is the reduced Planck's constant and $c$ is our beloved speed of light. If, as we have just seen, $\alpha$ varies with time, given its composite structure, a natural question arises: which component of $\alpha$ is varying? Since $\alpha$ is a ratio, the electric charge $e$ could be varying, while $\hbar c$ is constant, a possibility considered by \cite{bekenstein1982fine}. Equally, $e$ could be perfectly constant, while either $\hbar$ or $c$ would be varying. Finally, in principle, all three constants could be varying to give the observed variation in $\alpha$. The main difference between the first and second scenario, and also the main reason why someone would choose to fix $\hbar c$ while making $e$ varying, is that in the first case, Lorentz invariance would be preserved while breaking charge conservation. In the second case, Lorentz invariance would be clearly broken. For simplicity, we leave out the case of a varying $\hbar$, being sure that a varying $c$ would be sufficiently heretic.\\ Therefore, in the context of Varying Speed of Light theories, we will consider that a varying $\alpha$ is actually a varying $c$ with $e$ and $\hbar$ both constants. It is here that a crucial point arises: varying dimensional constants do not have any meaning physically \cite{albrecht1999time}. This is because dimensional constants can only be measured as a ratio to a unit of measurement. It only makes sense to have dimensionless constants varying, as these are pure numbers without dimensions. However, it is possible to consider varying $c$ models where the whole dimensionless fine structure constant is varying, while the ratio of the other dimensional constants is kept fixed. This will be the basis of our next developments, where we will consider a set of constants, labelled $\bm{\beta}$ appearing in the Einstein-Cartan action. We will postulate that these constants vary with respect to physical relational times as defined in section \eqref{UniIntro}, arising from the generalised unimodular term.\\
\\
To conclude, we briefly mention the last main reason why we believe that constants could vary. This is more of a heuristic motivation, extensively based on the more abstract interpretation of the meaning behind the laws of physics. As pointed out by \cite{damour2009equivalence} and \cite{damour2012theoretical}, the appearance of General Relativity in physics drastically changed our understanding of Reality. Neither because of its complete and extended model of gravity nor because of its successful tests, but because of the fundamental change it provided to our logic. It allowed us to change from a strictly static and rigid view of spacetime to a fluid and dynamic one. This freed our view of the world, allowing us to run in the wild landscape of a mutable Reality that led to so many new insights and discoveries. Likewise, we believe that there should not be any limitation on imagining a Universe where constants might change, leading to new and exciting models of Reality.

\subsection{The Speed of Light} \label{different_c}

Speed is conventionally a dimensional quantity ($[\frac{L}{T}]$) that relates perhaps the two most fundamental concepts of Nature: time and space. It is really a conversion factor between the two. The faster something moves through space, the more this space is converted into time, keeping track of the temporal interval required to complete that distance. As such, the concept of speed is intrinsically a variable one, making it a purely dynamical quantity associated with objects, rather than a fundamental one determining the nature of such objects. As an example, it would be illogical to presume that the speed at which an electron is moving is defining the essence of an electron. The electron is such no matter its state of motion, as this can change due to external influences, but its intrinsic properties like charge and mass are not only unique but also, and most importantly, unchangeable. It would be madness to define, to substantiate any particle by the speed at which it propagates in the Universe, as this would be subject to changes, thus not making it an absolute property unique to that particle.
\\
\\
This is not, however, the case for the photon, or, to be more abstract, for any particle lacking mass. Any massless particle travels at the speed of light, and any particle that is travelling at the speed of light must be massless \cite{einstein1905electrodynamics}. As surprising as it might be, what is colloquially called ``the speed of light" is the intrinsic property defining any massless particle, because of a simple but rather incredible occurrence: the speed of light is constant.
How exactly can something like the speed of a particle be always fixed, to the point of becoming an invariant between reference frames? How did it happen that a concept traditionally variable not only became constant but, as such, it became the fundamental property defining intrinsically the nature of a particle, thus ascending to the primary role of constant of Nature? To answer these questions, we need to ask where this speed first appeared, and we need to explore the deeper idea underlying the commonly known speed of light as the speed at which a photon, i.e., electromagnetic waves, propagate in spacetime.\\
\\
Everything began with the early experiments on magnetostatic and electrostatic. Electricity and Magnetism are characterised, as most physical phenomena, by two constants: the electric vacuum permittivity, $\epsilon_0$, and the magnetic vacuum permeability, $\mu_0$. The first one describes the electric field density forming as a result of a charge, while the second one models the strength of the magnetic field formed by moving charges. Interestingly, when combined, they have the same dimensions of a velocity \cite{griffiths2005introduction}. This remained a coincidence until Maxwell unified electricity and magnetism. If we consider the Maxwell's equations \cite{griffiths2005introduction}
\begin{equation}
    \nabla \cdot \bm{E} = -\frac{\rho}{\epsilon_0}
\end{equation}
\begin{equation}
    \nabla \cdot \bm{B} = 0
\end{equation}
\begin{equation}
    \nabla \times \bm{E} = -\frac{\partial \bm{B}}{\partial t} \label{maxwell_3}
\end{equation}
\begin{equation}
    \nabla \times \bm{B} = \mu_0 \biggl[\bm{J} + \epsilon_0 \frac{\partial \bm{E}}{\partial t}\biggl] \label{maxwell_4}
\end{equation}
where $\bm{J}$ is the electric current density, we can take the curl of equation \eqref{maxwell_3} and, using standard vector identities, we obtain 
\begin{equation}
    \nabla^2 \bm{E} = \frac{\partial }{\partial t} (\nabla \cross \bm{B})
\end{equation}
And, using substituting equation \eqref{maxwell_4} inside, we obtain
\begin{equation}
    \nabla^2 \bm{E} = \epsilon_0 \mu_0 \frac{\partial^2 \bm{E}}{\partial t^2}
\end{equation}
Following a similar procedure, we have the same equation for the magnetic field 
\begin{equation}
    \nabla^2 \bm{B} = \epsilon_0 \mu_0 \frac{\partial^2 \bm{B}}{\partial t^2}
\end{equation}
Conclusion? Both equations above have the form of the wave equation \cite{french2003vibrations}, i.e.,
\begin{equation}
    \nabla^2 \bm{x} = \frac{1}{v^2} \frac{\partial^2 \bm{x}}{\partial t^2}
\end{equation}
where $v$ is the speed of the propagation of the wave. By analogy with this equation, it is straightforward to see that 
\begin{equation}
    c_0 = \frac{1}{\sqrt{\epsilon_0 \mu_0}}
\end{equation}
where $c_0$ is now the speed at which every electromagnetic wave propagates in the vacuum. This value is constant, as it is made itself of two constants coming from electric and magnetic phenomena. Strangely enough, electric charges give rise to electromagnetic waves, the speed of which is constant. This is our first encounter with one of the many roles the speed of light takes: it is the fundamental constant of electromagnetism. As we will now see, following \cite{ellis2005c}, there are other roles this concept can take.
\\
\\
The speed of light appears as well in the Lorentz transformations as the spacetime constant, and, to differentiate it from the other roles, following \cite{ellis2005c}, we will label it as $c_{ST}$. If we consider the line element for Minkowski flat space
\begin{gather}
    ds^2 = g_{\mu \nu}dx^\mu dx^\nu \\
    \\
    = -c_{ST}^2 dt^2 + dx^2 + dy^2 + dz^2
\end{gather}
we immediately understand that here the speed of light acts as a fixed conversion factor between time and space coordinates, so that they are fully unified in a 4-vector, forming spacetime coordinates. Furthermore, by fixing the speed of light and assuming nothing travels faster like Einstein did, we can define a causal structure in spacetime. In fact, depending on the sign of the Minkowski line element, two spacetime events can be separated by the following intervals \cite{bohmer2016introduction}
\begin{equation}
    ds^2 < 0 \Leftrightarrow \text{Time-like interval}
\end{equation}
\begin{equation}
    ds^2 = 0 \Leftrightarrow \text{Light-like interval}
\end{equation}
\begin{equation}
    ds^2 > 0 \Leftrightarrow \text{Space-like interval}
\end{equation}
This shows how the speed of light, being a universal invariant constant between reference frames, fixes the causality of events in spacetime: physical phenomena could only be time-like or light-like, while space-like phenomena break causality, as they would propagate faster than light. Lastly, $c_{ST}$ also enters the famous $E = mc_{ST}^2$ formula, becoming a conversion factor between energy and mass. In the next chapters, we will see that $c_{ST}$ enters the gravitational metric $g_{\mu \nu}$ with the same role discussed here (as a causality constant relating space and time), so we will label it $c_g$ to differentiate it from the other types.\\
\\
The third different type of speed of light appears in General Relativity. When considering the Einstein Field Equations, we have that 
\begin{equation}
    G_{\mu \nu} = \frac{8 \pi G}{c_{FE}^4} T_{\mu \nu}
\end{equation}
where this time $c_{FE}$ is a conversion factor between the gravity part of the field equations and the matter content. As discussed in \cite{ellis2005c}, this speed of light is ultimately equivalent to $c_{ST}$ in the Newtonian limit. This equivalence is however only valid in the context of General Relativity, and it will still be important to differentiate it when considering the variability of constants. Finally, \cite{ellis2005c} also shows how the speed of light enters the linearised field equations, being the speed at which gravitational waves propagate. This turns out to be, once again, the same as $c_{ST}$, confirming that the speed of light is actually the maximum speed of propagation of information and thus causality.\\
\\
We have seen how the speed of light plays different roles in physics, ranging from the propagation's speed of electromagnetic waves to the structure constant of spacetime determining causality. By doing so, we understand that the concept of the speed of massless particles is fundamentally related to the maximum speed of propagation of information in general, whether this is via photons or gravitational waves. Having established its different roles, we can now properly differentiate them when considering the variation of $c$, depending on whether this speed of light comes from the Einstein-Hilbert action, from the gravitational metric or from the energy density $\rho$.

\section{Brans-Dicke: a scalar-tensor theory of gravity} \label{BD}

Following our discussion from section (\ref{UniIntro}), we now focus on which base action $S_0$ we could use. Thus far \cite{magueijo2023evolving} \cite{magueijo2021cosmological} \cite{magueijo2022connection}, $S_0$ has always been taken to be the Einstein-Cartan action, leading to a relatively small number of parameters that could be used as $\bm{\alpha}$ or $\bm{\beta}$. Furthermore, there is no reason why, \emph{a priori}, we should avoid using any other physical action as a base theory $S_0$, ranging from quantum field theory ones to the full standard model Lagrangian. These will potentially encounter renormalisation problems related to the constants used (e.g., the distinction between bare and physical mass), but could still be interesting fields of research. Nevertheless, here we will focus on purely gravitational theories, specifically investigating the outcomes of applying the varying constant formalism to the Brans-Dicke theory. There are two main reasons why we chose to use the Brans-Dicke action as $S_0$: its inclusion, by construction, of a varying gravitational constant (via Mach's principle) and the presence of the dimensionless parameter $\omega$. Therefore, we will review the foundational ideas and results of this theory, highlighting how its unique feature nicely connect with the idea of varying constants.

\subsection{Mach's principle}

Imagine to be floating in empty space. You are surrounded by never never-ending void, and the only available reference points are distant fixed stars visible in the background. Suddenly, you start to rotate around yourself, thus experiencing your arms and legs being pulled away from you. This is nothing but the centrifugal force caused by the rotation: your body's inertia is opposing to the rotation, hence the outward force. However, more interestingly, the fixed stars of the background are also rotating, prompting the question: why do you experience centrifugal force when the background stars are rotating? Is there any connection between you and far distant object that seems incapable of influencing you from a distance?
This thought experiment, inspired by \cite{weinberg1972gravitation}, is at the heart of the idea of Mach's principle and its consequences for General Relativity and the Brans-Dicke theory.\\
\\
Another way to formulate this idea is to think about Newton's famous "Bucket Experiment". If we imagine a bucket containing water rotating on itself, we would notice that after some time has passed, the surface of the water will have changed shape. It will go from being completely flat to having a paraboloidal shape, with the water touching the bucket's side higher than the one in the centre. This will be the result of the inertia of the water opposing the motion induced by the rotation. Once the bucket stops rotating, the water's rotation will also slow down, eventually reaching the flat state it originally started from. If, however, the bucket is still but the entire universe around is rotating, Newton's conclusion was that the water on the surface would still be flat, since it is not moving and the universe's rotation would affect the water's inertia.
Disagreeably to Newton, Mach claimed that the water in the bucket would still change shape \cite{mach1893science}, and thus the inertia would be affected, even when the bucket is not moving but the universe around it is.\\
\\
More fundamentally \cite{lichtenegger2008mach}, this thought experiment becomes crucial in the context of the dichotomy between absolute and relative space and time. In fact, considering two buckets filled with water, one rotating and the other not, how can we appreciate their difference? This would certainly be possible by noticing that the water's surface of the rotating bucket is different from the stationary one. But with respect to what? Only by comparing them to an absolute fixed space background, with respect to which one bucket is rotating and the other one is not, we would be able to tell the buckets' states. This evidence led Newton to believe that the rotation \emph{with respect to absolute space} is why the two buckets differ. However, Mach pointed out that an essential assumption in Newton's interpretation of the bucket experiment would radically change its outcomes. In fact, Mach argued that Newton's conclusions were only valid in a hypothetical empty universe, where any source of matter has been removed. Furthermore, these conclusions still hold true only if it is assumed that physical systems retain their properties even when put in full isolation from the rest. However, Mach's key intuition was that such conditions of pure empty space are, at best, ideal, and in the actual physical universe there is always matter present, undermining Newton's central assumption.
Following this line of thought, the two buckets can be compared not relative to absolute space, but rather with respect to the background fixed stars, an example of the matter in the universe. Therefore, this suggests that the difference in the inertia of the two buckets can be observed with respect to the masses of the background stars, establishing a link between the local inertia and the matter content of the universe.\\
This is exactly what we mentioned in the example before: because the inertial mass of an object and the matter content of the universe relate to each other, Mach believed that also the water in the fixed bucket would change its shape if the universe is rotating.\\
\\
Having developed an intuition on how the inertia of rotating objects related to distance references in their background, we can summarise Mach's principle more formally: the geometry of space-time, and hence, the inertial properties of every infinitesimal test particle, is required to be determined by the distribution of mass-energy throughout the universe \cite{marzke1964gravitation} (pag. $303$). On more heuristic and philosophical grounds, a universe subject to Mach's principle would not carry any notion of absolute space and time, but rather it would only allow the notion of quantities such as motion \emph{relatively} to some other reference. This idea is precisely the base of Einstein's theory of General Relativity. In fact, generalising Mach's principle, he postulated a theory of gravity where only relative motion is observable, thus disallowing any privileged reference frame. Furthermore, he postulated that inertia should be related to the gravitational interaction with matter only, and that the geometry and metric structure of the universe is entirely determined by its matter content \cite{raine1975mach}.\\
\\
However, Einstein's implementation of Mach's principle is only partial in his theory. It has been pointed out \cite{singleton2016global} that Mach's principle is, to some extent, conflicting with the other founding principle of General Relativity: the equivalence principle. While the former has somehow a global nature, as it relates the local inertia of a particle to the global energy-momentum tensor of the universe, the latter strictly applies locally, as tidal forces are negligible only locally, since they can differentiate gravitational acceleration from inertial. Moreover, going beyond this tension in the founding principles of the theory, the non-fully-Machian nature of General Relativity arises when considering some particular solutions to the field equations. Specifically, the Minkowski spacetime solution posses inertia but it does not contain any matter \cite{heckman1962relativistic}, the Gödel rotating universe \cite{ozsvath2001approaches} breaks the idea of relative motion introducing a preferred (and hence absolute) direction of motion, and finally the Taub-NUT model provides a curved solution which is singularity free, in contradiction with one formulation of Mach's principle \cite{vishwakarma2015machian}. Motivated by the desire to fully implement Mach's principle in General Relativity, Brans and Dicke decided to develop their theory of scalar-tensor gravity, as we will see in the next section.

\subsection{Varying $G$ and Mach's principle}

The main aim of the Brans-Dicke theory of gravity is to provide an extension of General Relativity fully inclusive of Mach's principle. To do so, a central observation is necessary: Dirac's Large Number Hypothesis \cite{dirac1938new}. As we have illustrated in section (\ref{constant_section}), Dirac noticed a deep connection between cosmological numbers and atomic ones, which lead to the following idea. A striking result is obtained when considering a hollow-mass sphere containing two particles attracting each other gravitationally \cite{sciama1957monthly}. An equation for the inertial force experienced by one of the two particles that is fixed in a rest reference frame can be obtained as 
\begin{equation}
    F_{inertia} = bmMa G^\alpha r^\beta c^\gamma \label{F_1}
\end{equation}
where $b$ is a dimensionless constant set to one, $m$ is the mass of the particle, $M$ is the mass of the spherical shell, $r$ is the radius of the shell and $a$ is the acceleration of the sphere relative to the particle \cite{dicke1959new}. Furthermore, $\alpha$, $\beta$ and $\gamma$ are constants as well. To match the dimensions on the LHS and on the RHS of equation (\ref{F_1}), we must have that $\alpha = 1$, $\beta = -1$ and $\gamma = -2$. Finally, equating equation (\ref{F_1}) to Newton's second law leads to
\begin{equation}
    \frac{G M}{R c^2} \sim 1 \label{G_1}
\end{equation}
where $G$ is the gravitational constant, $c^2$ is the speed of light in the vacuum and $M$ and $R$ are the total mass and radius of the observable universe. Interestingly, expressing (\ref{G_1}) in terms of $G$ leads to
\begin{equation}
    G \sim \frac{R c^2}{M} \label{G_2} 
\end{equation}
which in turn suggests a fundamental relationship between the gravitational constant $G$ and the total mass $M$ of the observable universe. Indeed, recasting (\ref{G_2}) as $G \sim G(M)$, where $G$ becomes effectively a function of the total mass $M$, we clearly see that the gravitational constant can be treated as a function of $M$. This result is exactly the desired implementation of Mach's principle. In fact, we have managed to express mathematically how the geometry of space-time should be determined by the mass distribution of the universe: the inertia experienced by an accelerated observer relative to distant matter is equivalent to a gravitational force acting on a fixed observer due to distant accelerated matter \cite{brans1961mach}.\\
\\
Additionally, relation (\ref{G_2}) suggests that either the ratio $\frac{R}{M} = constant$, or that the gravitational constant should be a locally variable quantity, determined by the mass distribution about the point. Discarding the first possibility due to an expanding universe which requires a changing $R$ implying matter-conservation violations to keep the ratio constant, we have finally arrived at the full implementation of Mach's principle which naturally leads to a variable $G$ dependent on the mass distribution of the Universe.

\subsection{Brans-Dicke Theory}

In this subsection, we review the work presented in \cite{brans1961mach}, highlighting the most relevant results for our future extension. Given the observations in the previous section, it is now natural to think of $G$ as a variable function, changing from position to position depending on the matter distribution. Since it is desirable that this new $G$-field is coordinate independent, it should be chosen to be a scalar field. Furthermore, considering where $G$ appears in the Einstein-Hilbert action (\ref{EHstart}) and setting $c = 1$, we can see that
\begin{equation}
    \phi := \frac{1}{G} \sim \frac{M}{R}
\end{equation}
where the scalar field $\phi$ has the correct dimensions of $G$, i.e., $[\phi] =\frac{M T^2}{L^3}$. We could, in the spirit of varying constants, not set $c=1$ above and actually see that a varying speed of light might lead to a varying $G$, while $M$ and $R$ are constant. However, we will not consider this case here, as it will be the object of future work \cite{bassani2023}.  It is now possible to precede and generalise action (\ref{EHstart}) including a scalar field of the variation of $G$ with respect to the coordinates.
We start from the Einstein-Hilbert action inclusive of a matter Lagrangian term
\begin{equation}
    S = \int{ d^4 x \: \sqrt{-g} \: \biggl[R + \frac{16 \pi G}{c^4} L_m \biggl] } \label{BD_1}
\end{equation}
where $L_m$ is the matter Lagrangian density containing all non-gravitational fields. 
We then multiply action (\ref{BD_1}) by $G^{-1}$, substitute in the Ricci scalar term $\phi$ as the varying gravitational constant, and include a general Lagrangian density for the scalar field, depending both on the field and its derivatives \cite{brans1961mach}
\begin{equation}
    S = \int d^4 x \: \sqrt{-g} \: \biggl [\phi R + \frac{16 \pi}{c^4}L_m + L_{\phi}(\phi, \partial_\mu \phi) \biggl] \label{BD_2}
\end{equation}
This new action contains the scalar field $\phi$ which is coupled to mass and geometry. This, therefore, leads to a non-purely geometrical theory of gravity, as part of gravitational phenomena are described by the scalar field coupled to the Ricci scalar. Additionally, given its scalar nature, we might expect that $\phi$ obeys some second-order wave equation such as 
\begin{equation}
    \Box{\phi} = \rho_{scalar} \label{scl}
\end{equation}
where $\rho_{scalar}$ acts as a scalar matter source term for (\ref{scl}). Therefore, following this and minimal dimensional consistency, we can arrive at a form for $L_\phi$ which would give us, when varied, the desired wave equation for $\phi$:
\begin{equation}
    L_\phi = - \frac{\omega}{\phi} g^{\mu \nu} \: \partial_\mu \phi  \: \partial_\nu \phi
\end{equation}
which can then be inserted into (\ref{BD_2}), leading to the full Brans-Dicke action
\begin{equation}
    S_{BD} = \frac{c^4}{16 \pi}\int{d^4 x \: \sqrt{-g} \: \biggl [\phi R - \frac{\omega}{\phi} g^{\mu \nu} \: \partial_\mu \phi  \: \partial_\nu \phi \biggl]} + S_m  \label{BD_3}
\end{equation}
where $S_m$ is a general matter action.\\
The parameter $\omega$ is a dimensionless coupling constant which can take any value, made such by including the factor $\phi^{-1}$. Also, it is possible to recover Einstein's theory in the limit where $\omega \rightarrow \infty$.
The Lagrangian density term introduced for the scalar field has the form of a kinetic term for $\phi$. This is actually the case, as this term will provide the dynamical evolution of the scalar field. Furthermore, the kinetic term's introduction assures the full diffeomorphism's invariance of the action and the conservation of the energy-momentum tensor via $\nabla_\mu T^{\mu \nu} = 0$ \cite{almeida2021quantum}. It is also possible to include a general potential $V(\phi)$ \cite{papagiannopoulos2017dynamical}, thus arriving at an action with full kinetic and potential terms for $\phi$, but, in the following, we will not include it.\\
\\
We can now proceed to derive the field equations of Brans-Dicke theory in the Jordan frame using action (\ref{BD_3}). Firstly, we derive the field equation for $\phi$, by varying (\ref{BD_3}) with respect to $\phi$ and $\partial_\mu \phi$
\begin{equation}
    \Box \phi = \frac{g^{\mu \nu} \: \partial_\mu \phi \: \partial_\nu \phi}{2 \phi} - \frac{\phi R}{2 \omega} \label{phi_1}
\end{equation}
where the d'Alambertian operator is defined as 
\begin{equation}
    \Box \phi = \frac{1}{\sqrt{-g}} \partial_\mu [\sqrt{-g} \: \partial^\mu \phi]
\end{equation}
Secondly, using equation (\ref{phi_1}), we can vary action (\ref{BD_3}) to obtain the modified field equations as 
\begin{equation}
    \frac{\delta S_{BD}}{\delta g^{\mu \nu}} = 0 \Leftrightarrow \frac{\delta S_g}{\delta g^{\mu \nu}} + \frac{\delta S_m}{\delta g^{\mu \nu}} = 0
\end{equation}
where $S_g$ is the gravity action inclusive of the Brans-Dicke kinetic term, while $S_m$ is a generic matter action giving the energy-momentum tensor as 
\begin{equation}
    \frac{\delta S_m}{\delta g^{\mu \nu}} = -\frac{1}{g}\frac{\delta g}{\delta g^{\mu \nu}} L_m - 2\frac{\delta L_m}{\delta g^{\mu \nu}} := T_{\mu \nu}
\end{equation}
where $g = det(g_{\mu \nu})$ and the variations above have been obtained using the chain rule. Thus, we arrive at the field equations:
\begin{equation}
    R_{\mu \nu} - \frac{1}{2}R g_{\mu \nu} - \frac{\omega}{\phi^2} \biggl ( \partial_\mu \phi \: \partial_\nu \phi -\frac{1}{2}g_{\mu \nu} \: \partial_\alpha \phi \: \partial^{\alpha} \phi \biggl) - \frac{1}{\phi} \biggl(\nabla_\mu \partial_\nu \phi - g_{\mu \nu} \Box \phi \biggl) = \frac{8 \pi}{\phi c^4}T_{\mu \nu} \label{FE_BD}
\end{equation}
where the third term on the RHS is obtained by including $g^{\mu \nu}$ explicitly in action (\ref{BD_3}) to account for a more general curved space-time. Obviously, the field equations (\ref{FE_BD}), as expressed here, should be interpreted as $\phi$ adding geometrical contributions to (\ref{FE_BD}) to model the full gravitational field (metric plus scalar field). Conversely, if the terms involving $\phi$ are moved to the RHS of the equations, we see that they present a relation between space-time geometry and a matter content not only given by the usual $T_{\mu \nu}$ but also by a scalar field $\phi$.
Finally, we can arrive at a more compact and evocative form for equation (\ref{phi_1}) by considering contraction of the field equations (\ref{FE_BD}) with $g^{\mu \nu}$. This leads to
\begin{equation}
    R = -\frac{8 \pi}{\phi c^4}T_{\mu \nu} - \frac{\omega}{\phi^2} + \frac{\omega}{\phi^2} \partial_\alpha \phi \: \partial^\alpha \phi + \frac{3}{\phi} \Box \phi
\end{equation}
which is then used for the Ricci scalar in equation (\ref{phi_1}) to obtain
\begin{equation}
    \Box \phi = \frac{8 \pi}{2 \omega + 3} T \label{final_phi}
\end{equation}
where $T = g^{\mu \nu}T_{\mu \nu}$ is the trace of the energy-momentum tensor.
Equation (\ref{final_phi}) clearly shows that the $\phi$ field has as source term the matter content of the solutions, described by the energy-momentum tensor. Therefore, it is reasonable to assume that the scalar field acts as an auxiliary geometrical contribution to the usual field equations, justifying its appearance on the LHS. Furthermore, equation (\ref{final_phi}) has the desired form of a wave equation for $\phi$, as we initially hoped for in (\ref{scl}), providing us with an expression for the evolution of $\phi$. Specifically, it is interesting to consider a solution for (\ref{final_phi}) found for a quasi-flat universe with matter content $\rho$:
\begin{equation}
    \phi = 8 \pi \frac{4 + 3 \omega}{6 + 4 \omega} \biggl(\frac{t}{t_0} \biggl)^{\frac{2}{4 + \omega}} \rho_0 t_0
\end{equation}
where $\rho_0$ is the matter density at a given time $t = t_0$. Noteworthily, this solution is compatible with (\ref{G_1}); when the matter density is small, $G = \phi^{-1}$ becomes large, which physically implies the disappearance of inertia which is precisely Mach's principle's prescription. Therein the inertia of a particle is directly related to the universe's matter content. This result justifies Brans and Dicke's intentions of including Mach's principle in a theory of gravitation by generalising Einstein's theory.\\
\\
Finally, efforts have been made to measure the value of the parameter $\omega$ to understand if, physically, the Brans-Dicke theory is distinguishable from General Relativity. The two theories are equivalent in the limit $\omega \rightarrow \infty$, so a high value of $\omega$ would suggest that Brans-Dicke is essentially equivalent to General Relativity. Recent observations made with the Cassini Saturn Probe \cite{bertotti2003test} set the value of $\omega$ at $\omega > 4 \times 10^4$, much greater than the value close to unity proposed by Brans and Dicke. This suggests that, at least in the Solar System, Brans-Dicke is indistinguishable from General Relativity, so the latter is preferred. However, the question of which theory is correct on cosmological scales is still open and more observations are required.

\section{Overview and Aims} \label{overview_section}

In this work, we will study three ideas that are, in appearance, separated, unifying them to develop several varying constants scenarios. To begin, we consider the speculation that Natural constants might not be that constant after all. We investigate what the Universe would look like if the value of a certain physical constant evolved with time. This possibility has been initially considered by Dirac in his Large Number Hypothesis \cite{dirac1938new} and has, ever since, attracted increasing interest, particularly in the cases of a varying $G$ and $\alpha$. More recently \cite{albrecht1999time}, Albrecht and Magueijo have developed a theory in which the speed of light can vary, called the Varying Speed of Light (VSL) theory. Besides providing an alternative explanation to the open cosmological problems,  VSL created a new conceptual landscape: if $c$ can vary along with $G$ and $\alpha$, imagination is really the only limit to which constants vary.\\
\\
It is precisely in this context that the second idea considered in this work arises. How can we provide a rigorous mathematical framework describing the evolution of multiple physical constants and their respective evolutions' times? The answer is promptly provided by the idea of unimodular gravity. Initially formalised by Einstein as a first attempt to solve the Cosmological Constant problem, it was extended by Henneaux and Teitelboim (HT) to include a more general framework for $\Lambda$ to be a mere constant on-shell. Following their idea \cite{magueijo2023evolving}, Magueijo generalised the HT formalism to allow any constant appearing in the Einstein-Cartan to vary. Furthermore, HT formalism provides excellent definitions of relational times \cite{magueijo2023evolving} that can be used as a parameter for the evolution of our constants. Therefore, the generalisation of unimodular gravity, as presented in section \eqref{UniIntro}, turns out to be the perfect mathematical platform to develop varying constants theories, rigorously exploring this idea and its consequences.\\
\\
We have now the heuristic, the \textit{why} of our work lying in the idea of varying constants and the method, the machinery, the \textit{how} built from the unimodular gravity formalism. All we are missing is the \textit{what}: what are we applying all of these ideas and mathematics to? For the idea of varying constants, one of the earliest attempts at formalising it in a theory of gravity was the Brans-Dicke theory. Hoping to develop an extension of General Relativity that would fully incorporate Mach's principle, Brans and Dicke proposed a varying gravitational constant  \cite{brans1961mach}, formalised as a scalar field called $\phi$. Motivated by their early insight, this dissertation extends the unimodular formalism to the Brans-Dicke theory, producing original results in the field of varying constants. Specifically, the aim of this work is an open-ended exploration of energy conservation in a Brans-Dicke cosmological model of the Universe. Therefore, the main results in this dissertation will be various energy conservation scenarios given by different varying constants or combinations of them. Furthermore, as part of a future publication \cite{bassani2023}, some of the results obtained in this dissertation will be analysed in greater detail, providing insights and possibly a solution to the Cosmological Constant problem.\\
\\
To conclude, this dissertation is structured as follows. In Chapter 1, we have introduced the necessary background to our results. We review the unimodular formalism and show how its generalisation provides a development platform for the idea of varying constants. We provide extensive motivations for the variation of natural constants, ranging from more speculative ones to actual experimental pieces of evidence. We also review the Brans-Dicke theory, developing the intuitions behind it, as well as its main results and observational constraints.\\
Secondly, in Chapter 2, we review the main applications of VSL and the cosmological problem it aims to solve. Furthermore, we provide an extensive derivation of the minisuperspace formalism for both the Einstein-Hilbert and Einstein-Cartan actions. Also, we present the foundational results obtained in \cite{magueijo2023evolving} applying the varying constant idea to energy conservation scenarios in the Universe.\\
Finally, in Chapter 3, we present the original results of this research. They include several scenarios of varying constants applied to Brans-Dicke cosmologies, leading to different energy conservation laws. We also lay the foundations of our future developments in the Cosmological Constant problem. Let the fun begin!

\clearpage{\pagestyle{empty}\cleardoublepage}

%%%%%%%%%%%%%%%%%%%%%%%%%%%%%%%%%%%%
%%%%%%%%%%%%%%%%%%%%%%%%%%%%%%%%%%%%
\chapter{Varying Constants: Developments}

In this chapter we present an additional motivation supporting the idea of varying constants and we provide examples of its implementation in the conservation of energy on cosmological scales. Firstly, we show how currently open problems in cosmology can be solved by the Varying Speed of Light theory. This provides not only a straightforward application of the idea of varying constants, but it also supports the heuristic behind why the constant of Nature should be varying. Secondly, we derive the Einstein-Hilbert and the Einstein-Cartan actions in minisuperspace (MSS), showing their equivalence when the torsion-free condition is assumed. This will give the Hamiltonian which will be used to obtain our results. Finally, we combine the MSS form of the Einstein-Cartan action with the unimodular generalisation presented in section \eqref{UniIntro} to obtain multiple energy conservation scenarios for different varying constants. This last section will be predominantly a review of \cite{magueijo2023evolving}, which will form the basis for our original results in Chapter 3.

\section{Open problems in Cosmology}

In this section, we present the main open problems in the standard Big Bang cosmology. These are the Horizon, the Flatness and the Cosmological Constant problems. They are normally explained with a period of accelerated expansion in the early stages of the Universe, called inflation. Inflationary cosmologies provide solid models to account for the Horizon and Flatness problems, while they can partially explain the Cosmological Constant one \cite{martin2012everything}. Among their main shortcomings is the requirement of a physical mechanism to drive inflation which is, to this date, still experimentally undetermined \cite{borde1996singularities}. We therefore present an alternative explanation to inflation, the Varying Speed of Light theory, where the speed of light is assumed to have transitioned from a high value in the early Universe to the lower one we observe today. This theory provides solid explanations to these problems, opening the possibility to alternative cosmological models.

\subsection{Horizon problem}

The Cosmological Principle \cite{thorne2000gravitation} postulates that the spacial distribution of matter in the Universe is homogeneous, i.e., equally distributed, and that, when considered from large scales, it appears to be isotropic, meaning that it should look the same from every direction. Homogeneity and isotropy are at the heart of modern cosmology and represent the central assumption behind the Friedmann-Lemaître-Robertson-Walker (FLRW) metric. Besides it being assumed as a first principle, and therefore not requiring, $\textit{a priori}$, any justifications, observations \cite{fixsen1994cosmic} have confirmed a very isotropic universe. In fact, measuring the Cosmic Microwave Background Radiation temperature of very distant regions of the universe, it was possible to establish an outstanding similarity between them, setting their temperature to $T_0 = 2.726 \pm 0.010$ \cite{fixsen1994cosmic}. This value turns out to be remarkably similar through different regions of the universe that are causally-disconnected. To illustrate this, lets remember that, according to the Big Bang cosmology, the Universe has, in this moment, a finite age, called $t_0$. If we consider the cosmological distance defined by this age, $ct_0$, we have the furthest distance some photons could have travelled to us. Obviously, this maximum distance stretches from Earth into every direction, but light coming from one direction to the Earth could not have had enough cosmic time to travel to the opposite direction. In other words, light from a region of the universe cannot affect (is not in causal contact) with another directly opposite region. Nevertheless, both regions appear to be perfectly homogeneous, showing a remarkable equivalence of their Cosmic Microwave Background Radiation temperature. \\
\\
Lets illustrate this issue in more detail. We consider a photon originating from the surface of last scattering, when the phase transition between a very hot to a colder universe happened, leading to the recombination era which ultimately allowed photons to start to prorate freely into the Universe. For simplicity, we will assume a radial motion, and, given our working assumption, we will use the FLRW metric to model the early homogeneous and isotropic Universe, which takes the form \cite{d2022introducing}
\begin{equation}
    ds^2 = - c^2 N(t)^2 dt^2 + a(t)^2 \biggl[ \: \frac{dr^2}{K(r)^2} + r^2(d\theta^2 + \sin{\theta}^2 d\phi^2) \biggl]
\end{equation}
where $a(t)$ is the scale factor, $K(r) = \sqrt{1-kr^2}$ and $k$ is the Universe's curvature, which we will assume to be flat, giving $K(r) = 1$. Given radial motion, the line element above simplifies to
\begin{equation}
    ds^2 = dt^2 + a^2 dr^2 = 0 \label{www}
\end{equation}
since for radial motion $d\theta = d\phi = 0$ and we have assumed $N(t) = 1$ and $c = 1$. Rearranging \eqref{www} for the distance travelled in a given time interval, we obtain 
\begin{equation}
    \int^{r_2}_{r_1} {dr} = \int^{t_2}_{t_1}{\frac{dt}{a(t)}} \label{dd}
\end{equation}
where $t_1$ is the time the photon was emitted and $t_2$ is the time when it is received, which in our case is the current age of the universe. Furthermore, considering the Robertson-Walker line element for the Universe's spacial slice only at a fixed time $t = t_0$ \cite{d2022introducing}, we have 
\begin{equation}
    d\sigma^2 = a_0^2 dr^2 \Leftrightarrow dr = \frac{d\sigma}{a_0}
\end{equation}
where $a_0 = a(t_0)$. Substituting this expression into \eqref{dd}, we obtain
\begin{equation}
    \Delta r = a(t_2) \int^{t_2}_{t_1}{\frac{dt}{a(t)}}
\end{equation}
where the integration of the 3D spacial line element $d\sigma$ gives the total distance travelled by the photon $\Delta r$ (i.e., its proper distance) and we choose the time $t_0$ to be the present age of the Universe coinciding with the photon's detection time $t_2$. The time $t_1$ is the emission of the photon, which could be at $t_1 = 0$, i.e., the Big Bang, or at $t_1 = t_ls$, i.e., when the photo left the surface of last scattering. In the first case,  this expression gives us the maximum proper distance a photon emitted at the Big Bang has travelled, and its called the particle horizon. 
To obtain an explicit expression for the maximum proper distance travelled, we assume a radiation dominated universe, as it was in its very early stages until approximately $47 \: 000$ years after the Big Bang, for which $a(t) = C t^{\frac{1}{2}}$, where $C$ is some proportionality constant. Therefore, if $t_2 = t_0 = t$, the current age of the universe and $t_1 = 0$, we get
\begin{equation}
    \Delta r = Ct^{\frac{1}{2}} \int^t_0 {\frac{du}{Cu^{\frac{1}{2}}}} = 2t \label{rad_era}
\end{equation}
On the other hand, if we would like to obtain the same expression for the particle horizon in the matter dominated era, we need to consider the scale factor for matter, which is $a(t) = D t^{\frac{2}{3}}$, where $D$ is once again a proportionality constant. This era started $47 \: 000$ years after the Big Bang, but it was not until $380 \: 000$ \cite{gorbunov2011introduction} years that photons could finally freely travel into the rest of the Universe. This event is called the surface of last scattering, marking the time from which we are receiving the first photons from the early Universe. Therefore, the maximum distance traveled by a photon from the surface of last scattering time until now is given by 
\begin{equation}
     \Delta r = Dt^{\frac{2}{3}} \int^t_0 {\frac{du}{Cu^{\frac{2}{3}}}} = 3t
\end{equation}
where, in this case, $t_1 = 0$ coincides with the time of last scattering and not with the Big Bang. Focusing on the surface of last scattering, we can re-insert the speed of light in expression \eqref{rad_era} to obtain the particle horizon for causal interaction in the early universe as \cite{bohmer2016introduction}
\begin{equation}
    \Delta r = 2ct \approx 0.27 Mpc
\end{equation}
This means that at the time of last scattering, causal interaction between photons could only happen if they were separated by a distance equal or inferior to $0.27 Mpc$. If, at that time, photons had a greater separation, they could have not possibly interacted, thus evolving independently until our time. However, as stated above, we observe that causally disconnected regions of the Universe appear to be extremely homogeneous. This is exactly the Horizon problem: how could causally disconnected regions of the Universe evolve independently to the point of having essentially the same radiation's temperature? As we will see in the next subsection, inflation will provide an answer, but not the only one.

\subsection{Flatness problem}

Another shortcoming of the current cosmological model is the Flatness problem. When defining the Universe's total energy density as \cite{d2022introducing}
\begin{equation}
    \Omega_{total} = \Omega + \Omega_{\Lambda}
\end{equation}
we can re-express the first Friedmann equation as 
\begin{equation}
    \Omega_{total} - 1 = \frac{k}{a^2 H^2} \label{fried_man}
\end{equation}
where $k$ is the curvature and $H = \frac{\dot{a}}{a}$ is the Hubble's parameter. If we have a flat universe (one for which $k=0$) we immediately see from \eqref{fried_man} that $\Omega_{total}$ must equal to one at all times for the first Friedmann equation to hold. This is the case for the current universe, where since
\begin{equation}
    \Omega_{m} + \Omega_{\Lambda} \approx 1
\end{equation}
the total energy density is close to the critical one. If, however, the curvature is not zero, $\Omega_{total}$ evolves with time because of the dependence of $a^2 H^2$ in \eqref{fried_man}. To understand better this time evolution, as we have seen above, we consider the dependence of the scale factor on cosmological time. If we consider either a radiation dominated or matter dominated universe, we find that 
\begin{equation}
    aH \propto t^{-\frac{1}{2}} \Leftrightarrow \Omega_{total} - 1 \propto t
\end{equation}
\begin{equation}
    aH \propto t^{-\frac{1}{3}} \Leftrightarrow \Omega_{total} - 1 \propto t^{\frac{2}{3}}
\end{equation}
where the first equation applies to the evolution of $\Omega_{total}$ in a radiation dominated Universe, while the second equation applies to a matter dominated one. In both cases, we notice that the total energy density increases with time. This time dependence implies that, since we have a total energy density close to unity today, it should have been the same in the early universe, otherwise, over time, it would have evolved to be a different value. In fact, estimates on nucleosyntesis constrain the value of the total energy density at \cite{bohmer2016introduction} 
\begin{equation}
    |\Omega_{total} - 1| \le 10^{-16}
\end{equation}
which indeed confirms the closeness to one. For earlier times, this value would be even smaller. This is the flatness problem: this initial condition on $\Omega_{total}$ so that the universe would evolve to have a curvature $k = 0$ is way too specific. If this condition would have been slightly different, the Universe would have evolved radically different from how we observe it today. Therefore, we conclude that such fine-tuned initial conditions are hard to explain in the current cosmological models.

\subsection{Cosmological constant problem} \label{cosmo_const_prob}

The last and most important open problem in cosmology is the Cosmological Constant one. It would be restrictive to regard it as a purely cosmological problem, as it arises from the comparison of measurements and predictions coming from very different fields of physics. In fact, as we have seen, General Relativity predicts the existence of a cosmological constant in the field equations as a geometrical contribution, responsible for the dynamical nature of the universe. This results in including the cosmological constant on the LHS of the field equations as
\begin{equation}
    G_{\mu \nu} + \Lambda g_{\mu \nu} = \frac{8 \pi G}{c^4}T_{\mu \nu} \label{eoo}
\end{equation}
where we can define the energy density of the cosmological constant to be 
\begin{equation}
    \rho = \frac{\Lambda c^2}{8 \pi G} \label{insertion}
\end{equation}
This is regarded as the the purely geometrical cosmological constant problem \cite{albrecht1999time}, and it is distinct from the vacuum energy problem we will see below.
\\
\\
On the other hand, following a different approach, it is possible to consider the cosmological constant as a source term for the field equations, thus moving $\Lambda$ on the RHS of equation \eqref{eoo}: this is the approach were the cosmological constant represents the vacuum energy density \cite{padilla2015lectures}. In this case, the cosmological constant becomes part of the energy momentum tensor giving the spacetime geometry due to the Universe's matter content, and it can be equivalently treated as the vacuum energy density $\rho_{vac}$ \cite{abbott1988mystery}. This vacuum energy density is made up of three main components: the bare cosmological constant, the quantum fluctuations and contributions form particles and interactions that are yet not accounted for in the Standard Model \cite{abbott1988mystery}. In fact, according to quantum field theory \cite{srednicki2007quantum}, the vacuum has a non-zero energy density due to particle creation and annihilation: this energy, as any other form of matter, creates curvature, modifying the geometrical structure of spacetime. Since this energy density contributes to curvature, it is measurable with cosmological observations, and, given its appearance in the theory, it is also theoretically predictable. However, there is a big inconsistency between these two values.
\\
\\
Following \cite{weinberg1989cosmological}, since the energy momentum tensor must respect Lorentz invariance, it must take the form 
\begin{equation}
    T_{\mu \nu} = -g_{\mu \nu} \rho_{vac}
\end{equation}
Comparing this result with \eqref{eoo}, we see that this requirement is equivalent to adding a new term to the total vacuum energy density as
\begin{equation}
    \rho_{total} = \rho_{vac} + \frac{\Lambda c^2}{8 \pi G}
\end{equation}
According to cosmological measurements \cite{carroll2001cosmological}, this value has an upper bound give by 
\begin{equation}
    |\rho_{total}| \le 10^{-48} \: GeV^4 \label{rho_exp}
\end{equation}
On the other hand, the theoretically predicted vacuum energy is the sum of different contributions, ranging from potential energies of scalar fields to zero-point fluctuations of each field theory, as well as a bare cosmological constant contributions \cite{carroll2001cosmological}. Assuming that quantum field theory is still valid at the Planck's scale, we obtain a theoretical prediction on the vacuum energy as
\begin{equation}
    \rho_{total} \sim 10^{72} \: GeV^4 \label{rho_teor}
\end{equation}
The ratio of \eqref{rho_teor} to \eqref{rho_exp}, the theoretical value to the experimental one, is indeed the cosmological constant problem
\begin{equation}
    \frac{\rho_{total}^{theory}}{\rho_{total}^{obs}} \sim 10^{120}
\end{equation}
This huge discrepancy of $120$ orders of magnitudes between theory and observations is the core of the cosmological constant problem as presented here.

\subsection{The solution: Varying Speed of Light}

The problems exposed above have all one aspect in common: they arise from the standard Big Bang cosmology, where inflation has been an attempt at solving them. While successful on many grounds, inflation requires a negative pressure fluid as well as its fundamental mechanism is still unknown \cite{borde1996singularities}. An alternative solution to the problems exposed above could be the Varying Speed of Light theory, firstly explored by Moffat \cite{moffat1993superluminary} and then extended by Albrecht and Magueijo \cite{albrecht1999time}. Before diving into the detail of how VSL solves the three problems mentioned above, we will briefly discuss how inflation is modelled and how it tackles these issues.
\\
\\
The theory of inflation was first proposed by Alan Guth while trying to explain the absence of observations of magnetic monopoles. Postulating the idea of inflation, it was quickly realised that it could solve the horizon and flatness problems \cite{guth1981inflationary}, as well as explain how a universe dominated by an attractive force (gravity) could be expanding \cite{turok2002critical}. 
Inflation could have plenty physical origins, but the common feature of all inflationary models is the presence of an exotic form of energy for which the pressure is negative, i.e., $p = -\rho$, thus having a repulsive-like force driving expansion against gravity.\\   A common example for inflation is a scalar field \cite{d2022introducing}. The action for a scalar field  with kinetic and potential energy is 
\begin{equation}
    S = \int {d^4 x \: \sqrt{-g} \: \biggl[\frac{1}{2}g^{\mu \nu} \nabla_\mu \phi \nabla_\nu \phi + V(\phi) \biggl]}
\end{equation}
From this we obtain the scalar field's pressure and energy density as
\begin{equation}
    p_{\phi} = \frac{1}{2}\dot{\phi}^2 - V(\phi)
\end{equation}
\begin{equation}
    \rho_{\phi} = \frac{1}{2}\dot{\phi}^2 + V(\phi)
\end{equation}
which combined give 
\begin{equation}
    \rho + 3p = 2 (\dot{\phi}^2 - V(\phi)) \label{subsub}
\end{equation}
where, if we assume the slow-roll approximation (the kinetic energy is small), the potential $V(\phi)$ dominates, giving indeed a negative pressure as $p \approx -\rho$.
On the other hand, if we consider the Einstein Fields Equations, assuming a perfect fluid as matter content, we obtain the second Friedmann equation 
\begin{equation}
    \Ddot{a} = -\frac{4 \pi G}{3} \biggl (\rho + \frac{3p}{c^2}\biggl)a \label{fried_man_2}
\end{equation}
In standard cosmology, we usually have a positive energy density and pressure, making the RHS of \eqref{fried_man_2} negative, so that the scale factor can only decrease over time. This scenario is however not the same if, in the slow-roll approximation, we use equation \eqref{subsub} with $V(\phi) \approx \rho_{\phi}$ such that equation \eqref{fried_man_2} becomes \cite{turok2002critical}
\begin{equation}
    \Ddot{a} \approx \frac{8 \pi G \rho}{3}a \equiv H_{I}^2 a
\end{equation}
where $H_{I} = \frac{8 \pi G \rho}{3}$ is the Hubble parameter at inflation's epoch. 
Solving this equation for the scale factor, we immediately get that
\begin{equation}
    a \propto e^{\pm H_{I}t}
\end{equation}
This solutions shows that the universe's scale factor is experiencing a period of exponential evolution, confirming that a matter content with negative pressure does indeed produce a repulsive force. We now review the three issues above in the context of VSL, highlighting how its proposed solutions relate to inflationary models.

\subsubsection{Solution to the Horizon Problem}

The first problem we consider is the Horizon one. Inflation provides a straightforward solution, assuming the universe underwent a period of accelerated expansion. If the early universe experienced a period in which with $\Ddot{a} > 0$, regions that are now observed to be causally disconnected were indeed in causal contact, justifying their extreme homogeneity in temperature we observe today.\\
Interestingly, either of two mutually exclusive possibilities would perfectly address the Horizon problem: either spacetime is expanding subliminally while the speed of light has the same value as now (i.e., inflation), or, equivalently, spacetime is not accelerating but the speed of light had a value several orders of magnitudes greater that it does now \cite{albrecht1999time}. We can therefore imagine that light travelled much faster in the early Universe then it does now, thus allowing causal contact for regions that now are disconnected because $c$ travels slower. Furthermore, the speed of light underwent a sharp phase transition at time $t_c$, where its value changed from $c^{-}$, the early universe value, to $c^{+}$, the value we measure today. If we express the universe's horizon size in terms of conformal time $\eta_c$, we obtain
\begin{equation}
    r_{h} = c^{-}\eta_c
\end{equation}
Then, if $c^-$ is indeed greater then $c^+$, their ratio compares to the one of the conformal time now $\eta_0$ and the conformal time at $t_c$ as
\begin{equation}
    \frac{c^-}{c^+} \gg \frac{\eta_0}{\eta_c} \label{c_plus}
\end{equation}
because, since we do not have inflation in  the early Universe, $\eta_0$ and $\eta_c$ are close in value. Expression \eqref{c_plus} implies that $r_h \gg r$, meaning that the entire observable Universe today has always been in causal constant, since its radius $r$ is smaller then the radius of the horizon. We have therefore shown how a varying speed of light with a phase transition can explain the Horizon problem.

\subsubsection{Solution to the Flatness Problem}

Our next topic is the Flatness problem. In the inflationary model, if we define a total density parameter as \cite{turok2002critical}
\begin{equation}
    \Omega = \frac{8 \pi G \rho}{3 H^2}
\end{equation}
where $H = \frac{\dot{a}}{a}$ is the Hubble parameter, we immediately see that as $a$ increases, for fixed $\rho$, $\Omega$ tends to unity, thus explaining why the current Universe's curvature is observed to be close to $k = 1$. A different solution can be found by allowing $c$ to vary in the Friedmann equations and in the conservation equation. These, inclusive of any $c$ factor usually set to one, are 
\begin{equation}
    \biggl(\frac{\dot{a}}{a} \biggl) = \frac{8 \pi G \rho}{3} - \frac{k c^2}{a^2}
\end{equation}
\begin{equation}
    \Ddot{a} = -\frac{4 \pi G}{3} \biggl (\rho + \frac{3p}{c^2}\biggl)a 
\end{equation}
\begin{equation}
    \dot{\rho} + 3\frac{\dot{a}}{a} \biggl(\rho + \frac{p}{c^2}\biggl) = -\rho \frac{\dot{G}}{G} + 3\frac{k c^2}{4 \pi G a^2 }\frac{\dot{c}}{c}
\end{equation}
where the first two equations are derived from the Einstein field equations in a FLRW metric, while the third one can either be the result of a combination of the two Friedmann equations or it can be directly derived from $\nabla_\mu T^{\mu \nu}$ assuming a perfect fluid. Now we can define $\epsilon = \Omega - 1$, where $\Omega = \frac{\rho}{\rho_c}$ is the total energy density as above, but expressed in terms of critical density $\rho_c$, which reads
\begin{equation}
    \rho_c = \frac{3}{8 \pi G} \biggl (\frac{\dot{a}}{a} \biggl)^2
\end{equation}
Therefore, considering the evolution of $\epsilon$ we obtain
\begin{equation}
    \dot{\epsilon} = (1+ \epsilon) \biggl (\frac{\dot{\rho}}{\rho_c} - \frac{\dot{\rho_c}}{\rho_c} \biggl)
\end{equation}
Finally, using the Friedmann equations for $\rho$ and the conservation equation for $\dot{\rho}$ we can get an explicit expression for $\epsilon$ as
\begin{equation}
    \dot{\epsilon} = \epsilon \frac{\dot{a}}{a}(1+ \epsilon)(1 + 3w) + 2 \frac{\dot{c}}{c}\epsilon \label{epsi_lon}
\end{equation}
where the equation of state is $p = w \rho c^2$. In this formulation of the Flatness problem, we need $\epsilon$ going to zero, such that $\Omega$ goes to one. If, as before, we assume a sharp phase transition for the value of the speed of light from $c^-$ to $c^+$, we see that
\begin{equation}
    \left|\frac{\dot{c}}{c}\right| \gg \frac{\dot{a}}{a} \label{a_c}
\end{equation}
This allows us to neglect the $\frac{\dot{a}}{a}$ term in equation \eqref{epsi_lon}, leading to
\begin{equation}
    \frac{\dot{\epsilon}}{\epsilon} = 2 \frac{\dot{c}}{c} \Leftrightarrow \log{\epsilon} = 2\log{c} \Leftrightarrow \epsilon \propto c^2 \label{drusillosaurissimo}
\end{equation}
This shows the dependency between the total energy density and the speed of light, confirming that a sharp change is $c$ would indeed drive $\epsilon$ to zero, giving $\Omega = 1$. This equivalently solves the Flatness problem in cosmology.

\subsubsection{Solution to the Cosmological Constant Problem}

Finally, we consider the Cosmological Constant problem. As we have seen in subsection \eqref{cosmo_const_prob}, this problem is two-fold, meaning that we can discriminate the geometrical one from the vacuum energy problem. While the first one can be solved using VSL theory, the second one is more complicated \cite{padilla2015lectures} and will require more thoughts and consideration. Therefore, following \cite{albrecht1999time}, we have a conservation equation for matter and cosmological constant content, of the form
\begin{equation}
    \dot{\rho_m} + 3\frac{\dot{a}}{a} \biggl(\rho_m + \frac{p_m}{c^2} \biggl) = -\dot{\rho_\Lambda} -\rho \frac{\dot{G}}{G} + \frac{3kc^2}{4 \pi G a^2 } \frac{\dot{c}}{c} \label{ue}
\end{equation}
We can now define
\begin{equation}
    \epsilon := \frac{\rho_\Lambda}{\rho_m} \label{epsi_lambd}
\end{equation}
for $\Lambda$ and, assuming that $\Lambda$ is actually constant, by taking the time derivative of \eqref{insertion} we obtain 
\begin{equation}
    \frac{\dot{\rho_\Lambda}}{\rho_\Lambda} = 2\frac{\dot{c}}{c} - \frac{\dot{G}}{G} \label{ddddd}
\end{equation}
We now would like an equation for the evolution of $\epsilon_\Lambda$. Taking the time derivative of \eqref{epsi_lambd} we have
\begin{align}
    \dot{\epsilon_\Lambda} = \frac{\dot{\rho_\Lambda}}{\rho_m} - \frac{\rho_\Lambda}{\rho_m^2}\dot{\rho_m} \nonumber\\ 
    =& \: \epsilon_\Lambda \biggl[2\frac{\dot{c}}{c} - \frac{\dot{G}}{G} -\frac{\dot{\rho_m}}{\rho_m}\biggl]
\end{align}
where in the second line we have used \eqref{ddddd} expressed for $\dot{\rho_\Lambda}$. Finally, using the conservation equation \eqref{ue} for $\dot{\rho_m}$, the equation of state and $\epsilon = \Omega - 1$ we arrive at 
\begin{equation}
    \dot{\epsilon_\Lambda} = \epsilon_\Lambda \biggl[2\frac{\dot{a}}{a}(1+w) +2\frac{\dot{c}}{c}\frac{1+\epsilon_\Lambda}{1 + \epsilon}\biggl] \label{eps_i_lo_nlmbd}
\end{equation}
In the standard inflationary cosmology, we of course have $\dot{c} = 0$. Current estimates and observations set $\epsilon_\Lambda$ very close to one \cite{martel1998likely}, but this cannot be explained by inflation. In fact, the solution $\epsilon_\Lambda = 0$ whether $w > -1$ or $w = -1$ would still allow $\epsilon_\Lambda$ to grow after inflation, thus not explaining its current value. However, this is not the case if we assume $\dot{c} \neq 0$. Once again, assuming a sharp phase change in $c$ such that \eqref{a_c} is also valid in this case, we can see that the second term in equation \eqref{eps_i_lo_nlmbd} dominates. The condition that 
\begin{equation}
    \frac{\dot{c}}{c} \ll 0
\end{equation}
would indeed drive $\epsilon_\Lambda$ to zero, thus implementing a mechanism to explain the small value of $\epsilon_\Lambda$ today.
Therefore, assuming condition \eqref{a_c}, we find that 
\begin{equation}
    \frac{\dot{\epsilon_\Lambda}}{\epsilon_\Lambda (1+ \epsilon_\Lambda)} = 2\frac{\dot{c}}{c}\frac{1}{1 + \epsilon}
\end{equation}
which, combined with equation \eqref{drusillosaurissimo} gives the evolution of $\epsilon_\Lambda$ in terms of $\epsilon$
\begin{equation}
    \frac{\epsilon_\Lambda}{1+ \epsilon_\Lambda} \propto \frac{\epsilon}{1 + \epsilon}
\end{equation}
Finally, if we assume, as initial conditions $\epsilon \approx 1$ and $\epsilon_\Lambda \approx 1$, we obtain 
\begin{equation}
    \epsilon_\Lambda \propto c^2
\end{equation}
which indeed solves the geometrical cosmological constant problem by showing that a  function $c(t)$ with a steep phase transition between $c^-$ and $c^+$ would lead $\epsilon_\Lambda$ to one, the present observed value.

\section{Minisuperspace Reduction} \label{MSS_section}

In this section, we provide explicit derivations of the Einstein-Cartan action in minisuperspace (MSS) for the Friedmann-Lemaître-Robertson-Walker (FLRW) metric. The same is done for the Einstein-Hilbert action. Finally, in the third subsection, we show how the Einstein-Cartan action is equivalent to the Einstein-Hilbert one in minisuperspace under the assumption that the torsion is set to zero. We also show to to derive the on-shell condition that relates the conjugate variables $a^2$ and $b$ in MSS. This section is essential as it will provide the starting point to construct the entire dynamics of varying constants. We will then extend these results using the Einstein-Cartan action for the Brans-Dicke, reducing it to minisuperspace in Chapter 3.

\subsection{Einstein-Cartan action in MSS} \label{EC_sect}

The Einstain-Cartan theory is an extension of General Relativity that allows to include torsion in spacetime, relating it to the density of intrinsic angular momentum of particles \cite{trautman2006einstein}. Furthermore, the Einstein-Cartan formalism of gravity is equivalent to the Einstein-Hilbert one, assuming that the tetrads are non-degenerate ($det \: e^a_\mu \neq 0$) and that the torsion is zero.
Therefore, we will proceed deriving the Einstein-Cartan (EC) action in minisuperspace. The EC action is \cite{gronwald1996gauge}

\begin{equation}
    S_{EC} = \frac{c^4}{32 \pi G} \int{\epsilon_{abcd}  \: e^a e^b R^{cd}} \label{EC_1}
\end{equation}
where $\epsilon_{abcd}$ is the Levi-Civita tensor, $e^a$ is the non-coordinate tetrad basis defined as
\begin{equation}
    e^a = e^a_\mu dx^\mu 
\end{equation}
and $R^{cd}$ is the Einstein-Cartan curvature given by 
\begin{equation}
    R^a_b = \frac{1}{2} R^a_{b \mu \nu} dx^\mu dx^\nu
\end{equation}
To proceed, we need to introduce Cartan's first and second equations, which will allow us to find explicit expressions for the tertad which, given line element (\ref{FLRW}), we can plug into action (\ref{EC_1}). Cartan's first equation, which gives the torsion 2-form, dropping the bases $dx^\mu$, is \cite{gronwald1996gauge}
\begin{equation}
    T^a = De^a = de^a + \Gamma^a_b e^b
\end{equation}
where $\Gamma^a_{\mu b} dx^\mu$ is the connection 1-form coming from the exterior derivative, satisfying the property $\Gamma_{a b} = - \Gamma_{b a}$. Assuming the torsion-free condition, i.e., $T^a = 0$, we get 
\begin{equation}
    de^a = -\Gamma^a_b e^b \label{first_C_E}
\end{equation}
On the other hand, we have the curvature 2-form given by Cartan's second equation as \cite{gronwald1996gauge}
\begin{equation}
    R^a_b = d\Gamma^a_b + \Gamma^a_c \Gamma^c_b \label{second_C_E}
\end{equation}
Finally, given the tetrads, we have, by definition,
\begin{equation}
    ds^2 = \eta_{a b} e^a \otimes e^b \label{metric_E_1}
\end{equation}
With these equations, the EC action can be reduced to minisuperspace for the Friedmann-Lemaître-Robertson-Walker (FLRW) metric. We start from the FLRW metric in 
$(t, r, \theta, \phi)$ coordinates \cite{d2022introducing}
\begin{equation}
    ds^2 = - c^2 N(t)^2 dt^2 + a(t)^2 \biggl[ \: \frac{dr^2}{K(r)^2} + r^2(d\theta^2 + \sin{\theta}^2 d\phi^2) \biggl] \label{FLRW}
\end{equation}
where $N(t)$ is the lapse function, $a(t)$ is the scale factor, $K(r) = \sqrt{1-kr^2}$ and $k = {-1, 0, 1}$ is the universe's curvature, which could be locally modelled by a 3-sphere ($\bm{\mathcal{S}^3}, k =+1$), by flat euclidean space ($\bm{\mathcal{E}^3}, k = 0$) and by hyperbolic space ($\bm{\mathcal{H}^3}, k =-1$). It is important to note the inclusion of the speed of light $c^2$ in the time component of the line element. This will be important in the next sections, where we require $c$ to vary. Hence, it is crucial to have it included in the Hamiltonian.\\
Therefore, using metric (\ref{FLRW}) and equation (\ref{metric_E_1}), we can obtain the tetrads as \cite{{bruno}}
\begin{equation}
    e^0 = cN dt
\end{equation}
\begin{equation}
    e^1 = \frac{a}{K} dr
\end{equation}    
\begin{equation}
    e^2 = a \: r d\theta
\end{equation}
\begin{equation}
    e^3 = a \: r \sin{\theta} d\phi
\end{equation}
We can then compute the exterior derivatives of these tetrads, giving
\begin{equation}
    de^0 = 0
\end{equation}
\begin{equation}
    de^1 = \frac{b}{ac}e^0 \wedge e^1
\end{equation}
\begin{equation}
    de^2 = \frac{b}{ac}e^0 \wedge e^2 + \frac{K}{ar}e^1 \wedge e^2
\end{equation}
\begin{equation}
    de^3 = \frac{b}{ac}e^0 \wedge e^3 + \frac{K}{ar}e^1 \wedge e^3 + \frac{\cot{\theta}}{ar}e^2 \wedge e^3
\end{equation}
Then, using equation (\ref{first_C_E}), we arrive at the expression for the connections 
\begin{equation}
    \Gamma^1_0 = \frac{b}{ac}e^1 = \frac{b}{Kc} dr\label{11}
\end{equation}
\begin{equation}
    \Gamma^2_0 = \frac{b}{ac}e^2 = \frac{b}{c} d\theta \label{44}
\end{equation}
\begin{equation}
    \Gamma^3_0 = \frac{b}{ac}e^3 = \frac{b}{c} r \sin{\theta} d\phi  \label{55}
\end{equation}
\begin{equation}
    \Gamma^2_1 = K d\theta \label{22}
\end{equation}
\begin{equation}
    \Gamma^3_1 = K \sin{\theta} d\phi
\end{equation}
\begin{equation}
    \Gamma^3_2 = \cos{\theta} d\phi \label{33}
\end{equation}
where $b = \frac{\dot{a}}{N}$ is the connection variable, as we will see later when we show how the EC action reduces to the EH action. Importantly, the $de^0$ term does not lead to any contribution and the expressions form (\ref{11}) to (\ref{44}) are given by the $de^1$, $de^2$ and $de^3$ terms which also give expressions from (\ref{22}) to (\ref{33}). In fact, as an example, if we take $de^2$ we see that
\begin{equation}
    de^2 = -\Gamma^2_b e^b = -\Gamma^2_0 e^0 - \Gamma^2_1 e^1 -\Gamma^2_2 e^2 -\Gamma^2_3 e^3
\end{equation}
The $\Gamma^2_2$ and $\Gamma^2_3$ terms do not contribute, whereas the $\Gamma^2_0$ will give 
\begin{equation}
    \Gamma^2_0 = \frac{b}{ac} e^2
\end{equation}
as one of the contributions, while the $\Gamma^2_1$ term will give expression (\ref{22}) as 
\begin{equation}
    -\Gamma^2_1 e^1 = \frac{K}{ar} e^1 \wedge e^2 \Leftrightarrow \Gamma^2_1 = K d\theta 
\end{equation}
Now we can use equation (\ref{second_C_E}) to obtain the curvature 2-form $R^a_b$. To begin, we compute the exterior derivative of expressions (\ref{11}) to (\ref{22})
\begin{equation}
    d\Gamma^1_0 = \frac{\dot{b}}{Nac^2}e^0 e^1
\end{equation}
\begin{equation}
    d\Gamma^2_0 = \frac{\dot{b}}{Nac^2}e^0 e^2 + \frac{b K}{r a^2 c}e^1 e^2
\end{equation}
\begin{equation}
    d\Gamma^3_0 = \frac{\dot{b}}{Na c^2}e^0 e^3 + \frac{b K}{r a^2 c}e^1 e^3 + \frac{b \cot{\theta}}{r a^2 c}e^2 e^3
\end{equation}
\begin{equation}
    d\Gamma^3_1 = - \frac{k}{a^2} e^1 e^3 + \frac{K \cot{\theta}}{r^2 a^2} e^2 e^3
\end{equation}
\begin{equation}
    d\Gamma^3_2 = - \frac{1}{a^2 r^2}e^2 e^3
\end{equation}
where $d\Gamma^2_1 = 0$ because $K$ is constant. Given these, we can use equation (\ref{second_C_E}) to obtain the curvature 2-forms
\begin{equation}
    R^0_i = \frac{\dot{b}}{Nac}e^0 e^i
\end{equation}
\begin{equation}
    R^i_j = \frac{b^2 + kc^2}{a^2}e^i e^j
\end{equation}
Finally, we expand the EC action as 
\begin{equation}
    \epsilon_{abcd}  \: e^a e^b R^{cd} = 2\epsilon_{0ijk}e^0 e^iR^{jk} + 2\epsilon_{ij0k}e^i e^j R^{0k}
\end{equation}
and using the equations for the curvature 2-forms we obtain
\begin{equation}
    \epsilon_{abcd}  \: e^a e^b R^{cd} = 2 \epsilon_{0ijk}e^0 e^i e^j e^k \biggl[\frac{\dot{b}}{Na} + \frac{b^2 + kc^2}{a^2}  \biggl]
\end{equation}
which, ultimately, plugged into action (\ref{EC_1}) gives the Einstein-Cartan action reduced to minisuperspace for the FLRW metric \cite{magueijo2020equivalence}
\begin{equation}
    S_{EC} = \frac{3 V_c}{8 \pi G} \int{dt \: [\dot{b}a^2 + Na (b^2 + kc^2)]} \label{EC_final_final}
\end{equation}
where $V_c = \int {dr d\theta d\phi \frac{r^2 \sin{\theta}}{K}}$ is the Universe's co-moving volume.
We now derive the Einstein-Hilbert action in minisuperspace and  show how it is equivalent to the Einstein-Cartan action we found above.

\subsection{Einstein-Hilbert action in MSS}

We now reduce the Einstein-Hilbert action to minisuperspace for the Friedmann-Lemaître-Robertson-Walker (FLRW) metric. This will allow us to then use the Hamiltonian formalism to derive the canonical variable's equations of motion, as well as any other dynamics of the system.
We begin from the Einstein-Hilbert action with $\Lambda = 0$ for simplicity \cite{d2022introducing}
\begin{equation}
    S_{EH} = \frac{c^4}{16 \pi G} \int{d^4 x \: \sqrt{-g}R} \label{agg_EH}
\end{equation}
We now need to derive expressions for the Ricci scalar and for the determinant of the metric in action (\ref{agg_EH}) using the FLRW metric (\ref{FLRW}). We start form the Ricci scalar, which can be computed via 
\begin{equation}
    R = g^{\mu \nu} R_{\mu \nu} = g^{tt}R_{tt} + g^{rr}R_{rr} + g^{\theta \theta}R_{\theta \theta} + g^{\phi \phi}R_{\phi \phi} \label{Ricci-scl}
\end{equation}
where the summations only applies to the diagonal components because the metric is zero everywhere else and where $g^{\mu \nu}$ are the components of the inverse metric and $R_{\mu \nu}$ is the Ricci tensor defined \cite{bohmer2016introduction} as 
\begin{equation}
    R_{\mu \nu} = \partial_\rho \Gamma^{\rho}_{\mu \nu} - \partial_\nu \Gamma^{\rho}_{\mu \rho} + \Gamma^{\sigma}_{\mu \nu}\Gamma^{\rho}_{\sigma \rho} - \Gamma^{\sigma}_{\mu \rho} \Gamma^{\rho}_{\sigma \nu} \label{Ricci_tnr}
\end{equation}
where the $\Gamma^{\rho}_{\mu \nu}$ are the Christoffel symbols. Following (\ref{Ricci_tnr}), we obtain the four non-zero components of the Ricci tensor as
\begin{equation}
    R_{tt} = -\frac{3}{N^2 c^2}\frac{\Ddot{a}}{a}
\end{equation}\\
\begin{equation}
    R_{rr} = \frac{1}{N^2} \biggl [\frac{a \Ddot{a} + 2 \dot{a}^2 + kc^2 N^2}{1 -kr^2} \biggl]
\end{equation}\\
\begin{equation}
    R_{\theta \theta} = \frac{r^2}{N^2}  [ a \Ddot{a} + 2 \dot{a}^2 +kc^2 N^2]
\end{equation}\\
\begin{equation}
    R_{\phi \phi} = \frac{r^2 \sin^2{\theta}}{N^2} [a \Ddot{a} + 2 \dot{a}^2 + kc^2 N^2]
\end{equation}\\
Adding these components following (\ref{Ricci-scl}), we arrive at the expression for the Ricci scalar
\begin{equation}
    R = 6\frac{k c^2}{a^2} + \frac{6}{N^2} \biggl [\frac{\Ddot{a}}{a} + \frac{\dot{a}^2}{a^2} \biggl] \label{ricci_ricci}
\end{equation}\\
where in all the above we have set $\dot{N} = 0$ assuming a constant lapse function. This result agrees with \cite{borowiec2022scalar}, including the speed of light factor. Additionally, we can calculate the determinant of the metric, thus obtaining\\
\begin{equation}
    \sqrt{-g} = \frac{r^2 \sin^2{\theta}}{\sqrt{1- kr^2}}cNa^3 \label{sq_met}
\end{equation}
At this point, it is important to remember that we may write action (\ref{agg_EH}) as
\begin{equation}
    S_{EH} = \frac{\kappa}{2} \int {d^4 x \: \sqrt{-g} R}
\end{equation}
where $\kappa = \frac{c^4}{8 \pi G}$. As it is normally done, we can set $\kappa = 1$ and so, using this form of the action, combining (\ref{ricci_ricci}) and (\ref{sq_met}) we arrive at
\begin{equation}
    S_{EH} = 3 V_c \int{dt \: \frac{1}{N}\dot{a}^2 a + \frac{1}{N} \Ddot{a}a^2 + Na kc^2} \label{EH_beginning}
\end{equation}
where $V_c = {dr d\theta d\phi}\frac{c\:r^2 \sin^2{\theta}}{K}$ is the spacial co-moving volume expressed in terms of polar coordinates. It will be convenient, however, to include the factor of $\frac{1}{8}$ from the constant $k$ into action (\ref{EH_beginning}). In fact, this factor will become important in the next section, where we will show how the Einstein-Cartan action can lead to the Einstein-Hilbert one with the addition of a boundary term. Therefore, the final form of the EH action in MSS that we will use is
\begin{equation}
    S_{EH} = \frac{3 V_c}{8 \pi G} \int{dt \: \biggl[\frac{1}{N}\dot{a}^2 a + \frac{1}{N} \Ddot{a}a^2 + Na kc^2 \biggl]} \label{eh_22}
\end{equation}
From this, we will compare our results in this section with the ones in subsection (\ref{EC_sect}) to show how they are equivalent.

\subsection{From Einstein-Cartan to Einstein-Hilbert in MSS}

From the previous subsections, we have shown how to derive the EC and the EH actions in MSS for the FLRW metric. We now wish to understand mainly two features of these results: how does the connection $b$ appears in the EC action and how can we obtain the EH action from it. In fact, having assumed the torsion-free condition as we did, it is expected that the two actions are equivalent \cite{gronwald1996gauge}.
We begin by considering the EC action
\begin{equation}
    S_{EC} = \frac{3 V_c}{8 \pi G} \int{dt \: [ \dot{b}a^2 + Na (b^2 + kc^2)]} \label{act_EC}
\end{equation}
which is now integrated by parts to obtain an $\dot{a}$ variable, hopefully bringing us closer to the EH action. The results in 
\begin{equation}
    S_{EC} = \frac{3 V_c}{8 \pi G} \int{dt \: [-\dot{(a^2)}b + Na (b^2 + kc^2)] + boundary \: term} \
\end{equation}
where we stress the importance of including the boundary term. In fact, this will turn out to be crucial to obtain the EH action. As it is usually done, we might simply assume that the boundary term vanishes at infinity. This, however, especially in the context of quantum cosmology, is not a good choice \cite{vilenkin1986boundary}. Therefore, instead of this assumption, we add to action (\ref{act_EC}) a counter-term, such that it is the same as the boundary term but with opposite sign:
\begin{equation}
    S_{EC} = \frac{3 V_c}{8 \pi G} \int{dt \: [\dot{b}a^2 + Na (b^2 + kc^2)]} + \frac{3 V_c}{8 \pi G} [a^2 b]^{t_f}_{t_i} \label{bt_ghy}
\end{equation}
This addition will automatically cancel the boundary term once action (\ref{act_EC}) is integrated by parts, removing the necessity to assume vanishing terms at infinity. Furthermore, it is a perfectly valid term from a dynamical point of view, as it does not affect the equations of motion in any way: they are in fact the same with or without it. This additional term is known as the Gibbons-Hawking-York (GHY) boundary term \cite{york1986boundary}.
We will further explain how this procedure is key to obtain the EH action, but 
we firstly focus on the first question, i.e., how $b$ appears in the EC action and how is it related to the other canonical variable $a^2$.
Therefore, including the GHY boundary term and integrating by parts, action (\ref{bt_ghy}) now reads
\begin{equation}
    S_{EC} = \frac{3 V_c}{8 \pi G} \int{dt \: [-\dot{(a^2)}b + Na (b^2 + kc^2)] }\label{bt_no}
\end{equation}
Varying this action,
\begin{equation}
    \frac{\delta S_{EC}}{\delta b} = 0 \Leftrightarrow - \dot{(a^2)} + 2Nab = 0 \label{aa}
\end{equation}
and, expressing equation (\ref{aa}) for $b$, we obtain 
\begin{equation}
    2a \dot{a} = 2Nab \Leftrightarrow b = \frac{\dot{a}}{N}
\end{equation}
which is exactly the expression for $b$ we used when deriving the EC action in MSS. Firstly, this condition only holds on-shell, meaning that it is relevant at the level of the equations of motion. Secondly, we can now provide some intuition on what the conjugate variables $a^2$ and $b$ really are. The scale factor is, in the FLRW, directly linked to the metric $g_{\mu \nu}$ of our spacetime manifold. On the other hand, the connection $\Gamma$ contains first derivatives of the metric. Projecting this link in MSS, we see that $b$ is indeed given by first derivatives of the scale factor $a$. Therefore we conclude that in MSS the scale factor represents the metric and its conjugate $b$ the connection. Hence the name connection we used in the previous subsection for $b$.
\\
\\
Finally, we can show the equivalence of the EC and EH actions. To begin, using the fundamental theorem of calculus and the expression for $b$, we can write the GHY boundary term as 
\begin{equation}
    \frac{3 V_c}{8 \pi G} [a^2 b]^{t_f}_{t_i} = \frac{3 V_c}{8 \pi G} \frac{\partial}{\partial t} \int^{t_f}_{t_i}{dt \: \biggl[ a^2 \frac{\dot{a}}{N}\biggl]} =  \frac{3 V_c}{8 \pi G} \int^{t_f}_{t_i}{dt \: \biggl[2a \frac{\dot{a}^2}{N} + \frac{a^2 \Ddot{a}}{N} \biggl]} \label{ftc}
\end{equation}
Then, plugging $b = \frac{\dot{a}}{N}$ inside action (\ref{bt_ghy}) and including expression (\ref{ftc}), we get
\begin{equation}
    S_{EC} = \frac{3 V_c}{8 \pi G} \int{dt \: \biggl[\frac{-\dot{a}\dot{(a^2)}}{N} + \frac{a \dot{a}^2}{N} + Nakc^2 + 2a \frac{\dot{a}^2}{N} + \frac{a^2 \Ddot{a}}{N} \biggl]}
\end{equation}
which, when simplified, gives 
\begin{equation}
    S_{EC} = \frac{3 V_c}{8 \pi G} \int{dt \: \biggl[\frac{1}{N}\dot{a}^2a + \frac{1}{N}\Ddot{a}a^2 + Nakc^2\biggl]} \equiv S_{EH}
\end{equation}
which is exactly the Einstein-Hilbert action we found in equation (\ref{eh_22}). Therefore, we now can appreciate the importance of the GHY boundary term: without it, it would not be possible to link the EC action to the EH action. The same case would arise if we were to just assume that the boundary term vanishes at infinity.\\
To conclude, we can easily see that, starting from the EH action (\ref{eh_22}), if we substitute the on shell condition $b = \frac{\dot{a}}{N}$ we get 
\begin{equation}
    S_{EH} = \frac{3 V_c}{8 \pi G} \int{dt \: \dot{b}a^2 + Na (b^2 +kc^2)} \equiv S_{EC} \label{EH_finall}
\end{equation}
which is indeed the Einstein-Cartan action reduced to minisuperspace for the FLRW metric. This result shows that the EC and EH actions are essentially equivalent assuming the torsion-free condition and using the GHY boundary term.

\section{Energy Conservation} \label{first_stone}

In this section, we present the current developments in the field of varying constant theories. These arise as a natural extension of VSL when additional constant appearing in the EC action are assumed to be varying. Furthermore, unlike VSL, these theories see a full implementation of the unimodular formalism applied to minisuperspace, which allows a far more general treatment of the subject. Specifically, after the  EC action's reduction to MSS done in section \eqref{MSS_section}, it is possible to generalise the unimodular formalism form section \eqref{UniIntro} to account for a vector of constants $\bm{\alpha}$, giving relational physical times which provide the evolution of another set of constants, called $\bm{\beta}$. The aim of this extension is, however, different from VSL. In fact, by postulating multiple varying constants, we consider the overall energy conservation in the Universe, developing models in which we can have net energy violation or net energy conservation, depending on the constants involved. These models allow us to explore more the consequences of braking Local Lorentz Invariance, as much as they provide interesting insights on possible solutions to the Cosmological Constant problem (more will be explained in the last two chapters). Therefore, in the next subsections, we provide the basic results we will use to extend this formalism to the Brans-Dicke theory in Chapter 3.

\subsection{Full action and constants}

Having the EC action in minisuperspace, we can now include matter in out treatment and show how to derive the equations of motion from the Hamiltonian direclty.
From before, we have the EC action in MSS, which we now write as 
\begin{equation}
    S_{EC} = V_c \int{dt \: \alpha_2 \biggl[\dot{b}a^2 + Na(b^2 + kc_g^2)  \biggl]} \label{starting_point}
\end{equation}
We now label the speed of light that was appearing in action (\ref{EC_final_final}) as $c^2$ differently, specifically as $c^2_g$. This is because, as we have discussed in subsection (\ref{different_c}), we should differentiate the speed of light originating from the gravity metric (hence $c^2_g$) from other speeds of light, like the one from a matter content (labelled $c^2_m$). Secondly, we have introduced the constant $\alpha_2$ \cite{magueijo2023evolving}, which is defined as 
\begin{equation}
    \alpha_2 = \frac{3 c^2_p}{8 \pi G_P} \label{alpha_22}
\end{equation}
where $c^2_P$ and $G_P$ are the speed of light and the gravitational constant originating from Planck's mass. In fact, going back to the EC action in MSS, we can re-write it as 
\begin{equation}
     S_{EC} = \frac{3 V_c M_P^2}{\hbar c_P} \int{dt \: [\dot{b}a^2 + Na (b^2 + kc^2_g)]}
\end{equation}
where $M_P^2 = \frac{\hbar c_P}{8 \pi G_P}$ is the Planck's mass. For dimensional reasons, we defined $\hbar$ such that it gives a $c_P^2$ factor in the numerator, hence obtaining the squared speed of light in the expression (\ref{alpha_22}) for $\alpha_2$. We remark that in action (\ref{EC_final_final}) we really have two distinct and different speeds of light. One, $c^2_P$ is the speed of light coming from the Planck's mass, which is itself appearing in front of the EC  action together with $G_P$ because of the required dimensions of the action. This is ultimately the reason why we use the subscript $P$ for the speed of light and the gravitational constant. On the other hand, we have the speed of light $c^2_g$ coming from the FLRW metric, which is clearly a gravitational-type constant, since it is originating from a gravitational metric.\\
Continuing, we include the unimodular action, which, according to the procedure outlined in section (\ref{UniIntro}), should be combined with the base theory $S_0$, which is the EC action in this case. This gives 
\begin{equation}
    S_U = V_c \int{dt \: \dot{\bm{\alpha}} \: \bm{T_\alpha}}
\end{equation}
where we have integrated by parts action (\ref{starting}) to have an expression where there is a dynamical constant multiplying a conjugate time. As we will see shortly, $\bm{\alpha}$ is a vector of constants appearing in the EC action. The inclusion of the unimodular action will allow us to create the relational physical times giving the constants' evolution.
Finally, we wish to include a generic matter action in our analysis. Considering a perfect fluid Lagrangian \cite{iyer1997lagrangian} \cite{brown1993action} and reducing it to MSS as in \cite{gielen2020singularity} we arrive at
\begin{equation}
    S_m = V_c \int{dt \: \alpha_3 [\dot{m_i}\psi_i - Na^3 \rho]}
\end{equation}
where $\alpha_3$ is another constant appearing in from of the matter action, taking the form
\begin{equation}
    \alpha_3 = \frac{G_M}{G_P}
\end{equation}
Interestingly, as pointed out in \cite{brown1993action}, the canonical pair $\dot{m_i} \psi_i$ in the matter action above forms already a unimodular-like term, with $m_i$ acting as a constant of motion.\\
Once again, we differentiate the gravitational constant originating from the Planck's mass (and therefore related to the gravity action) from $G_M$, which is defining the gravitational coupling to matter, as explained in subsection \eqref{different_c}. Here, we allow the matter action to accommodate for different types of fluids, hence the index $i$ allows to move from one species to the other. Furthermore, $m_i$ is the constant of motion dual to the canonical variable $\psi_i$ defined in \cite{magueijo2023evolving}. Finally, the total energy density $\rho$ is defined as 
\begin{equation}
    \rho = \rho_\Lambda + \rho_M = \rho_\Lambda + \sum_i \rho_i
\end{equation}
where the $\rho_i$ are the different fluids' densities given by 
\begin{equation}
    \rho_i = \frac{m_i}{a^{3(1+w_i)}} \label{rho_substitution}
\end{equation}
with $w_i$ being the constant relating the pressure to the density in the equation of state.\\
Having defined all the actions that will contribute to our final theory, it is important to comment on their constants. In fact, the constants $\alpha_2$ and $\alpha_3$ we found for the EC and matter actions are all part of the $\bm{\alpha}$ vector appearing in the unimodular action. This vector is then
\begin{equation}
    \bm{\alpha} = (\alpha_1, \alpha_2, \alpha_3, ...) = \biggl (\rho_\Lambda, \frac{3c^2_P}{8 \pi G_P}, \frac{G_M}{G_P}, ... \biggl) \label{alpa}
\end{equation}
where we have included also $\alpha_1 = \rho_\Lambda$ in case we wish to consider a purely unimodular time dependence. Furthermore, we could have additional constants appearing in the $\bm{\alpha}$ vector, as we will do in Chapter 3, so $\bm{\alpha}$ is not limited to these only. These constants will give rise to their relational times defined above as the $\bm{T_\alpha}$. On the other hand, the set of $\bm{\beta}$ constants will the one varying with respect to these clocks as we have outlined in section (\ref{constant_section}).
\\
We can now add these three actions together, arriving at the final form of the full action
\begin{equation}
    S_{full} = V_c \int{dt \: \alpha_2 \dot{b}a^2 + \alpha_3 \dot{m_i}\psi_i + Na\alpha_2 (b^2 + kc_g^2) - Na^3 \alpha_3 \rho + \dot{\rho_\Lambda}T_\Lambda} \label{action_1}
\end{equation}
where, to begin with, we will consider the cosmological constant $\Lambda$ as our relational time. Before deriving the Hamiltonian and the equations of motion, it is crucial to stress a key point. In the action above, the vacuum energy $\rho_\Lambda$ is included in the matter content together with the other matter species. Therefore, it appears in the matter Lagrangian together with $\alpha_3$. However, this case would be different if we were to consider a geometrical Cosmological Constant, where we would have a $\alpha_2$ factor appearing in front of the matter Lagrangian.

\subsection{The Hamiltonian and the Dynamics} \label{fixed_alpha}

Using action (\ref{action_1}) from the previous subsection, we can directly infer the Hamiltonian of the system. In fact, action (\ref{action_1}) is, explicitly, the integral of the Lagrangian 
\begin{equation}
    S = \int{dt L}
\end{equation}
and we can relate the Lagrangian of a system to its Hamiltonian via a Legendre transform \cite{lowenstein2012essentials}
\begin{equation}
    L(q, \dot{q}, t) =  \sum_{i=1}^{n} p_i \dot{q}^i - H(p, q, t) \label{def_h_l}
\end{equation}
where the $p_i$ are the generalised momenta and the $\dot{q_i}$ are the generalised velocities. The generalised momenta are defined as
\begin{equation}
    p_i = \frac{\partial L}{\partial \dot{q}_i}
\end{equation}
so we immediately see that action (\ref{action_1}) is in the form required by (\ref{def_h_l}) to obtain the Hamiltonian. This therefore leads to 
\begin{equation}
    H = - Na \alpha_2 (b^2 + kc_g^2) +  \alpha_3 \rho Na^3 \label{H_1}
\end{equation}
Firstly, due to $\alpha_2$ in front of the canonical pair in action (\ref{action_1}), the conjugate of $b$ needs to be defined as $A^2 = a^2 \alpha_2$ instead of simply $a^2$. This distinction is very important when evaluating the Poisson's brackets for $H$, unless $\alpha_2$ is kept constant in the canonical pair. Doing so completely brakes Local Lorentz Invariance and generates two different theories, as we will see in the next subsections and in Chapter 3 for Brans-Dicke. For now, we keep $\alpha_2$ fixed, while still writing the Hamiltonian (\ref{H_1}) in terms of $A^2$ as
\begin{equation}
    H = - NA \sqrt{\alpha_2} (b^2 + kc_g^2) + \alpha_3 \alpha_2^{-\frac{3}{2}} \rho NA^3 \label{H_2}
\end{equation}
We obtain an Hamilton constraint on the dynamics of our variable by varying action (\ref{action_1}) with respect to N, 
\begin{equation}
    \frac{\delta S_{full}}{\delta N} = 0 \Leftrightarrow b^2 + kc_g^2 = \frac{\alpha_3}{\alpha_2} \rho a^2 \label{const_11}
\end{equation}
The Poisson's brackets for two functions $f(p_i, q_i, t)$ and $g(p_i, q_i, t)$ are defined as \cite{lowenstein2012essentials}
\begin{equation}
    \{f, g \} = \sum_{i=1}^{N} \biggl( \frac{\partial f}{\partial q_i} \frac{\partial g}{\partial p_i} - \frac{\partial f}{\partial p_i} \frac{\partial g}{\partial q_i}   \biggl) \label{hamilton}
\end{equation}
Applying this definition to Hamiltonian (\ref{H_2}), we arrive at the first Hamilton's equation, the equation of motion for $a$
\begin{equation}
    \dot{a} = \{a, H\} = \frac{\partial a}{\partial b} \frac{\partial H}{\partial A^2} - \frac{\partial a}{\partial A^2} \frac{\partial H}{\partial b} = - \frac{\partial a}{\partial A^2} \frac{\partial H}{\partial b}
\end{equation}
where the first term vanishes because the variable $a$ does not depends on $b$. Furthermore, using the trick 
\begin{equation}
    \frac{\partial a}{\partial A^2} = \frac{\partial a}{\partial (a^2 \alpha_2)} = \frac{1}{\alpha_2}\frac{1}{2}\frac{1}{a} \frac{\partial a}{\partial a}
\end{equation}
since, in this case, $\alpha_2$ is kept fixed, we arrive at 
\begin{equation}
    \dot{a} = \{a, H\} = Nb \label{a_dot_1}
\end{equation}
Moreover, using once again (\ref{hamilton}) for $b$ we have
\begin{equation}
    \dot{b} = \{b, H\} = \frac{\partial b}{\partial b}\frac{\partial H}{\partial A^2} - \frac{\partial b}{\partial A^2}\frac{\partial H}{\partial b} = \frac{1}{2a \alpha_2}\frac{\partial H}{\partial a}
\end{equation}
where again the second term vanishes because $b$ does not depend on $A^2$. Therefore, using the Hamilton constraint (\ref{const_11}) we arrive at
\begin{equation}
    \dot{b} = \{b, H\} = -\frac{\alpha_3}{2\alpha_2} (\rho +3p)Na \label{b_dot_1}
\end{equation}
where we have also used (\ref{rho_substitution}) for $\rho$ in the matter term of the Hamiltonian, such that we obtain a term of the form 
\begin{equation}
   - (1 + 3w)\rho = -\rho -3w\rho = - (\rho + 3p) \label{rho_plus_p}
\end{equation}
where we invoke the equation of state of a perfect fluid $p = w\rho$.
\\
\\
It is also possible to obtain the equations of motion following the Lagrangian approach. This implies using the Euler-Lagrange equation for the $a^2$ and the $b$ variables. Starting from the $a$ equation, we first need to integrate the $\dot{b}a^2 \alpha_2$ term in action (\ref{action_1}) by parts. This allows us to have an explicit $\dot{a}$ term appearing in the action. The integration by parts leads to
\begin{equation}
    S_{full} = V_c \int{dt \: [ -2a\dot{a}b \alpha_2 + \alpha_3 \dot{m_i}\psi_i + Na\alpha_2 (b^2 + kc_g^2) - Na^3 \alpha_3 \rho + \dot{\rho_\Lambda}T_\Lambda]}
\end{equation}
Therefore, the Euler-Lagrange equation for $a$ is 
\begin{equation}
    \frac{\partial L}{\partial b} - \frac{d}{dt}\frac{\partial L}{\partial \dot{b}} = 0
\end{equation}
resulting in
\begin{equation}
    -2a\dot{a} \alpha_2 + 2bNa\alpha_2 = 0 \Leftrightarrow \dot{a} = Nb
\end{equation}
which is consistent with equation (\ref{a_dot_1}) and it also gives the relation between $\dot{a}$ and $b$ we found when considering the EC and EH actions.
Continuing, we obtain the equation of motion for $b$ applying the Euler-Lagrange  equation to the non-integrated by parts action this time. This is because we require an explicit $\dot{b}$ dependence for the equation of motion. Therefore, using 
\begin{equation}
    \frac{\partial L}{\partial a^2} - \frac{d}{dt}\frac{\partial L}{\partial \dot{a^2}} = \frac{1}{2a}\frac{\partial L}{\partial a} - \frac{d}{dt}\biggl [\frac{1}{2 \dot{a}}\frac{\partial L}{\partial \dot{a}} \biggl] = 0
\end{equation}
leads to 
\begin{equation}
    \dot{b}\alpha_2 + \frac{N \alpha_2}{2a}(b^2 + kc_g^2) -\frac{3}{2} \rho \alpha_3 Na = 0 \Leftrightarrow \dot{b} = -\frac{\alpha_3}{2\alpha_2} (\rho + 3p) Na
\end{equation}
where, once again, we have used the Hamilton's constraint (\ref{const_11}) and equation (\ref{rho_plus_p}).
This alternative approach proves that the two results are equivalent, it does not require defining $A^2$ when $\alpha_2$ is not fixed and it will be useful in the next chapter to check that results agree.
\\
\\
Having derived the full dynamics of the theory, we can consider the energy conservation equation. Just like in General Relativity we can obtain the conservation equation from the Bianchi Identities, in the MSS treatment we derive the same result combining the equations of motion with the Hamilton's constraint. This gives us the energy conservation in a Universe described by the FLRW metric, where we assumed a matter content modeled by a perfect fluid, including also the cosmological constant. To begin with, we take the time derivative of the Hamilton's constraint. We do so because it describes the time evolution of any field in the theory, so it is expected that dotting it will produce the desired time dependence of the energy density. Doing so, we include the possibility for $c_g^2$ to vary generically with parameter time $t$, as it is part of the set of constants $\bm{\beta}$ which can do so. For the moment, we do not specify which of the $\bm{\alpha}$ are providing the relational time $c_g^2$ is varying with respect to, as these can be many different ones, as we will see later.\\
Proceeding, we dot Hamilton's constraint, obtaining 
\begin{equation}
    2b \dot{b} + k \frac{dc_g^2}{dt} = \frac{\alpha_3}{\alpha_2} \dot{\rho}a^2 + 2\frac{\alpha_3}{\alpha_2} \rho a \dot{a} 
\end{equation}
where $\alpha_2$ and $\alpha_3$ are, for the moment, kept constant, thus not contributing the the conservation equation.
Then, using the equations of motion (\ref{a_dot_1}) and (\ref{b_dot_1}) for $b$ and $\dot{b}$ in the equation above and rearranging we have
\begin{equation}
    \dot{\rho} + 3\frac{\dot{a}}{a}(\rho + p) = \frac{k \alpha_2}{\alpha_3 a^2}\frac{dc_g^2}{dt} \label{cons_frst}
\end{equation}
This energy ``conservation" equation seems to imply, instead, energy violation, as we clearly have a energy source term given by a varying $c_g^2$. This is an expected result, as by braking Local Lorentz Invariance, we are braking the symmetric properties of our equations, hence generating source terms. However, this is not the full picture, because, depending on the constant $\bm{\alpha}$ chosen and the parameter $\bm{\beta}$, the clock might absorb all the source energy.\\
To see this, we consider a $c_g^2$ dependent on the clock defined by the cosmological constant (in the literature called the unimodular time), such that $c_g^2 = c_g^2 (T_{\Lambda})$, where we have $\bm{\beta} = c_g^2$ and $\bm{\alpha} = \rho_{\Lambda}$. Assuming this dependence provides us with two extra Hamilton's equations which give the dynamics of the unimodular time and its conjugate variable $\rho_\Lambda$
\begin{equation}
    \dot{T_\Lambda} = \{T_\Lambda, H\} = -\frac{\partial H}{\partial \rho_\Lambda} = - \alpha_3 Na^3 \label{time_lmbd}
\end{equation}
\begin{equation}
    \dot{\rho_\Lambda} = \{\rho_\Lambda, H\} = \frac{\partial H}{\partial T_\Lambda} = - k \alpha_2 \frac{dc_g^2}{dT_\Lambda}Na \label{varying_lmbd}
\end{equation}
Using the chain rule, equation (\ref{varying_lmbd}) can be re-expressed as 
\begin{equation}
    \dot{\rho_\Lambda} = - k \alpha_2 \frac{dc_g^2}{dt}\frac{dt}{dT_\Lambda}Na
\end{equation}
where we can now plug in equation (\ref{time_lmbd}) obtaining 
\begin{equation}
    \dot{\rho_\Lambda} = \frac{k\alpha_2}{\alpha_3 a^2} \frac{dc_g^2}{dt} \label{rho_lmbd}
\end{equation}
Equation (\ref{rho_lmbd}) suggests that the source term due to a varying $c_g^2$ in equation (\ref{cons_frst}) is entirely given by the vacuum energy of the system.

\subsection{Varying $\alpha_2$ and $\alpha_3$ theory} \label{joaito}

In the previous subsection, we have assumed that $\alpha_3$ is not varying and that also $\alpha_2$ is fixed. This, however, is not a strict requirement, and allowing their variability will include new terms arising in the equations of motion and, as a consequence, also in the conservation equation.
We begin by considering the full action of our theory 
\begin{equation}
    S_{full} = V_c \int{dt \: \alpha_2 \dot{b}a^2 + \alpha_3 \dot{m_i}\psi_i + Na\alpha_2 (b^2 + kc_g^2) - \alpha_3 \rho Na^3+ \dot{\alpha_2}T_R + \dot{\alpha_3} T_N} \label{action_222}
\end{equation}
where we have removed the pure unimodular term since we would like to consider the variations given by $\alpha_2$ and $\alpha_3$. Instead, we have included the Ricci $T_R$ and Newton $T_N$ times, associated with our two varying constants. This action is the same as (\ref{action_1}), but now $\alpha_2$ is not fixed anymore, leading to a different equation of motion for $\dot{a}$. It is essential now that the Hamiltonian is expressed in terms of the new canonical variable $A^2 = a^2 \alpha_2$ as
\begin{equation}
     H = - NA \sqrt{\alpha_2} (b^2 + kc_g^2) + \alpha_3 \alpha_2^{-\frac{3}{2}} \rho NA^3 \label{H_A}
\end{equation}
where now the canonical pair is $(b, A^2)$ and not $(b, a^2)$ anymore as before.\\
Following the Poisson's brackets, we calculate the equation of motion form $A^2$, given by 
\begin{equation}
     \dot{(A^2)} = \{A^2, H\} = \frac{\partial A^2}{\partial b}\frac{\partial H}{\partial A^2} - \frac{\partial A^2}{\partial A^2}\frac{\partial H}{\partial b} = -\frac{\partial H}{\partial b}
\end{equation}\\
which, computing the LHS and the RHS leads to 
\begin{equation}
    2a\dot{a}\alpha_2 + a^2 \dot{\alpha_2} = 2bNA \sqrt{\alpha_2}= 2bNa\alpha_2 
\end{equation}
eventually giving 
\begin{equation}
    \dot{a} + \frac{\dot{\alpha_2}}{2 \alpha_2} = Nb \label{equal_to_KTC}
\end{equation}\\
On the other hand, the equation of motion for $b$ is
\begin{equation}
    \dot{b} = \{b, H\} = \frac{\partial b}{\partial b}\frac{\partial H}{\partial A^2} - \frac{\partial b}{\partial A^2}\frac{\partial H}{\partial b} = \frac{\partial H}{\partial A^2} = \frac{1}{2a\alpha_2} \frac{\partial H}{\partial a}
\end{equation}
where we have used the definition of $A^2$ and we are now using the Hamiltonian expressed in terms of $a$ instead. Furthermore, using equation (\ref{rho_plus_p}) to substitute $\rho$, we arrive at the final form
\begin{equation}
    \dot{b} = -\frac{\alpha_3}{2 \alpha_2}(\rho +3p)Na
\end{equation}
\\
\\
It is also possible to derive the same equations following the Lagrangian approach. To begin with, we integrate action (\ref{action_222}) by parts, assuming that $\alpha_2$ is not constant, hence obtaining
\begin{equation}
    S_{full} = V_c \int{dt \: -ba^2 \dot{\alpha_2} - 2a\dot{a}b \alpha_2 + \alpha_3 \dot{m}_i \psi_i + Na\alpha_2 (b^2 +kc_g^2) - \alpha_3 \rho Na^3 +\dot{\alpha_2}T_R + \dot{\alpha_3} T_N} 
\end{equation}
Then, writing the Euler-Lagrange equation for $a$
\begin{equation}
    \frac{\partial L}{\partial b} - \frac{d}{dt}\frac{\partial L}{\partial \dot{b}} = 0
\end{equation}
we obtain
\begin{equation}
    -a^2 \dot{\alpha_2} -2a\dot{a}\alpha_2 + 2Nab\alpha_2 = 0 \Leftrightarrow \dot{a} + \frac{\dot{\alpha_2}}{2 \alpha_2}a = Nb
\end{equation}\\
On the other hand, using the Euler-Lagrange equation for $b$ on the original action (\ref{action_222}),
\begin{equation}
    \frac{\partial L}{\partial a^2} - \frac{d}{dt}\frac{\partial L}{\partial \dot{a}^2} = \frac{1}{2a}\frac{\partial L}{\partial a} - \frac{d}{dt}\biggl[ \frac{1}{2 \dot{a}}\frac{\partial L}{\partial \dot{a}}\biggl] = 0
\end{equation}\\
invoking the Hamilton's constraint and equation (\ref{rho_plus_p}) for $\rho$, we finally get 
\begin{equation}
    \alpha_2 \dot{b} + \frac{N \alpha_2}{2a}(b^2 + kc_g^2) -\frac{3\alpha_2}{2}\rho Na^3 = 0 \Leftrightarrow \dot{b} = -\frac{\alpha_3}{2 \alpha_2} (\rho + 3p) Na
\end{equation}\\
\\
Once again, as before, we dot Hamilton's constraint, assuming $\alpha_2$ and $\alpha_3$ to be varying. This leads to
\begin{equation}
    2b\dot{b} +k\frac{dc_g^2}{dt} = \frac{\dot{\alpha_2}}{\alpha_2} \rho a^2 -\frac{\alpha_3 \rho a^2}{\alpha_2^2}\dot{\alpha_2} + \frac{\alpha_3}{\alpha_2}a^2 \dot{\rho} + 2\frac{\alpha_3}{\alpha_2}\rho a \dot{a}
\end{equation}
Finally, using the equations of motion above for $b$ and $\dot{b}$, we find the conservation equation
\begin{equation}
    \dot{\rho} + 3\frac{\dot{a}}{a}(\rho + p) = -\frac{\dot{\alpha_3}}{\alpha_3}\rho + \frac{\dot{\alpha_2}}{\alpha_2}\frac{\rho -3p}{2} + \frac{k \alpha_2}{\alpha_3 a^2}\frac{dc_g^2}{dt}
\end{equation}
\\
We notice that, besides the $\dot{\alpha_3}$ term which we would have expected having allowed $\alpha_3$ to vary, we also get the second term on the RHS for a varying $\alpha_2$. As before, we have additional equations coming from the Hamiltonian which give the dynamics of the clock and its constant. In this case, since we assume $\alpha_2$ and $\alpha_3$ to vary, we have four equations in total: one pair for the Ricci time, and the second pair for the Newton time. These are the times $c_g^2$ depends upon, as
\begin{equation}
    \dot{T_R} = \{T_R, H\} = -\frac{\partial H}{\partial \alpha_2} = \frac{\alpha_3}{\alpha_2}\frac{\rho - 3p}{2}Na^3 \label{rho_snd_exp}
\end{equation}
\begin{equation}
    \dot{\alpha_2} = \{\alpha_2, H\} = \frac{\partial H}{\partial T_R} = - \alpha_2 k \frac{\partial c_g^2}{\partial T_R}Na
\end{equation}
where in the Ricci time equation we have used the Hamilton's constraint. On the other hand, the equations for $T_N$ and $\alpha_3$ are 
\begin{equation}
    \dot{T_N} = \{T_N, H\} = -\frac{\partial H}{\partial \alpha_3} = -\rho Na^3
\end{equation}
\begin{equation}
    \dot{\alpha_3} = \{\alpha_3,H\} = \frac{\partial H}{\partial T_N} = -\alpha_2 k \frac{\partial c_g^2}{\partial T_N}Na
\end{equation}
where we stress once again that, taking these derivatives, we have used Hamiltonian (\ref{H_A}) expressed in terms of $A^2$. To show that equation (\ref{rho_snd_exp}) is indeed a conservation equation, we consider that $c_g^2$ has a double dependence on $T_R$ and $T_N$, which implies that
\begin{equation}
    \frac{dc_g^2}{dt} = \frac{\partial c_g^2}{\partial T_R}\dot{T_R} + \frac{\partial c_g^2}{\partial T_N}\dot{T_N}
\end{equation}\\
which, using the Ricci and the Newton times equations, becomes 
\begin{equation}
    \frac{dc_g^2}{dt} = \frac{\alpha_2}{\alpha_3}\frac{\rho -3p}{2} \frac{\partial c_g^2}{\partial T_R}Na^3 - \rho \frac{\partial c_g^2}{\partial T_N}Na^3 \label{bo}
\end{equation}
Therefore, plugging the equations for $\alpha_2$ and $\alpha_3$ into equation (\ref{rho_snd_exp}) and also using equation (\ref{bo}), we can show that
\begin{equation}
    \dot{\rho} + 3\frac{\dot{a}}{a}(\rho + p) = 0
\end{equation}
This result shows how a gravitational parameter, $c_g^2$, depending on a gravitational clock does not lead to any net energy violation. It occurs because, if the clock is gravitational, there is no matter component absorbing the constant's variation, so energy is fully conserved. We will investigate this pattern further once a more complete picture is build, extending these results to Brans-Dicke theory.

\clearpage{\pagestyle{empty}\cleardoublepage}

%%%%%%%%%%%%%%%%%%%%%%%%%%%%%%%%%%%%
%%%%%%%%%%%%%%%%%%%%%%%%%%%%%%%%%%%%
\chapter{Brans-Dicke and Evolving Laws}

In our final chapter, we present the original results achieved in this research by applying varying constants to energy conservation in cosmology. To do so, we use the Brans-Dicke action coupled with the unimodular one. The results are constructed around two main bifurcations: a non-dynamical Brans-Dicke theory and a dynamical one. In the first case, our results mimic the ones obtained in \cite{magueijo2023evolving}, with the substitution $\alpha_2 \rightarrow \phi$. However, new results are obtained when considering a fixed scalar field $\phi_0$ in the canonical term, such that the $\phi$ multiplying the Hamiltonian is still varying. On the other hand, when including the Brans-Dicke kinetic term, we develop several scenarios of varying constants. All of these produce new results, with particular interest given by the $\omega = \omega (T_\Lambda)$ and $c_g^2 = c_g^2 (T_\omega)$ scenarios. These are specifically relevant for phenomenology, also providing an insight in the Cosmological Constant problem, as we will study in \cite{bassani2023}. Therefore, to summarise, in section \eqref{mss_bd} we reduce the Brans-Dicke action to minisuperspace, showing how to include the kinetic term in the full Hamiltonian. In section \eqref{no_kt} we derive the equations of motion for a non-dynamical Brans-Dicke theory, focusing on the $\phi_0$ cases. Finally, section \eqref{yes_kt} develops the multiple scenarios arising form the dynamical Brans-Dicke theory, highlighting the main phenomenological results.

\section{Brans-Dicke in MSS} \label{mss_bd}

In this short section, we derive the form of the Kinetic Brans-Dicke action in minisuperspace. As we will see, we will have two cases, one where we set $\omega$, the Brans-Dicke parameter, to zero and the other one where we use the full Brans-Dicke action. It is therefore necessary, for the ladder case,  to have the full Brans-Dicke action in MSS. This will allow us to infer the Brans-Dicke kinetic Hamiltonian, which will be at the heart of our original results. Finally, we will highlight how including or excluding the $c$ factor appearing in the Brans-Dicke Hamiltonian can lead to different results, providing an additional parameter that could be varying, as we will explore in our future work \cite{bassani2023}.

\subsection{Reducing the Brans-Dicke action to MSS} \label{BD_act_der_sub}

Our starting point is the Brans-Dicke action, which should be reduced to minisuperspace to implement the formalism outlined in the previous chapter. Firstly, in the Brans-Dicke action, the Ricci scalar term is essentially equivalent to the Einstein-Hilbert action, except for the constant multiplying $R$. In fact, if we compare the EH action with the Brans-Dicke one for $\omega = 0$ (non-dynamical Brans-Dicke)
\begin{equation}
    S_{EH} = \alpha_2 \int{d^4x \: \sqrt{-g}R}
\end{equation}
\begin{equation}
    S_{\omega = 0} = \int{d^4x \sqrt{-g}\: \phi R}
\end{equation}
we immediately notice that the scalar field $\phi$ plays the role of $\alpha_2$ once we do not assume a dynamical scalar field. Of course, the scalar field is nothing but a variable  gravitational constant, while $\alpha_2$ includes also the speed of light, but up to a constant (including the numerical factor and $\pi$) the two theories are equivalent. Therefore, it is straightforward to reduce the Brans-Dicke action to MSS assuming $\omega = 0$. We simply take action (\ref{starting_point}), replace $\alpha_2 \rightarrow \phi$ and adjust the constants to get
\begin{equation}
    S_{\omega=0} = \frac{3V_c}{8 \pi} \int{dt \: \phi \biggl[\dot{b}a^2 + Na(b^2 + kc_g^2)  \biggl]}
\end{equation}
or, more simply, by including the pre-factor $\frac{3}{8 \pi}$ in the definition of $\phi$, we get the final form of the Brans-Dicke action
\begin{equation}
    S_{\omega = 0} = V_c \int{dt \: \phi \biggl[\dot{b}a^2 + Na(b^2 + kc_g^2)  \biggl]} \label{bd_11}
\end{equation}
Action (\ref{bd_11}) will be the starting point of the original results we will introduce in this chapter. Importantly, like in the varying $\alpha_2$ case, when deriving the Hamilton's equations, it is $A^2 = a^2 \phi$ to be used as a canonical variable. This is even more relevant here because $\phi$, unlike $\alpha_2$, is inherently a variable in the Brans-Dicke theory, modelling the change of the gravitational constant from spacetime position to spacetime position.\\ Since the only difference between action (\ref{bd_11}) and the full Brans-Dicke is the kinetic term, we might just consider it and understand how to obtain the Brans-Dicke Hamiltonian in MSS from it. We begin by considering the kinetic term as
\begin{equation}
    -\frac{\omega}{\phi}\partial_\mu \phi \: \partial^\mu \phi = -\frac{\omega}{\phi} [g^{tt} \partial_t \phi \: \partial_t \phi +g^{ij}\partial_i \phi \: \partial_j \phi] = -\frac{\omega}{\phi} g^{tt} \dot{\phi}^2
\end{equation}
because the scalar field $\phi$ is exclusively time-dependent, so that all its spacial derivatives vanish. Including the usual $\sqrt{-g}$ gives the full kinetic term as
\begin{equation}
    S_{K.T.} = \int{d^4 x \: \sqrt{-g} \: \biggl[-\frac{\omega}{\phi}g^{tt} \dot{\phi}^2\biggl]}
\end{equation}
Furthermore, from the FLRW metric, we have that $g^{tt} = - \frac{1}{c^2N^2}$, and that
\begin{equation}
    \sqrt{-g} = \frac{r^2 \sin^2{\theta}}{\sqrt{1- kr^2}}cNa^3
\end{equation}
These two expressions combined lead to the standard form \cite{frion2019affine}
\begin{equation}
    S_{K.T.} = V_c \int{dt \: Na^3 \frac{\omega}{\phi}\frac{\dot{\phi}^2}{N^2}} = V_c \int{dt \: \dot{\phi}^2 \frac{\omega}{\phi}\frac{a^3}{Nc^2}}
\end{equation}
where the remaining factors coming from $\sqrt{-g}$ have been absorbed into $V_c$.
Following the procedure outlined in \cite{vilenkin1994approaches}, we obtain the canonical momentum conjugate to $\phi$ as
\begin{equation}
    \pi_\phi = \frac{\delta S_{K.T.}}{\delta \dot{\phi}} = 2 \omega \frac{\dot{\phi}}{\phi Nc}a^3
\end{equation}
At this point it is important to notice the extra factor of $c$ multiplying $N$ in the conjugate momentum expression and in the kinetic action in MSS. If we leave it in the kinetic term, it adds another constant that can be varying. However, in this work, for simplicity, we absorb the $c$ factor into the parameter $\omega$, while its full inclusion in the theory will be the object of next research \cite{bassani2023}.
Using the Fourier transform on the Lagrangian as described in (\ref{def_h_l}), we can write
\begin{equation}
    S_{K.T.} = V_c \int{dt \: \omega \frac{\dot{\phi}^2}{\phi}\frac{a^3}{N}} \overset{!}{=}  V_c \int{dt \: \dot{\phi}\pi_\phi - H_{BD}}
\end{equation}
such that, re-expressing in terms of $H_{BD}$ we obtain
\begin{equation}
    H_{BD} (\phi, \pi_\phi) = \dot{\phi}\pi_\phi - \omega \frac{\dot{\phi}^2}{\phi}\frac{a^3}{N} \label{kinetic_ham_11}
\end{equation}
Now,  following again the prescriptions given by \cite{vilenkin1994approaches}, we find an expression for $\dot{\phi}$ by inverting the conjugate momentum equation, such that
\begin{equation}
    \pi_\phi = \frac{2\omega}{\phi}\frac{\dot{\phi}}{N}a^3 \Leftrightarrow \dot{\phi} = \frac{\phi \pi_\phi N}{2\omega a^3} \label{conj_mom_def}
\end{equation}
We can use the expression above for $\dot{\phi}$ into the Hamiltonian (\ref{kinetic_ham_11}), leading to 
\begin{equation}
    H_{BD}(\phi, \pi_\phi) = \frac{\phi \pi_\phi^2 N}{2\omega a^3} - \frac{\omega}{\phi}\frac{\phi^2 \pi_\phi^2 N^2}{4 \omega^2 a^6}\frac{a^3}{N} = \frac{1}{4}\frac{\phi \pi_\phi^2}{\omega a^3}N
\end{equation}
This is finally the Hamiltonian coming from the Brans-Dicke kinetic term which, which, together with the canonical pair $\dot{\phi}\pi_\phi$, creates the kinetic Brans-Dicke Lagrangian. Therefore, combining it with action (\ref{bd_11}), we arrive at the full Brans-Dicke action in minisuperspace
\begin{equation}
    S_{BD} = V_c \int{dt \: \biggl[\phi \dot{b}a^2 + \dot{\phi}\pi_\phi + \phi Na (b^2 + kc_g^2) -\frac{1}{4}\frac{\phi \pi_\phi^2 }{\omega a^3}N\biggl]}
\end{equation}
Concluding, we can extend the purely gravitational action to include a coupling to matter and the generalised unimodular term, as we did in Chapter 2. The unimodular term will allow us to have relational times for the constants evolution, while the matter action will provide the additional $\alpha_3$ parameter. Doing so, we arrive at the complete unimodular Brans-Dicke action coupled to matter (assuming a perfect fluid as before):
\begin{equation}
    S = V_c \int{dt \: \biggl[\phi \dot{b}a^2 + \dot{\phi}\pi_\phi + \alpha_3 \dot{m}_i \psi_i + \phi Na (b^2 + kc_g^2) -\alpha_3 \rho Na^3 -\frac{1}{4}\frac{\phi \pi_\phi^2 }{\omega a^3}N + \bm{\dot{\alpha}} \bm{T_\alpha}\biggl]} \label{alpha_drusonis}
\end{equation}\\
This theory will be the starting point of all the following results. It is also the most general action we will use for Brans-Dicke, so each specific scenario will differ mostly because of the unimodular term, depending on which constant we are varying and with respect to which time. These could be very simple or also very intricate combinations as we will see in the following. Also, we will explore a bifurcation in the theory given by only fixing $\phi$ in the conjugate pair term. This will be the object of study in the next sections and subsections.

\section{Non-Dynamical Brans-Dicke} \label{no_kt}

In this section, we present the simplified case of a non-dynamical scalar field in the Brans-Dicke theory. Accordingly, $\omega = 0$, so we cannot invoke any dependence on this parameter. Importantly, this scenario is essentially equivalent to the one covered in section \eqref{first_stone}, with the substitution $\alpha_2 \rightarrow \phi$, such that we include a dynamical $G$, as required by the Brans-Dicke theory. However, we investigate a new theory where Local Lorentz Invariance is fully broken by a constant scalar field $\phi_0$ multiplying the canonical term while the scalar field appearing in the Hamiltonian is left varying. This theory is different from the fixed $\alpha_2$ one analysed in subsection \eqref{fixed_alpha}, as it leads to different equations of motion. Therefore, we derive the equations of motion and the most general conservation equations for the $\phi_0$ and full $\phi$ cases. We then present some varying constants scenarios for the $\phi_0$ case, excluding any full $\phi$ scenarios, since they are equivalent to previous sections.

\subsection{The Equations of Motion for the $\phi_0$ scenario} \label{craken}

This subsection derives the equations of motion and the general conservation equation for a non-dynamical Brans-Dicke theory, with a decoupled scalar field $\phi_0$. Importantly, given the fixed scalar field in the canonical pair, this theory shows a totally broken Local Lorentz Invariance, as pointed out in \cite{magueijo2023evolving}.  The general action in MSS is 
\begin{equation}
    S = V_c \int{ dt \: \phi_0 \: \dot{b} a^2 + \alpha_3 \dot{m}_i \psi_i  + \phi Na(b^2 + kc_g^2) - \alpha_3 \rho Na^3} + S_U
\end{equation}
where $S_U$ is the unimodular action that will provide the evolution times and $\dot{m}_i \psi_i$ is the canonical term from the generic matter action.
The Hamiltonian, obtained via Legendre transform of the Lagrangian in the action above, is
\begin{equation}
    H = - \phi Na(b^2 +kc_g^2) + \alpha_3 \rho Na^3
\end{equation}
Following the standard requirement \cite{vilenkin1994approaches} for the Hamilton's constraint, we obtain 
\begin{equation}
    H \overset{!}{=} 0 \Leftrightarrow b^2 + kc_g^2 = \frac{\alpha_3 \rho a^2}{\phi}
\end{equation}
which we use to derive the conservation equation. However, before proceeding, we need to obtain the equations of motion for the canonical pair $(b, a^2)$. In this case, since the scalar field $\phi$ is multiplying the canonical pair, it is better to express the Hamiltonian in terms of a new canonical variable $A^2 = \phi_0 a^2$. Furthermore, when evaluating the Poisson's brackets, we will use $A^2$ instead of $a^2$. This naturally leads 
\begin{equation}
    \dot{(A^2)} = \{A^2, H\} = \frac{\partial A^2}{\partial b}\frac{\partial H}{\partial A^2} -\frac{\partial A^2}{\partial A^2}\frac{\partial H}{\partial b} =-\frac{\partial H}{\partial b}
\end{equation}
giving
\begin{equation}
    2a\dot{a}\phi_0 = 2bNa\phi \Leftrightarrow \dot{a} = \frac{\phi}{\phi_0}Nb
\end{equation}
which is the equation of motion for $a$. On the other hand, the same procedure can be applied to derive the equation of motion for the other conjugate variable $b$
\begin{equation}
    \dot{b} = \{b, H\} = \frac{\partial b}{\partial b}\frac{\partial H}{\partial A^2}-\frac{\partial b}{\partial A^2}\frac{\partial H}{\partial A^2} = \frac{\partial H}{\partial A^2} = \frac{1}{2a\phi_0}\frac{\partial H}{\partial a}
\end{equation}
giving 
\begin{align}
    \dot{b} &= \frac{1}{2a \phi_0 }\frac{\partial H}{\partial a} \\ \nonumber\\
    &= \frac{1}{2a \phi_0} \biggl[-N\phi (b^2 + kc_g^2) - 3w \frac{\alpha_3 m}{a^{3w +1}}N \biggl] \\ \nonumber\\ 
    &= \frac{1}{2a \phi_0} \biggl[-\alpha_3 \rho Na^2 -3w\frac{\alpha_3 m}{a^{3w+1}}N\biggl]\\ \nonumber\\ 
    &= -\frac{1}{2}\frac{\alpha_3 \rho}{\phi_0}Na -\frac{3}{2}w \frac{\alpha_3\rho}{\phi_0}Na \\ \nonumber\\
    &= \frac{\alpha_3}{\phi_0}\frac{(-\rho -3w \rho)}{2}Na \\ \nonumber\\
    &= -\frac{\alpha_3}{2 \phi_0}(\rho + 3p)Na
\end{align}
where we have used the Hamilton's constraint for $(b^2 + kc_g^2)$ and the definition of $\rho$ from \eqref{rho_substitution}. As a side note, we remark that it is possible to obtain the same results using the Lagrangian approach, as we showed in subsection \eqref{joaito}.
Now we can derive the conservation equation from the condition that the Hamiltonian must be conserved with time since, if $H = 0$, then it is also true that  $\dot{H} = 0$. Therefore, taking the time derivative of the Hamilton's constraint, we obtain 
\begin{equation}
    2b \dot{b} + k \frac{d c_g^2}{dt} = \frac{\dot{\alpha_3} \rho a^2}{\phi} + \frac{\alpha_3 \dot{\rho} a^2}{\phi} + 2\frac{\alpha_3 \rho a \dot{a}}{\phi} -\frac{\alpha_3 \rho a^2}{\phi^2}\dot{\phi} 
\end{equation}
which, rearranged for the standard form of the conservation equation and using the equations of motion for the canonical pair, gives
\begin{equation}
    \dot{\rho} + 3\frac{\dot{a}}{a}(\rho + p) = -\frac{\dot{\alpha_3}}{\alpha_3}\rho + \frac{\dot{\phi}}{\phi}\rho + \frac{k \phi }{\alpha_3 a^2}\frac{dc_g^2}{dt}
\end{equation}
Interestingly, we have a $\dot{\phi}$ term behaving like the $\dot{\alpha_3}$ one. This term is new compared to previous theories \cite{magueijo2023evolving} and it arises from totally braking Local Lorentz Invariance, because of the constant scalar field in the canonical term.

\subsection{The Equations of Motion for the full $\phi$ scenario}

While the results in the previous subsection are different from the scenario in subsection \eqref{fixed_alpha} (covered in \cite{magueijo2023evolving}) because of a varying $\phi$ in the Hamiltonian but a fixed one in the canonical pair, the following scenario is fully equivalent to subsection \eqref{joaito}. The only difference between the two is that, instead of $\alpha_2$ which is postulated to vary, we have the scalar field $\phi$, which is naturally varying by construction of the non-kinetic Brans-Dicke action. However, the equations of motion and the conservation equation are the same in both scenarios. We therefore include a brief summary of the results obtained to show how the scalar field treatment is equivalent and to provide a few examples of energy conservation or violation. Beginning from the Brans-Dicke action with $\omega = 0$ in MSS
\begin{equation}
    S = V_c \int{ dt \: \phi \: \dot{b} a^2 + \alpha_3 \dot{m}_i \psi_i  + \phi Na(b^2 + kc_g^2) - \alpha_3 \rho Na^3}) + S_U
\end{equation}
we immediately have the Hamiltonian 
\begin{equation}
    H = -\phi Na (b^2 + kc_g^2) + \alpha_3 \rho Na^3 
\end{equation}
which gives also the Hamilton's constraint 
\begin{equation}
    b^2 + kc_g^2 = \frac{\alpha_3 \rho a^2}{\phi}
\end{equation}
Simply following the procedure in subsection \eqref{joaito}, with the substitution $\alpha_2 \rightarrow \phi$ and the necessary adjustment for $A^2 = \phi a^2$, we obtain the equations of motion for $a$ and $b$, which read
\begin{equation}
    \dot{a} + \frac{\dot{\phi}}{2 \phi} = Nb
\end{equation}
\begin{equation}
    \dot{b} = -\frac{\alpha_3}{2 \phi}(\rho + 3p) Na
\end{equation}
Dotting the Hamilton's constraint, rearranging and using the equations of motion we arrive at the general conservation equation 
\begin{equation}
    \dot{\rho} + 3\frac{\dot{a}}{a}(\rho + p) = -\frac{\dot{\alpha_3}}{\alpha_3}\rho + \frac{\dot{\phi}}{\phi}\frac{\rho -3p}{2} + \frac{k \phi}{\alpha_3 a^2}\frac{dc_g^2}{dt}
\end{equation}
From this point onward, it would be just a matter of choice or interest to consider which constants take part to the $\bm{\alpha}$ and which ones to the $\bm{\beta}$. As extensively illustrated in \cite{magueijo2023evolving}, we might have scenarios where $c_g^2$ depends on the Ricci time $T_{\phi}$, such that $c_g^2 = c_g^2 (T_\phi)$. This does not lead to any net energy violation because we are considering a gravitational parameter depending on a gravitational clock, following a pattern we will analyse later.\\ We could, as well, consider the speed of light depending on the unimodular time, such that $c_g^2 = c_g^2 (T_\Lambda)$. In this case, we do obtain a net energy violation, as the variation in $c_g^2$ is entirely absorbed by the unimodular clock $T_\Lambda$, leading all the energy in the vacuum energy density $\rho_\Lambda$.\\
Other more exotic scenarios include a matter speed of light $c_m^2$ depending on both the Ricci and the Newton times. This setup leads, as well, to net energy violation, because we are considering a matter parameter depending on a gravitational clock. This situation is the exactly parallel of the previous one, where we had a gravitational parameter depending on a matter clock.

\subsection{Fixed $\phi_0$ with $c_g^2 = c_g^2(T_\phi)$}

The first interesting theory to consider is when the speed of light depends on the Ricci-like time $T_{\phi}$. In this case, we have that $\bm{\alpha} = \phi$ and $\bm{\beta} = c_g^2$, such that our action reads
\begin{equation}
    S = V_c \int{ dt \: \phi_0 \: \dot{b} a^2 + \alpha_3 \dot{m_i}\psi_i + \phi Na(b^2 + kc_g^2) - \alpha_3 \rho Na^3} + \dot{\phi} T_\phi
\end{equation}
This leads to the following Hamiltonian:
\begin{equation}
    H = -\phi Na(b^2 + k c_g^2) + \alpha_3 \rho Na^3
\end{equation}
The additional equations of motions for $\phi$ and $T_{\phi}$ are obtained by computing the Poisson's brackets:
\begin{equation}
     \dot{\phi} = \{ \phi, H\} = \frac{\partial H}{\partial T_\phi} = -\phi Nak\frac{d c_g^2}{d T_\phi}
\end{equation}
\begin{equation}
    \dot{T}_\phi = \{T_\phi, H \}  = - \frac{\partial H}{\partial \phi} = Na(b^2 +kc_g^2) = \frac{\alpha_3 \rho Na^3}{\phi}
\end{equation}
where, in the $T_{\phi}$ equation, we have used Hamilton's constraint.
From the previous derivation, the most general conservation equation for which $\phi$ and $c_g^2$ are varying is
\begin{equation}
    \dot{\rho} + 3\frac{\dot{a}}{a} (\rho + p) = \frac{\dot{\phi}}{\phi} \rho + \frac{k \phi}{a^2 \alpha_3} \frac{dc_g^2}{dt}
\end{equation}
Using the equation of motion for $\phi$ in the first term and the $\dot{T_\phi}$ equation in the second one after applying the chain rule, we arrive at the standard conservation equation
\begin{equation}
    \dot{\rho} + 3\frac{\dot{a}}{a} (\rho + p) = 0
\end{equation}
thus not having, as expected, any energy violation.
Crucially, in the case of a scalar field coupled to the canonical pair $\dot{b}a^2$ as in the Brans-Dicke theory, the conjugate variable to $\dot{b}$ is $A^2 = \phi a^2$. However, this change only applies to theories where the $\phi$ is dynamical in both the canonical pair and in the Hamiltonian. Unlike these theories, in our case we have a non-dynamical $\phi_0$ coupled to the canonical pair. It is therefore redundant to express the Hamiltonian in terms of $A^2$, so we can simply use the standard Hamiltonian in terms of $a^2$ when taking $\phi$ derivatives.

\subsection{Fixed $\phi_0$ with $c_m^2 = c_m^2(T_\phi, T_N)$}

In this scenario, we consider the energy violation arising from a matter parameter depending on both a gravitational and matter clock. This is the case when $c_m^2$, the matter speed of light, is part of the $\bm{\beta}$ and its clocks are given by $\phi$ and $\alpha_3$. This is possible because, from \cite{magueijo2023evolving}, we can specify that $\rho_i = \rho_i (n_i, c_m)$, thus giving the derivatives of $c_m^2$ when it is varying. For completeness, we postulate the same time dependency for $c_g^2$, as this is a source term in the conservation equation and doing so we can substitute in it expressions from the equations of motion of the relational times $T_{\phi}$ and $T_N$. Our starting action is 
\begin{equation}
    S = V_c \int{dt \: \phi_0 \dot{b}a^2 + \alpha_3 \dot{m_i}\psi_i + \phi Na (b^2 + kc_g^2) -\alpha_3 \rho Na^3 + \dot{\phi} T_\phi + \dot{\alpha_3} T_N}
\end{equation}
directly giving the Hamiltonian 
\begin{equation}
    H = -\phi Na (b^2 +kc_g^2) + \alpha_3 \rho Na^3 
\end{equation}
From subsection \eqref{craken} we have the most general conservation equation, which, in this case, contains all three source terms:
\begin{equation}
    \dot{\rho} + 3 \frac{\dot{a}}{a} (\rho + p) = - \frac{\dot{\alpha_3}}{\alpha_3} \rho  + \frac{\dot{\phi}}{\phi} \rho + \frac{\phi k}{\alpha_3 a^2}\frac{dc_g^2}{dt}
\end{equation}
We can now obtain the additional Hamilton's equations. In each pair, the first equation provides the variation of the constant of motion of the clock (it is the equation of motion of the constant, if we allow this oxymoron), while the second one is the equation of motion of the clock itself:
\begin{equation}
    \dot{\phi} =\{\phi, H\} = \frac{\partial H}{\partial T_\phi} = - \phi k Na \frac{\partial c_g^2}{\partial T_\phi} + \alpha_3 Na^3 \frac{\partial \rho}{\partial c_m^2}\frac{\partial c_m^2}{\partial T_\phi}
\end{equation}
\begin{equation}
    \dot{T_\phi} = \{ T_{\phi}, H\} = - \frac{\partial H}{\partial \phi} = \frac{\alpha_3 \rho}{\phi}Na^3
\end{equation}
\begin{equation}
    \dot{\alpha_3} = \{\alpha_3, H\} = \frac{\partial H}{\partial T_N} = - \phi k Na \frac{\partial c_g^2}{\partial T_N} + \alpha_3 Na^3 \frac{\partial \rho}{\partial c_m^2}\frac{\partial c_m^2}{\partial T_N}
\end{equation}
\begin{equation}
    \dot{T_N} = \{ T_N, H\} = -\frac{\partial H}{\partial \alpha_3} = - \rho Na^3
\end{equation}
Finally, to substitute the equations of motion for the $\dot{c_g}$ term, we use the chain rule, giving  
\begin{equation}
    \frac{dc_g^2}{dt} = \frac{\partial c_g^2}{\partial T_\phi}\dot{T_\phi} + \frac{\partial c_g^2}{\partial T_N}\dot{T_N}
\end{equation}
Therefore, using this expression for $\dot{c_g}^2$ and the equations of motion above we obtain 
\begin{equation}
    \dot{\rho} + 3 \frac{\dot{a}}{a} (\rho + p) = \rho Na^3\frac{\partial \rho}{\partial c_m^2}\biggl[\frac{\alpha_3}{\phi}  \frac{\partial c_m^2}{\partial T_\phi} - \frac{\partial c_m^2}{\partial T_N} \biggl]
\end{equation}
This result is particularly important in sight of what we will develop in the next section. In fact, including the scalar field's kinetic term will enable us to obtain a similar energy violation equation. However, unlike the one given here, we will have extra new terms arising from the time ticked by the Brans-Dicke clock $T_\omega$.

\section{Dynamical Brans-Dicke} \label{yes_kt}

We now consider the complete case of a dynamical scalar field in the Brans-Dicke theory. Multiple scenarios will be developed, particularly focusing  on the ones where the Brans-Dicke parameter $\omega$ is either part of the $\bm{\alpha}$ or of the $\bm{\beta}$. As in the non-dynamical section, we follow the bifurcation arising from the fully broken Local Lorentz Invariance due to $\phi_0$. Therefore, we derive the equations of motion and the most general conservation equation for both cases. Interestingly, the equations of motion are different in the two cases, but lead to the same conservation equation. This is not a surprise, as the conservation equation derives from the field equations, which are the same in both cases. As a result, the differentiation between these two cases is redundant whenever we are not considering a dependency on the Ricci time $T_\phi$ given by the scalar field $\phi$. So, without further ado, let's dive into the wild ocean of Pandora's Box theories!

\subsection{The Equations of Motion for the $\phi_0$ scenario} \label{de_cou_pled}

This subsection will consider the simplified case where the $\phi$ multiplying the canonical pair $\dot{b}a^2$ is kept fixed, but the other terms are left variable. Importantly, this scenario is not to be confused with the one explained in subsection \eqref{fixed_alpha}. In fact, there the equivalent of $\phi$ (i.e., $\alpha_2$) is kept fixed for all the terms appearing in the Lagrangian. Conversely, in this case, only the $\phi$ multiplying the first canonical pair is fixed, while the others can vary. This leads to a full action of the form 
\begin{equation}
    S = V_c \int{dt \: \biggl[\phi_0 \dot{b}a^2 + \dot{\phi}\pi_\phi + \alpha_3 \dot{m}_i \psi_i + \phi Na (b^2 + kc_g^2) -\alpha_3 \rho Na^3 -\frac{1}{4}\frac{\phi \pi_\phi^2 }{\omega a^3}N}\biggl] \label{phi_zero_act}
\end{equation}
which is the usual Brans-Dicke action inclusive of kinetic term, coupled to matter (for a generic perfect fluid) and reduced to MSS.
The Hamiltonian, obtained in the usual way, is therefore 
\begin{equation}
    H = -\phi Na(b^2 +kc_g^2) +\alpha_3 \rho Na^3 +\frac{1}{4}\frac{\phi \pi_\phi^2}{\omega a^3}N
\end{equation}
leading to the Hamilton's constraint 
\begin{equation}
    H \overset{!}{=} 0 \Leftrightarrow b^2 + kc_g^2 = \frac{\alpha_3 \rho a^2}{\phi} + \frac{1}{4}\frac{\pi_\phi^2}{\omega a^4}
\end{equation}
Following the standard procedure, we derive the $\dot{a}$ equation of motion 
\begin{equation}
    \dot{(A^2)} = \{A^2, H\} = \frac{\partial A^2}{\partial b}\frac{\partial H}{\partial A^2} -\frac{\partial A^2}{\partial A^2}\frac{\partial H}{\partial b} =-\frac{\partial H}{\partial b}
\end{equation}
giving
\begin{equation}
    2a\dot{a}\phi_0 = 2bNa\phi \Leftrightarrow \dot{a} = \frac{\phi}{\phi_0}Nb \label{aaaa}
\end{equation}\\
On the other hand, the equation of motion for $b$ is obtained via \\
\begin{equation}
    \dot{b} = \{b, H\} = \frac{\partial b}{\partial b}\frac{\partial H}{\partial A^2}-\frac{\partial b}{\partial A^2}\frac{\partial H}{\partial A^2} = \frac{\partial H}{\partial A^2} = \frac{1}{2a\phi_0}\frac{\partial H}{\partial a}
\end{equation}
leading to 
\begin{align}
    \dot{b} &= \frac{1}{2a \phi_0 }\frac{\partial H}{\partial a} \\ \nonumber\\
    &= \frac{1}{2a \phi_0} \biggl[-N\phi (b^2 + kc_g^2) - 3w \frac{\alpha_3 m}{a^{3w +1}}N - \frac{3}{4}\frac{\phi \pi_\phi^2}{\omega a^4}N\biggl] \\ \nonumber\\ \label{sss}
    &= \frac{1}{2a \phi_0} \biggl[-\alpha_3 \rho Na^2 -\frac{1}{4}\frac{\phi \pi_\phi^2 }{\omega a^4}N -3w\frac{\alpha_3 m}{a^{3w+1}}N -\frac{3}{4}\frac{\phi \pi_\phi^2}{\omega a^4}N\biggl]\\ \nonumber\\ 
    &= -\frac{1}{2}\frac{\alpha_3 \rho}{\phi_0}Na -\frac{3}{2}w \frac{\alpha_3\rho}{\phi_0}Na -\frac{1}{8}\frac{\phi \pi_\phi^2}{\phi_0 \omega  a^5}N -\frac{3}{8}\frac{\phi \pi_\phi^2}{\phi_0 \omega a^5}N \\ \nonumber\\
    &= \frac{\alpha_3}{\phi_0}\frac{(-\rho -3w \rho)}{2}Na -\frac{1}{2}\frac{\phi \pi_\phi^2}{\phi_0 \omega a^5}N \\ \nonumber\\ \label{ttt}
    &= -\frac{\alpha_3}{2 \phi_0}(\rho + 3p)Na -\frac{1}{2}\frac{\phi \pi_\phi^2}{\phi_0 \omega a^5}N
\end{align}
where we have used the Hamilton's constraint in equation \eqref{sss} and the definition of $\rho$ from \eqref{rho_substitution} in equation \eqref{ttt}. Remarkably, due to the $\phi_0$ factor in the canonical pair, this equation of motion is different from the one we will obtain in the next section. Specifically, we lack the cancellation of $\phi$ in the momentum term.\\
\\
Finally, we need to obtain an equation of motion for the conjugate momentum $\pi_\phi$. We apply the same procedure, evaluating its related Poisson's brackets, giving 
\begin{equation}
    \dot{\pi_\phi} = \{\pi_\phi, H\} = \frac{\partial \pi_\phi}{\partial \phi}\frac{\partial H}{\partial \pi_\phi} - \frac{\partial \pi_\phi}{\partial \pi_\phi}\frac{\partial H}{\partial \phi} = -\frac{\partial H}{\partial \phi}
\end{equation}
and, computing the derivatives, we get
\begin{align}
    \dot{\pi_\phi} &= -\frac{\partial H}{\partial \phi} \\ \nonumber \\
    &= Na(b^2 +kc_g^2) -\frac{1}{4}\frac{\pi_\phi^2}{\omega a^3}N \\ \nonumber \\
    &= Na \frac{\alpha_3 \rho a^2}{\phi} + \frac{1}{4}\frac{\pi_\phi^2}{\omega a^3}N -\frac{1}{4}\frac{\pi_\phi^2}{\omega a^3}N \\ \nonumber \\
    &= \frac{\alpha_3 \rho}{\phi} Na^3 \label{half_mom}
\end{align}
where we have used the Hamilton's constraint in the $(b^2 + kc_g)$ term.\\
\\
The equations of motion above can be equivalently derived using the Lagrangian approach. We include the derivation to prove that the equations are indeed correct. Begin by considering action \eqref{phi_zero_act}: integrating by parts the canonical pair $\dot{b}a^2$ and eliminating the boundary term gives
\begin{equation}
    S = V_c \int{dt \: \biggl[-2a \dot{a}b \phi_0 + \dot{\phi}\pi_\phi + \alpha_3 \dot{m}_i \psi_i + \phi Na (b^2 + kc_g^2) -\alpha_3 \rho Na^3 -\frac{1}{4}\frac{\phi \pi_\phi^2 }{\omega a^3}N}\biggl]
\end{equation}
Then, using the Euler-Lagrange equation
\begin{equation}
    \frac{\partial L}{\partial b} - \frac{d}{dt}\frac{\partial L}{\partial \dot{b}} = 0
\end{equation}
we obtain
\begin{equation}
    \dot{a} = \frac{\phi}{\phi_0}Nb
\end{equation}\\
which is consistent with \eqref{aaaa}. On the other hand, using the Euler-Lagrange equation for $b$ on action \eqref{phi_zero_act}\\
\begin{equation}
    \frac{\partial L}{\partial a^2} - \frac{d}{dt}\frac{\partial L}{\partial \dot{a^2}} = \frac{1}{2a}\frac{\partial L}{\partial a} - \frac{d}{dt}\biggl [\frac{1}{2 \dot{a}}\frac{\partial L}{\partial \dot{a}} \biggl] = 0
\end{equation}
we arrive at 
\begin{equation}
    \dot{b} = -\frac{\alpha_3}{2 \phi_0}(\rho + 3p)Na -\frac{1}{2}\frac{\phi \pi_\phi^2}{\phi_0 \omega a^5}N
\end{equation}\\
which is as well consistent with the equation of motion previously derived using the Hamiltonian approach.
\\
\\
We now take the time derivative of  Hamilton's constraint and use the equations of motion to derive a conservation equation for our system. Doing so, we  follow the most general approach were all the constants appearing are assumed to be varying. This procedure will give the most general energy conservation equation, which we will simplify in different scenarios where only one constant is varying. Dotting Hamilton's constraint we obtain\\
\begin{equation}
    2b\dot{b} + k\frac{dc_g^2}{dt} = \frac{\dot{\alpha_3}\rho a^2}{\phi} + \frac{2a \dot{a}\alpha_3 \rho}{\phi} + \frac{\dot{\rho}\alpha_3 a^2}{\phi} -\frac{\alpha_3 \rho a^2}{\phi^2}\dot{\phi} + \frac{1}{2}\frac{\pi_\phi \dot{\pi_\phi}}{\omega a^4} -\frac{1}{4}\frac{\pi_\phi^2}{\omega^2 a^4}\dot{\omega} -\frac{\pi_\phi^2}{\omega a^5}\dot{a}
\end{equation}\\
Then, using the equation for $b$ coming from \eqref{aaaa}, the equation for $\dot{b}$ in the LHS, and multiplying both sides by $\frac{\phi}{\alpha_3 a^2}$, we obtain
\begin{multline}
    -(\rho + 3p)\frac{\dot{a}}{a} -\frac{\phi \pi_\phi^2}{\alpha_3 \omega a^7}\dot{a} +\frac{k \phi}{\alpha_3 a^2}\frac{dc_g^2}{dt} = \frac{\dot{\alpha_3}}{\alpha_3}\rho +2\frac{\dot{a}}{a}\rho + \dot{\rho} -\frac{\dot{\phi}}{\phi}\rho \\ \\ + \frac{1}{2}\frac{ \phi \pi_\phi \dot{\pi_\phi}}{\alpha_3 \omega a^6} -\frac{1}{4}\frac{\phi \pi_\phi^2}{\alpha_3 \omega^2 a^6}\dot{\omega} - \frac{\phi \pi_\phi^2}{\alpha_3 \omega a^7}\dot{a}
\end{multline}\\
Rearranging this expression, we obtain the energy conservation equation as\\
\\
\begin{equation}
    \dot{\rho} + 3\frac{\dot{a}}{a}(\rho +p) = -\frac{\dot{\alpha_3}}{\alpha_3}\rho +\frac{\dot{\phi}}{\phi}\rho -\frac{1}{2}\frac{\phi \pi_\phi \dot{\pi_\phi}}{\alpha_3 \omega a^6} + \frac{1}{4}\frac{\phi \pi_\phi^2}{\alpha_3 \omega^2 a^6}\dot{\omega} + \frac{k \phi}{\alpha_3 a^2}\frac{dc_g^2}{dt} \label{drusillo_oneee}
\end{equation}\\
\\
We can now use the $\dot{\pi_\phi}$ equation to further simplify this expression, as well as the definition of $\dot{\phi}$ given by the conjugate momentum, leading to\\ 
\begin{equation}
    \dot{\rho} + 3\frac{\dot{a}}{a}(\rho +p) = -\frac{\dot{\alpha_3}}{\alpha_3}\rho +\frac{1}{4}\frac{\phi \pi_\phi^2}{\alpha_3 \omega^2 a^6}\dot{\omega} + \frac{k \phi}{\alpha_3 a^2}\frac{dc_g^2}{dt}
\end{equation}

\subsection{The Equations of Motion for the full $\phi$ scenario} \label{master}

In this section, we derive the equations of motion for the canonical pair $(a^2, b)$ as well as for $\pi_\phi$, the conjugate momentum to $\phi$. We begin by considering the Brans-Dicke action reduced to MSS as done in subsection (\ref{BD_act_der_sub}). This action, inclusive of the matter part, but exclusive of any unimodular term, is 
\begin{equation}
    S = V_c \int{dt \: \biggl[\phi \dot{b}a^2 + \dot{\phi}\pi_\phi + \alpha_3 \dot{m}_i \psi_i + \phi Na (b^2 + kc_g^2) -\alpha_3 \rho Na^3 -\frac{1}{4}\frac{\phi \pi_\phi^2 }{\omega a^3}N}\biggl] \label{beta_drusonis}
\end{equation}
It is straightforward to extract the usual gravitational Hamiltonian, as well as the matter and pure kinetic Brans-Dicke Hamiltonian, which read
\begin{equation}
    H = - \phi Na (b^2 + kc_g^2) + \alpha_3 \rho Na^3 + \frac{1}{4}\frac{\phi \pi_\phi^2 }{\omega a^3}N \label{gamma_drusonis}
\end{equation}
We can obtain the Hamilton's constraint by setting the Hamiltonian equal to zero, such that 
\begin{equation}
    H \overset{!}{=} 0 \Leftrightarrow b^2 + kc_g^2 = \frac{\alpha_3 \rho a^2}{\phi} + \frac{1}{4}\frac{\pi_\phi^2}{\omega a^4}
\end{equation}
Alternatively, we could obtain the Hamilton's constraint directly from the action above by considering its variation with respect to the lapse function, so that
\begin{equation}
    \frac{\delta S}{\delta N} \overset{!}{=} 0 \Leftrightarrow b^2 + kc_g^2 = \frac{\alpha_3 \rho a^2}{\phi} + \frac{1}{4}\frac{\pi_\phi^2}{\omega a^4} \label{muccone}
\end{equation}
Recalling that the canonical pair is changed by the presence of $\phi$ in the first term of action (\ref{beta_drusonis}), care must be taken when evaluating the derivatives leading to the equations of motion. To begin with, we define $A^2 = a^2 \phi$ as the canonical variable to $b$, such that the canonical pair is now $(A^2, b)$. Then, we re-express the Hamiltonian in terms of this new canonical variable, obtaining
\begin{equation}
    H = -NA\sqrt{\phi}(b^2 +kc_g^2) + \frac{\alpha_3 \rho A^3N}{\phi^{\frac{3}{2}}} + \frac{1}{4}\frac{\phi^{\frac{5}{2}} \pi_\phi^2}{\omega A^3}N \label{HHH}
\end{equation}
We continue evaluating the $\dot{a}$ equation in the usual fashion, using Hamiltonian (\ref{HHH})
\begin{equation}
    \dot{(A^2)} = \{A^2, H\} = \frac{\partial A^2}{\partial b}\frac{\partial H}{\partial A^2} - \frac{\partial A^2}{\partial A^2}\frac{\partial H}{\partial b} = -\frac{\partial H}{\partial b}
\end{equation}\\
\\
leading to 
\begin{equation}
    \dot{a} + \frac{\dot{\phi}}{2\phi}a = Nb
\end{equation}\\
which we immediately recognise to be equivalent to equation (\ref{equal_to_KTC}) in the case of a non Brans-Dicke theory. This is not a coincidence: Brans-Dicke theory is indeed a theory with a variable $\alpha_2$, with the main difference given by the addition of a kinetic term for the dynamics of the scalar field $\phi$ (or $\alpha_2$ in the analogy).\\
Secondly, we compute the $\dot{b}$ equation, re-writing the Hamiltonian in terms of $a$ and expressing $\rho$ as 
\begin{equation}
    \rho_i = \frac{m_i}{a^{3(1+w_i)}}
\end{equation}
as we did before using (\ref{rho_substitution}) for a generic fluid with different matter contents. The Hamiltonian is therefore
\begin{equation}
    H = -Na\phi (b^2 + kc_g^2) + \frac{\alpha_3 m}{a^{3w}}N + \frac{1}{4}\frac{\phi \pi_\phi^2}{\omega a^3}N
\end{equation}
and the equation for $\dot{b}$ reads
\begin{equation}
    \dot{b} = \{b, H\} = \frac{\partial b}{\partial b}\frac{\partial H}{\partial A^2} - \frac{\partial b}{\partial A^2}\frac{\partial H}{\partial b} = \frac{\partial H}{\partial A^2} = \frac{1}{2a\phi} \frac{\partial H}{\partial a}
\end{equation}
leading to
\begin{align}
    \dot{b} &= \frac{1}{2 \phi a}\frac{\partial H}{\partial a} \\ \nonumber\\
    &= \frac{1}{2 \phi a} \biggl[-N\phi (b^2 + kc_g^2) - 3w \frac{\alpha_3 m}{a^{3w +1}}N - \frac{3}{4}\frac{\phi \pi_\phi^2}{\omega a^4}N\biggl] \\ \nonumber\\ \label{qq}
    &= \frac{1}{2 \phi a} \biggl[-\alpha_3 \rho Na^2 -\frac{1}{4}\frac{\phi \pi_\phi^2 }{\omega a^4}N -3w\frac{\alpha_3 m}{a^{3w+1}}N -\frac{3}{4}\frac{\phi \pi_\phi^2}{\omega a^4}N\biggl]\\ \nonumber\\ 
    &= -\frac{1}{2}\frac{\alpha_3 \rho}{\phi}Na -\frac{3}{2}w \frac{\alpha_3\rho}{\phi}Na -\frac{1}{8}\frac{\pi_\phi^2}{\omega a^5}N -\frac{3}{8}\frac{\pi_\phi^2}{\omega a^5}N \\ \nonumber\\
    &= \frac{\alpha_3}{\phi}\frac{(-\rho -3w \rho)}{2}Na -\frac{1}{2}\frac{\pi_\phi^2}{\omega a^5}N \\ \nonumber\\ \label{ee}
    &= -\frac{\alpha_3}{2 \phi}(\rho + 3p)Na -\frac{1}{2}\frac{\pi_\phi^2}{\omega a^5}N
\end{align}   
where we have used the Hamilton's constraint in equation (\ref{qq}) and re-inserted the expression for $\rho$ in equation (\ref{ee}).\\
\\
The final piece of information we need to determine the entire dynamics of the theory is the equation of motion for $\pi_\phi$, the canonical variable of $\phi$. We begin by considering action \eqref{beta_drusonis} and the resulting Hamiltonian \eqref{gamma_drusonis}. Expressing the Hamiltonian in terms of the redefined canonical variable $A^2$ and using \eqref{rho_substitution} for $\rho$, we obtain
\begin{equation}
    H = -NA\sqrt{\phi} (b^2 + kc_g^2) + \frac{\alpha_3 m}{A^{3w}}\phi^{\frac{3}{2}w}N + \frac{1}{4}\frac{\phi^{\frac{5}{2}} \pi_\phi^2}{\omega A^3}N
\end{equation}
The Poisson's brackets for $\dot{\pi_\phi}$, given the canonical pair $(\phi, \pi_\phi)$ are 
\begin{equation}
    \dot{\pi_\phi} = \{\pi_\phi, H\} = \frac{\partial \pi_\phi}{\partial \phi}\frac{\partial H}{\partial \pi_\phi} - \frac{\partial \pi_\phi}{\partial \pi_\phi}\frac{\partial H}{\partial \phi} = -\frac{\partial H}{\partial \phi}
\end{equation}
which, when evaluated, give
\begin{align}
    \dot{\pi_\phi} &= -\frac{\partial H}{\partial \phi} \\ \nonumber \\
    &= \frac{1}{2}NA\phi^{-\frac{1}{2}} (b^2 +kc_g^2) -\frac{3}{2}w\frac{\alpha_3 m}{A^{3w}}\phi^{\frac{3w-2}{2}}N -\frac{5}{8}\frac{\pi_\phi^2}{\omega A^3}\phi^{\frac{3}{2}}N \\ \nonumber \\
    &= \frac{1}{2}Na \biggl[\frac{\alpha_3 \rho a^2}{\phi} + \frac{1}{4}\frac{\pi_\phi^2}{\omega a^4} \biggl] -\frac{3}{2}w\frac{\alpha_3 m}{a^{3w} \phi^{\frac{3}{2}w}}\phi^{\frac{3-2w}{2}}N - \frac{5}{8}\frac{\pi_\phi^2}{\omega a^3}N \\ \nonumber \\
    &= \frac{\alpha_3 \rho}{2\phi}Na^3 -\frac{3}{2}w\frac{\alpha_3 m a^3}{a^{3(1+w)}\phi}N -\frac{1}{2}\frac{\pi_\phi^2}{\omega a^3}N \\ \nonumber \\
    &= \frac{\alpha_3 \rho}{2\phi}Na^3 -\frac{3}{2}\frac{w \rho \alpha_3 a^3}{\phi}N  -\frac{1}{2}\frac{\pi_\phi^2}{\omega a^3}N \\ \nonumber \\
    &= -Na \phi \biggl[\biggl(\frac{-1 +3w}{2}\biggl)\frac{\rho \alpha_3 a^2}{\phi^2} +\frac{1}{2}\frac{\pi_\phi^2}{\omega a^4}\frac{1}{\phi}\biggl] \\ \nonumber \\
    &= -Na \phi \biggl[\biggl(\frac{-\rho + 3p}{2}\biggl)\frac{\alpha_3 a^2}{\phi^2} + \frac{1}{2}\frac{\pi_\phi^2}{\omega a^4}\frac{1}{\phi}\biggl] \label{pi_eqn}
\end{align}
\\
As before, it is possible to obtain the same equations of motion following the Lagrangian approach, but, for simplicity, we will only include the Hamiltonian derivation.
With this equation of motion, we proceed to derive the conservation equation from the Hamilton's constraint. In fact, given the Hamiltonian formalism we are using, it is possible to derive it directly from the constraint and from the equations of motion for the different variables. In a normal theory without varying constants we would recover the usual Friedmann equations, since we are working in a FLRW MSS reduction. However, assuming varying constants, we will get additional terms arising from all the $\bm{\beta}$ we wish to use. In this case, trying to get the most general form of energy conservation, we assume all constants to be varying, but this will be different in the next sections. In fact, ultimately, we can have different combinations of constants for different scenarios as we will see. Therefore, taking the time derivative of \eqref{muccone} we obtain
\begin{equation}
    2b\dot{b} + k\frac{dc_g^2}{dt} = \frac{\dot{\alpha_3}\rho a^2}{\phi} + \frac{2a \dot{a}\alpha_3 \rho}{\phi} + \frac{\dot{\rho}\alpha_3 a^2}{\phi} -\frac{\alpha_3 \rho a^2}{\phi^2}\dot{\phi} + \frac{1}{2}\frac{\pi_\phi \dot{\pi_\phi}}{\omega a^4} -\frac{1}{4}\frac{\pi_\phi^2}{\omega^2 a^4}\dot{\omega} -\frac{\pi_\phi^2}{\omega a^5}\dot{a} \label{mucchina}
\end{equation}
Now, considering the LHS of equation \eqref{mucchina}, we can plug in the equation of motion for $\dot{b}$ and the expression for $b$ coming from the $\dot{a}$ equation, giving, in the $N=1$ gauge,
\begin{align}
    2b\dot{b} + k\frac{dc_g^2}{dt} &=  2\biggl[\dot{a}+\frac{\dot{\phi}}{2\phi}a\biggl] \biggl[-\frac{\alpha_3}{2 \phi}(\rho +3p)a -\frac{1}{2}\frac{\pi_\phi^2}{\omega a^5}\biggl] \\  \nonumber\\
    &= -\frac{\alpha_3}{\phi}(\rho +3p)a\dot{a} - \frac{\pi_\phi^2}{\omega a^5}\dot{a} -\frac{\alpha_3 \dot{\phi}}{2 \phi^2}(\rho +3p)a^2 - \frac{\pi_\phi^2 \dot{\phi}}{2 \omega \phi a^4} \\ \nonumber \\
    &= -(\rho +3p)\frac{\dot{a}}{a} - \frac{\phi \pi_\phi^2}{\alpha_3 \omega a^7}\dot{a} - \frac{\dot{\phi}}{2 \phi}(\rho + 3p) - \frac{\pi_\phi^2 \dot{\phi}}{2 \alpha_3 \omega a^6}
\end{align}
where in the last line we have multiplied every term by $\frac{\phi}{\alpha_3 a^2}$. Doing the same on the RHS of equation \eqref{mucchina} above gives
\begin{equation}
    \frac{\dot{\alpha_3}}{\alpha_3}\rho +2\frac{\dot{a}}{a}\rho + \dot{\rho} - \frac{\dot{\phi}}{\phi}\rho +\frac{1}{2}\frac{\pi_\phi \dot{\pi_\phi} \phi}{\alpha_3 \omega a^6} -\frac{1}{4}\frac{\phi \pi_\phi^2}{\alpha_3 \omega^2 a^6}\dot{\omega} -\frac{\phi \pi_\phi^2}{\alpha_3 \omega a^7}\dot{a} 
\end{equation}
Now, equating the LHS and the RHS and rearranging leads to 

\begin{gather}
    \dot{\rho} +3\frac{\dot{a}}{a}(\rho + p) = -\frac{\dot{\alpha_3}}{\alpha_3}\rho +\frac{\dot{\phi}}{\phi}\rho -\frac{\dot{\phi}}{\phi}\frac{\rho + 3p}{2} -\frac{\dot{\phi} \pi_\phi^2}{2 \alpha_3 \omega a^6} -\frac{1}{2}\frac{\pi_\phi \dot{\pi_\phi} \phi}{\alpha_3 \omega a^6} +\frac{1}{4}\frac{\phi \pi_\phi^2}{\alpha_3 \omega^2 a^6}\dot{\omega} + \frac{k \phi}{\alpha_3 a^2}\frac{dc_g^2}{dt} \label{this_one} \\ \nonumber \\ \label{rrr}
    = -\frac{\dot{\alpha_3}}{\alpha_3}\rho + \frac{\pi_\phi}{2 \omega a^3}\rho -\frac{\pi_\phi}{2 \omega a^3}\frac{\rho + 3p}{2} - \frac{1}{4}\frac{\phi \pi_\phi^3}{\alpha_3 \omega^2 a^9} + \frac{1}{4}\frac{\phi \pi_\phi^2}{\alpha_3 \omega^2 a^6}\dot{\omega} +\frac{k \phi}{\alpha_3 a^2}\frac{dc_g^2}{dt} -\frac{1}{2}\frac{\pi_\phi \dot{\pi_\phi} \phi}{\alpha_3 \omega a^6} \\ \nonumber \\ \label{bbbb}
    = -\frac{\dot{\alpha_3}}{\alpha_3}\rho + \frac{\pi_\phi}{2 \omega a^3}\biggl[\rho -\frac{\rho}{2}-\frac{3}{2}p\biggl]+\frac{1}{4}\frac{\phi \pi_\phi^2}{\alpha_3 \omega^2 a^6}\dot{\omega} -\frac{1}{4}\frac{\pi_\phi}{\omega a^3}(\rho -3p) +\frac{k \phi}{\alpha_3 a^2}\frac{dc_g^2}{dt} \\ \nonumber \\
    = -\frac{\dot{\alpha_3}}{\alpha_3}\rho +\frac{1}{4}\frac{\phi \pi_\phi^2}{\alpha_3 \omega^2 a^6}\dot{\omega} + \frac{k \phi}{\alpha_3 a^2}\frac{dc_g^2}{dt} \label{general_full_phi}
\end{gather}
where we have used the definition of $\dot{\phi}$,
\begin{equation}
    \pi_\phi = 2 \omega \frac{\dot{\phi}}{\phi}a^3 \Leftrightarrow \dot{\phi} = \frac{\phi \pi_\phi}{2 \omega a^3}
\end{equation}
for \eqref{rrr} and we have also used \eqref{pi_eqn}, the equation of motion for $\pi_\phi$, in line \eqref{bbbb}, always setting the gauge such that $N = 1$.
\\
\\
This is the most general form of the energy conservation equation, assuming that all the constants are varying and using the equations of motion for $a$, $b$ and $\phi$ to eliminate the extra factors, as well as the definition of the conjugate momentum $\pi_\phi$. From this general equation, it will be possible to consider specific scenarios where only one of the $\bm{\beta}$ is varying with respect to one or multiple times. On the other hand, we could also have multiple $\bm{\beta}$ varying with respect to one single time or multiple times (the same one or different ones): imagination is really the limit.

\subsection{Full $\phi$ with $\omega = \omega(T_\Lambda)$} \label{omega_starrt}

In this scenario, we assume a Brans-Dicke coupling $\omega$ varying with respect to the unimodular time $T_{\Lambda}$. This means that, in the choice of our constants, we have $\bm{\beta} = \omega$ and $\bm{\alpha} = \rho_\Lambda$ giving the unimodular time as its canonical variable. In this case, all the other constants are kept fixed. An important observation to make is that the case where $\phi = \phi_0$ in the canonical pair does not lead to any different result. This is not, however, the case in the next subsection, as we will see. Therefore, to consider the dependence $\omega = \omega(T_\Lambda)$ we start from the full Brans-Dicke action, coupled to matter and inclusive of the unimodular term
\begin{equation}
    S = V_c \int{dt \: \phi \dot{b} a^2 + \dot{\phi} \pi_\phi + \alpha_3 \dot{m}_i \psi_i + \phi Na (b^2 + kc_g^2) -\alpha_3 \rho Na^3 - \frac{1}{4} \frac{\phi \pi_\phi^2}{a^3 \omega}N + \dot{\rho_{\Lambda}} T_\Lambda}
\end{equation}
In this scenario, we simply assume that $\alpha_3$ does not have any time dependence. This implies that our general equation for energy conservation \eqref{general_full_phi} becomes
\begin{equation}
    \dot{\rho} +3\frac{\dot{a}}{a}(\rho + p) =  \frac{1}{4}\frac{\phi \pi_\phi^2}{\alpha_3 \omega^2 a^6}\dot{\omega} + \frac{k \phi}{\alpha_3 a^2}\frac{dc_g^2}{dt} \label{wewe}
\end{equation}
Furthermore, since we assume that only $\omega$ has a time dependence, the second term including the varying $c_g$ disappears, simplifying equation \eqref{wewe} to
\begin{equation}
    \dot{\rho} +3\frac{\dot{a}}{a}(\rho + p) =  \frac{1}{4}\frac{\phi \pi_\phi^2}{\alpha_3 \omega^2 a^6}\dot{\omega}
\end{equation}
Finally, since we are considering the presence of the cosmological constant as a matter content, the full conservation equation needs to be inclusive of a cosmological constant density term, therefore giving 
\begin{equation}
    \dot{\rho} +3\frac{\dot{a}}{a}(\rho + p) + \dot{\rho_{\Lambda}} =  \frac{1}{4}\frac{\phi \pi_\phi^2}{\alpha_3 \omega^2 a^6}\dot{\omega} \label{wwwww}
\end{equation}
where $\rho$ and $p$ give the usual matter content density and pressure. Now, we wish to obtain an equation for $\dot{\rho_{\Lambda}}$ as this will give us the full conservation equation. We can obtain this from the additional Hamilton's equations for the canonical pair $\rho_{\Lambda} T_{\Lambda}$. Starting from the Hamiltonian 
\begin{equation}
    H = -\phi Na(b^2 + kc_g^2) + \alpha_3 \rho Na^3 + \frac{1}{4}\frac{\phi \pi_\phi^2}{\omega a^3}N \label{saladino_H}
\end{equation}
we have the two Hamilton's equations 
\begin{equation}
    \dot{\rho_{\Lambda}} = \{\rho_{\Lambda}, H \} = \frac{\partial H}{\partial T_{\Lambda}} = - N\frac{1}{4}\frac{\phi \pi_\phi^2}{\omega^2 a^3} \frac{\partial \omega}{\partial T_{\Lambda}}
\end{equation}
\begin{equation}
    \dot{T_{\Lambda}} = \{T_{\Lambda}, H \} = - \frac{\partial H}{\partial \rho_{\Lambda}} = - \alpha_3 Na^3
\end{equation}
Thus, using the equation for $T_{\Lambda}$ into the one for $\rho_{\Lambda}$, we obtain 
\begin{equation}
    \dot{\rho_{\Lambda}} = - N\frac{1}{4}\frac{\phi \pi_\phi^2}{\omega^2 a^3} \frac{d \omega}{dt}\dot{T_{\Lambda}}^{-1} = \frac{1}{4}\frac{\phi \pi_\phi^2}{\alpha_3 \omega^2 a^6}\dot{\omega}
\end{equation}
We immediately see that $\dot{\rho_{\Lambda}}$ has the same form as the source term in equation \eqref{wwwww}, therefore we can substitute it in it and confirm that indeed 
\begin{equation}
    \dot{\rho} +3\frac{\dot{a}}{a}(\rho + p) = 0
\end{equation}
This result shows that there is no net energy violation solely given by a unimodular time-dependent $\omega$, confirming that energy is conserved for all the other matter contents.\\
Furthermore, can combine the equation of motion for $\rho_{\Lambda}$ with the definition of $\pi_\phi$ from equation \eqref{conj_mom_def} (where we choose the gauge $N = 1$) to obtain an explicit equation for the evolution of the vacuum energy density with $\omega$. This substitution leads to 
\begin{equation}
    \dot{\rho}_\Lambda = \frac{\dot{\phi}^2}{\phi}\dot{\omega}
\end{equation}
where, for simplicity, we have set $\alpha_3 = 1$. This solution is valid for any $\omega$ depending on $T_{\Lambda}$ and suggests that the vacuum energy density increases with increasing $\omega$. Further interpretations are certainly needed, and will be provided in our future research \cite{bassani2023}.

\subsection{Full $\phi$ with $\omega = \omega( T_N)$ and $\alpha_3 = \alpha_3 (T_\omega)$}

We can now consider the scenario where $\bm{\alpha} = \alpha_3 $ and $\bm{\beta} = \omega$, such that the Brans-Dicke parameter is dependent on the Newton time $T_N$. As it is evident from the title, this scenario can also be recast as the $\alpha_3$ parameter depending on $T_\omega$: both these scenarios are equivalent, leading to the same results. The action we start from is 
\begin{equation}
    S = V_c \int{dt \: \phi \dot{b} a^2 + \dot{\phi} \pi_\phi + \alpha_3 \dot{m}_i \psi_i + \phi Na (b^2 + kc_g^2) -\alpha_3 \rho Na^3 - \frac{1}{4} \frac{\phi \pi_\phi^2}{a^3 \omega}N + \dot{\omega} T_N}
\end{equation}
If we wish to consider a varying $\omega$ only, the general conservation equation reduces to 
\begin{equation}
    \dot{\rho} +3\frac{\dot{a}}{a}(\rho + p) = \frac{1}{4}\frac{\phi \pi_\phi^2}{\alpha_3 \omega^2 a^6}\dot{\omega} \label{lalalala}
\end{equation}
where, since $\omega = \omega(T_N)$ only, we can use the chain rule on $\dot{\omega}$ thus obtaining
\begin{equation}
    \frac{d \omega}{dt} = \frac{\partial \omega}{\partial T_N}\dot{T_N} = -\rho Na^3 \frac{\partial \omega}{\partial T_N}
\end{equation}
where, in the last step, we have use the $T_N$ equation of motion as given by 
\begin{equation}
    \dot{T_N} = \{T_N, H\}=  - \frac{\partial H}{\partial \alpha_3} = - \rho Na^3
\end{equation}
Equipped with these, we can then plug everything inside the conservation equation thus arriving at 
\begin{equation}
    \dot{\rho} +3\frac{\dot{a}}{a}(\rho + p) = -\frac{\rho}{4}\frac{\phi \pi_\phi^2 N}{\alpha_3 \omega^2 a^3}\frac{\partial \omega}{\partial T_N}
\end{equation}
which shows that a gravitational parameter depending on a matter clock does lead to an overall energy violation, given by the source term of the constant varying with respect to the matter clock. The same situation would be true if we were to revert the dependency, postulating that $\alpha_3 = \alpha_3 (T_\omega)$. In this parallel scenario, the conservation equation would be
\begin{equation}
    \dot{\rho} +3\frac{\dot{a}}{a}(\rho + p) = -\frac{\dot{\alpha_3}}{\alpha_3}\rho \label{rain}
\end{equation}
since, this time around, $\omega$ is constant. The relevant equation of motion would be the $\dot{T_\omega}$ equation, reading 
\begin{equation}
    \dot{T_\omega} = \{T_\omega, H\} = -\frac{\partial H}{\partial \omega} = \frac{1}{4}\frac{\phi \pi_\phi^2 N}{\omega^2 a^3}
\end{equation}
Using this equation of motion in $\dot{\alpha_3}$ leads to 
\begin{equation}
    \dot{\rho} +3\frac{\dot{a}}{a}(\rho + p) = -\frac{\rho}{4}\frac{\phi \pi_\phi^2 N}{\alpha_3 \omega^2 a^3}\frac{\partial \alpha_3}{\partial T_\omega}
\end{equation}
which is exactly what we expected: this conservation equation has the same pre-factor as equation \eqref{rain}, with the only difference being the varying constant. In the first case, $\omega$ varies and $\alpha_3$ provides the clock, while in the second one $\alpha_3$ varies and $\omega$ provides the clock. They both leads to the same net energy violation, but due to different varying source terms.\\
To appreciate better this equivalence, we derive a relationship between $\omega$ and $\alpha_3$. We take the equation of motion for $\omega$ (we could have also picked the one for $\alpha_3$),
\begin{equation}
    \dot{\omega} = \{\omega, H\} = \frac{\partial H}{\partial T_\omega} = \rho Na^3 \frac{\partial \alpha_3}{\partial T_\omega}
\end{equation}
where $\alpha_3$ is assumed to depend on $T_\omega$. Then, using the chain rule on the dependency of $\alpha_3$ on $T_\omega$ and invoking the definition of $\pi_\phi$ we obtain 
\begin{equation}
    \dot{\omega} = \frac{\phi \rho}{\dot{\phi}^2}\dot{\alpha_3} 
\end{equation}
where we have set $N = 1$. This equation could have been equivalently obtained starting from the $\alpha_3$ equation of motion and following the same procedure. Finally, we note that, when considering both $\alpha_3$ and $\omega$ varying in the same scenario, their source terms contributions to the energy violation cancel out, giving the usual overall energy conservation equation
\begin{equation}
    \dot{\rho} +3\frac{\dot{a}}{a}(\rho + p) = 0
\end{equation}

\subsection{Fixed $\phi_0$ with $\omega = \omega(T_\phi)$} \label{subsection_omega}

In this subsection, we explore the main idea of Brans-Dicke theory in the context of varying constants. In fact, having the additional constant $\omega$, we can make it dependent on $\phi$, creating a model of Brans-Dicke parameter dependent on Ricci time. To begin with, we consider the simplified case where the scalar field factor multiplying the fist canonical pair is fixed constant, i.e., $\phi = \phi_0$. This case is apparently different from the next one where the equations of motion are different, but, as we will see, they are actually the same. It happens so because ultimately the conservation equation is the same one, as it derives directly from the conservation of the energy-momentum tensor. Furthermore, it is important to notice that, as in the next subsection, the $\phi$ equation of motion needs to be modified to account for the extra dependency of $\omega$ on $\pi_\phi$. Finally, given the equivalence of $T_\phi$ and $\pi_\phi$, we do not need to include the unimodular term in the action, as the canonical pair from the Brans-Dicke Hamiltonian suffice. Therefore, the action we will use is
\begin{equation}
    S = V_c \int{dt \: \phi_0 \dot{b} a^2 + \dot{\phi} \pi_\phi + \alpha_3 \dot{m}_i \psi_i + \phi Na (b^2 + kc_g^2) -\alpha_3 Na^3 \rho - \frac{1}{4} \frac{\phi \pi_\phi^2}{\omega a^3}N}
\end{equation}
From subsection \eqref{de_cou_pled} we have the general conservation equation which we take in the form given by \eqref{drusillo_oneee} 
\begin{equation}
      \dot{\rho} + 3\frac{\dot{a}}{a}(\rho +p) = \frac{\dot{\phi}}{\phi}\rho -\frac{1}{2}\frac{\phi \pi_\phi \dot{\pi_\phi}}{\alpha_3 \omega a^6} + \frac{1}{4}\frac{\phi \pi_\phi^2}{\alpha_3 \omega^2 a^6}\dot{\omega} 
\end{equation}
where we have also set $\dot{\alpha_3} = \dot{c_g^2} = 0$ because, in this particular scenario, we are interested only in a varying $\omega$.
We can now compute the $\phi$ equation of motion, which reads 
\begin{equation}
    \dot{\phi} = \{\phi, H\} = \frac{\partial H}{\partial \pi_\phi} = \frac{1}{2}\frac{\phi \pi_\phi}{\omega a^3}N -\frac{1}{4}\frac{\phi \pi_\phi^2}{\omega^2 a^3}\frac{d \omega}{d \pi_\phi}N
\end{equation}
where, importantly, we notice the extra term for the $\omega$ dependency. On the other hand, the $\pi_\phi$ equation of motion is the same one as \eqref{half_mom}, reading
\begin{equation}
    \dot{\pi_\phi} = \{\pi_\phi, H\} = -\frac{\partial H}{\partial \phi} = \frac{\alpha_3 \rho}{\phi} Na^3
\end{equation}
Finally, expressing $\dot{\omega}$ in terms of $T_\phi$ as
\begin{equation}
    \frac{d \omega}{dt} = \frac{\partial \omega}{\partial T_\phi} \dot{T_\phi} = \frac{\partial \omega}{\partial T_\phi} \dot{\pi_\phi} = \frac{\alpha_3 \rho}{\phi} Na^3 \frac{\partial \omega}{\partial T_\phi}
\end{equation}
we obtain the final form of the conservation equation
\begin{equation}
    \dot{\rho} + 3\frac{\dot{a}}{a}(\rho +p) = 0
\end{equation}
which shows that energy is indeed conserved. This is an expected result because, as we will see, gravitational parameters depending on gravitational clocks do not lead to any net energy violation. As we now see, the full $\phi$ case leads to the same results.

\subsection{Full $\phi$ with $\omega = \omega (T_\phi)$}

In this subsection, we will explore an interesting and unusual scenario, where the Brans-Dikce parameter $\omega$ depends on the conjugate momentum of the scalar field, such that $\omega = \omega (\pi_\phi)$. The first observation we make is that the $\omega$ dependence on a Ricci-like time as $T_\phi$ is perfectly equivalent to its dependence on $\pi_\phi$. In fact, given that $\pi_\phi$ is the conjugate momentum to $\phi$ in the Hamiltonian formalism and that $T_\phi$ is also the conjugate of $\phi$, we see that they are equivalent, meaning that $\pi_\phi$ is indeed the relational Ricci time given by $\phi$. Secondly, precisely because of this equivalence, the unimodular term in the action represent an interesting but redundant addition. In fact, we are not required to add it to include the Ricci time as it is already provided by the canonical pair $\dot{\phi} \pi_\phi$ coming from the Hamiltonian. This is, however, an interesting occurrence, which we will potentially explore in our future works \cite{bassani2023}. Therefore, the action we will in our derivations is
\begin{equation}
    S = V_c \int{dt \: \phi \dot{b} a^2 + \dot{\phi} \pi_\phi + \alpha_3 \dot{m}_i \psi_i + \phi Na (b^2 + kc_g^2) -\alpha_3 Na^3 \rho - \frac{1}{4} \frac{\phi \pi_\phi^2}{a^3 \omega}N}
\end{equation}
Before proceeding, a crucial \textit{caveat} must be addressed. We might be tempted to just use the general conservation equation \eqref{general_full_phi} with only the $\dot{\omega}$ term as only $\omega$ is varying, but this would be a mistake. In fact, when deriving this equation, we have used the equation for $\dot{\phi}$ obtained by inverting the definition of the conjugate momentum $\pi_\phi$. However, when doing so, we were assuming that only $\phi$ depends on $\pi_\phi$ and that no other terms in the Hamiltonian contribute. This case is different: we have an additional dependency coming from $\omega$, which implies an extra term in the $\dot{\phi}$ equation of motion. As a result, we need to re-derive the conservation equation, accounting for this extra term. We start from its general form \eqref{this_one}, where, after setting $\dot{\alpha_3} = \dot{c_g^2} = 0$ in this particular case and without using the standard $\dot{\phi}$ equation, we have 
\begin{equation}
     \dot{\rho} +3\frac{\dot{a}}{a}(\rho + p) = \frac{\dot{\phi}}{\phi}\rho -\frac{\dot{\phi}}{\phi}\frac{\rho + 3p}{2} -\frac{\dot{\phi} \pi_\phi^2}{2 \alpha_3 \omega a^6} -\frac{1}{2}\frac{\pi_\phi \dot{\pi_\phi} \phi}{\alpha_3 \omega a^6} +\frac{1}{4}\frac{\phi \pi_\phi^2}{\alpha_3 \omega^2 a^6}\dot{\omega}
\end{equation}
It is precisely at this point that we go back to Hamiltonian \eqref{gamma_drusonis} to derive the $\dot{\phi}$ equation, this time accounting for the fact that, in this particular case, $\omega = \omega (\pi_\phi)$ 
\begin{equation}
    \dot{\phi} = \{\phi, H\} = \frac{\partial H}{\partial \pi_\phi} = \frac{1}{2}\frac{\phi \pi_\phi}{\omega a^3}N -\frac{1}{4}\frac{\phi \pi_\phi^2}{\omega^2 a^3}\frac{d \omega}{d \pi_\phi}N \label{sala}
\end{equation}
while the $\dot{\pi_\phi}$ equation of motion takes the usual form as
\begin{equation}
    \dot{T_\phi} = \{T_\phi, H\} = -\frac{\partial H}{\partial \phi} = -Na \phi \biggl[\biggl(\frac{-\rho + 3p}{2}\biggl)\frac{\alpha_3 a^2}{\phi^2} + \frac{1}{2}\frac{\pi_\phi^2}{\omega a^4}\frac{1}{\phi}\biggl] = \dot{\pi_\phi} \label{dino}
\end{equation}
We may now use equations \eqref{sala} and \eqref{dino} in the conservation equation above, where we also express $\dot{\omega}$ in terms of $\dot{\pi_\phi}$ using the chain rule. This leads to 
\begin{align}
    \dot{\rho} +3\frac{\dot{a}}{a}(\rho + p) = -\frac{(-\rho + 3p)}{8}\frac{\pi_\phi^2 N}{\omega^2 a^3}\frac{d \omega}{d \pi_\phi} -\frac{1}{8}\frac{\phi \pi_\phi^3 N}{\alpha_3 \omega^3 a^9}\frac{d \omega}{d \pi_\phi} +\frac{\rho}{2}\frac{\pi_\phi N}{\omega a^3} \\ \nonumber\\ -\frac{\rho}{4}\frac{\pi_\phi^2 N}{\omega^2 a^3}\frac{d \omega}{d \pi_\phi}  - \frac{(\rho + 3p)}{4}\frac{\pi_\phi N}{\omega a^3} + \frac{(\rho + 3p)}{8}\frac{\pi_\phi^2 N}{\omega^2 a^3}\frac{d \omega}{d \pi_\phi} -\frac{1}{4}\frac{\phi \pi_\phi^3 N}{\alpha_3 \omega^2 a^9} \\ \nonumber \\+ \frac{1}{8}\frac{\phi \pi_\phi^3 N}{\alpha_3 \omega^3 a^9}\frac{d \omega}{d \pi_\phi} +\frac{(-\rho + 3p)}{4}\frac{\pi_\phi N}{\omega a^3} +\frac{1}{4}\frac{\phi \pi_\phi^3 N}{\alpha_3 \omega^2 a^9} = 0
\end{align}
which simplifies to the standard conservation equation. This scenario shows two important results. Firstly, once again, we confirm that a gravitational parameter depending on a gravitational clock does not lead to any net energy violation. Secondly, we are reminded that any scenario including a dependency on the Ricci clock should be treated with additional care, as extra terms, which are normally not included, will arise in the general conservation equation. 

\subsection{Full $\phi$ for $c_g = c_g(T_{\omega})$}

We could, of course, change the pattern by considering $\omega$ as part of the $\bm{\alpha}$ constants giving times, while a different constant is part of the $\bm{\beta}$. In this case, we will consider the scenario where $\bm{\alpha} = \omega$, $T_{\bm{\alpha}} = T_{\omega}$ and $\bm{\beta} = c_g$, while the other constants appearing in the conservation equation are fixed, i.e., $\dot{\alpha_3} = \dot{\omega} = 0$. Therefore, our starting point is the general Brans-Dicke action with the $\omega$ unimodular term, reading
\begin{equation}
     S = V_c \int{dt \: \phi \dot{b} a^2 + \dot{\phi} \pi_\phi + \alpha_3 \dot{m}_i \psi_i + \phi Na (b^2 + kc_g^2) -\alpha_3 Na^3 \rho - \frac{1}{4} \frac{\phi \pi_\phi^2}{a^3 \omega}N + \omega \dot{T_\omega}}
\end{equation}
while the conservation equation for a varying $c_g$ only is 
\begin{equation}
    \dot{\rho} +3\frac{\dot{a}}{a}(\rho + p) = \frac{k \phi}{\alpha_3 a^2}\frac{dc_g^2}{dt}
\end{equation}
The Hamiltonian is the same one as \eqref{saladino_H} and we can use it to derive the Hamilton's equation for $T_\omega$. Also, since we are not computing any derivative with respect to $\phi$ that would require the use of $A^2$, the $\phi_0$ and $\phi$ cases lead to equivalent results. Using the Hamilton's equation, we obtain 
\begin{equation}
    \dot{T_\omega} = \{T_\omega, H\} = -\frac{\partial H}{\partial \omega} = \frac{1}{4}\frac{\phi \pi_\phi^2}{\omega^2 a^3}N
\end{equation}
The conservation equation, given the dependence of $c_g$ on $T_\omega$ becomes 
\begin{equation}
    \dot{\rho} +3\frac{\dot{a}}{a}(\rho + p) = \frac{k \phi}{\alpha_3 a^2}\frac{\partial c_g^2}{\partial T_\omega} \dot{T_\omega}
\end{equation}
which finally gives, using the $T_\omega$ equation of motion and the definition of $\pi_\phi$, 
\begin{equation}
    \dot{\rho} +3\frac{\dot{a}}{a}(\rho + p) = \frac{k a \dot{\phi^2}}{\alpha_3 N} \frac{\partial c_g^2}{\partial T_\omega}
\end{equation}\\
Furthermore, since $T_\omega$ is canonical to $\omega$ and we also have its equation of motion, we consider the inclusion of the $\dot{\omega}$ term in the conservation equation, thus obtaining a direct relation between the evolution of $\omega$ and $c_g$. To do so, we consider the more general conservation equation where also $\omega$ is varying:
\begin{equation}
    \dot{\rho} +3\frac{\dot{a}}{a}(\rho + p) = \frac{1}{4}\frac{\phi \pi_\phi^2}{\alpha_3 \omega^2 a^6} \dot{\omega} + \frac{k \phi}{\alpha_3 a^2}\frac{dc_g^2}{dt}
\end{equation}
The equation of motion for $T_\omega$ is the same one as above, but we additionally have the $\omega$ Hamilton's equation, reading
\begin{equation}
    \dot{\omega} = \{\omega, H\} = \frac{\partial H}{\partial T_\omega} = -k \phi Na \frac{\partial c_g^2}{\partial T_\omega}
\end{equation}
Substituting this inside the conservation equation and expanding the time dependence of $c_g$ in terms of $T_\omega$, we arrive at 
\begin{equation}
    \dot{\rho} +3\frac{\dot{a}}{a}(\rho + p) = -\frac{1}{4}\frac{\phi^2 k \pi_\phi^2}{\alpha_3 \omega^2 a^5}\frac{\partial c_g^2}{\partial T_\omega}N + \frac{k a \dot{\phi}^2}{\alpha_3 N}\frac{\partial c_g^2}{\partial T_\omega}
\end{equation}
which ultimately simplifies to 
\begin{equation}
    \dot{\rho} +3\frac{\dot{a}}{a}(\rho + p) = 0 \label{dubi}
\end{equation}
using the definition of $\pi_\phi$. Furthermore, we can derive a relationship between the Brans-Dicke parameter and the speed of light, such that 
\begin{equation}
    \dot{\omega} = -\frac{k}{a^2}\frac{\phi^2}{\dot{\phi}^2}\dot{c}_g^2 \label{evol}
\end{equation}

\subsection{Full $\phi$ with $\omega = \omega (T_{c_g})$} \label{mucchinella_drusina}

We might, of course, think about reverting the logic used in the scenario above. In fact, we could consider a fixed speed of light as part of the $\bm{\alpha}$ giving a relational time for the evolution of $\omega$ which becomes part of the $\bm{\beta}$. This example just shows how the definitions of constants and times can be easily interchangeable, crating interesting mixes of concepts: in some cases a constant could be a time for evolution, in other cases it varies with respect to a different time. Therefore, it is clear how the idea of ``evolution with respect to what'' creates a circular argument where different quantities can take roles that are dual to each others. Having established this, we proceed by considering our starting point, the action:
\begin{equation}
    S = V_c \int{dt \: \phi \dot{b} a^2 + \dot{\phi} \pi_\phi + \alpha_3 \dot{m}_i \psi_i + \phi Na (b^2 + kc_g^2) -\alpha_3 Na^3 \rho - \frac{1}{4} \frac{\phi \pi_\phi^2}{a^3 \omega}N + c_g^2 \dot{T_{c_g^2}}}
\end{equation}
where we notice that we keep $c_g^2$ as in the unimodular term as it appears in the Hamiltonian for dimensional reasons.
Given the dependency of $\omega$ on the gravitational speed of light, the conservation equation contains the source term given by $\dot{\omega}$ and the one given by $\dot{c_g}^2$, i.e., 
\begin{equation}
    \dot{\rho} +3\frac{\dot{a}}{a}(\rho + p) = \frac{1}{4}\frac{\phi \pi_\phi^2}{\alpha_3 \omega^2 a^6}\dot{\omega} + \frac{k \phi}{\alpha_3 a^2}\frac{dc_g^2}{dt}
\end{equation}
where we have included the $\dot{c_g}^2$ term because of the equation of motion for $c_g$ coming from the Hamiltonian.
The equations of motion for the canonical pair $(c_g^2, T_{c_g^2})$ are
\begin{equation}
    \dot{c_g^2} = \{c_g^2, H\} = \frac{\partial H}{\partial T_{c_g^2}} = -\frac{1}{4}\frac{\phi \pi_\phi^2 N}{ \omega^2 a^3}\frac{\partial \omega}{\partial T_{c_g^2}}
\end{equation}
\begin{equation}
    \dot{T_{c_g^2}} = \{T_{c_g^2}, H\} = -\frac{\partial H}{\partial c_g^2} = \phi Na k
\end{equation}
The time equation can be substituted inside $\dot{\omega}$, giving, while using also the definition of $\pi_\phi$, 
\begin{equation}
    \dot{\rho} +3\frac{\dot{a}}{a}(\rho + p) = \frac{ka}{\alpha_3 N}\dot{\phi}^2 \frac{\partial \omega}{\partial T_{c_g^2}} + \frac{k \phi}{\alpha_3 a^2}\dot{c_g^2}
\end{equation}
which, when using the $\dot{c_g}^2$ equation in the second term, finally simplifies to 
\begin{equation}
    \dot{\rho} +3\frac{\dot{a}}{a}(\rho + p) = 0
\end{equation}
as expected. This result is not a coincidence: it is, in fact, the same obtained in \eqref{dubi} in the previous section. We might have expected this, since these two scenarios differ only in whether the $\omega$ and the $c_g^2$ are playing the role of a constant or of a time. In both cases, since they are both gravitational parameters, the change in one is compensated by the change in the other, thus producing no net energy violation. We can see how these two parameters are not just related, but they are really dual to each others, in the sense that they absorb/emit exactly each other's variations. To confirm that this is indeed the case, we can express $\dot{c_g^2}$ in terms of $\dot{\omega}$, arriving at 
\begin{equation}
    \dot{c_g^2} = - \frac{a^2 \dot{\phi}^2}{k N^2 \phi^2}\dot{\omega}
\end{equation}
which is exactly the same as equation \eqref{evol} above. The interpretation of this law is that as the speed of light undergoes the sharp phase transition from the early Universe until now, $\omega$ increases until the large value observed today. Of course, the change in order of magnitudes in $c_g$ has to have been very large according to \cite{albrecht1999time}, larger then the change in $\omega$. However, to account for that, we could consider more closely the pre-factor and $\dot{\phi}$, as we will investigate in \cite{bassani2023}.

\subsection{Full $\phi$ with $c_m^2 = c_m^2 (T_\omega)$}

In this subsection we enrich the marvellous landscape of varying constants considering a further dependence on our new parameter, $\omega$. We analyse the scenario where $\bm{\alpha} = \omega$ and $\bm{\beta} = c_m^2$. Furthermore, since the addition of a varying $c_g^2$ does not affect the overall conservation equation, we postulate that also $c_g^2 = c_g^2 (T_\omega)$ for completeness. The action we begin with is 
\begin{equation}
    S = V_c \int{dt \: \phi \dot{b} a^2 + \dot{\phi} \pi_\phi + \alpha_3 \dot{m}_i \psi_i + \phi Na (b^2 + kc_g^2) -\alpha_3 Na^3 \rho - \frac{1}{4} \frac{\phi \pi_\phi^2}{a^3 \omega}N + \dot{\omega}{T_\omega}}
\end{equation}
which gives us the usual Hamiltonian 
\begin{equation}
     H = -\phi Na(b^2 + kc_g^2) +\alpha_3 \rho Na^3 + \frac{1}{4}\frac{\phi \pi_\phi^2}{\omega a^3}N
\end{equation}
The conservation equation for a varying $c_g^2$ and $c_m^2$ dependent on $T_\omega$ includes two source terms,
\begin{equation}
    \dot{\rho} +3\frac{\dot{a}}{a}(\rho + p) = \frac{1}{4}\frac{\phi \pi_\phi^2}{\alpha_3 \omega^2 a^6}\dot{\omega} + \frac{k \phi}{\alpha_3 a^2}\frac{dc_g^2}{dt} \label{const_cons}
\end{equation}
To find explicit equations for $\dot{\omega}$ and $\dot{c_g^2}$ above, we derive the Hamilton's equations as
\begin{equation}
    \dot{\omega} = \{\omega, H\} = \frac{\partial H}{\partial T_\omega} = -\phi Na k \frac{\partial c_g^2}{\partial T_\omega} +\alpha_3 Na^3 \frac{\partial \rho}{\partial c_m^2}\frac{\partial c_m^2}{\partial T_\omega} \label{muccone}
\end{equation}
\begin{equation}
    \dot{T_\omega} = \{T_\omega, H\} = -\frac{\partial H}{\partial \omega} = \frac{1}{4}\frac{\phi \pi_\phi^2}{\omega^2 a^3}N
\end{equation}
We can now substitute the $\dot{\omega}$ equation in \eqref{const_cons} and the $\dot{T_\omega}$ one in the $\dot{c_g^2}$ source term to obtain
\begin{equation}
    \dot{\rho} +3\frac{\dot{a}}{a}(\rho + p) = \frac{\dot{\phi}^2}{\phi}a^3 \frac{\partial \rho}{\partial c_m^2}\frac{\partial c_m^2}{\partial T_\omega}
\end{equation}
where the $\dot{c_g^2}$ term has cancelled the first contribution to $\dot{\omega}$ and we have used the definition of $\pi_\phi$ and we have set $N = 1$. 
\\
\\
We can go beyond this result, exploring the relationship between the Brans-Dicke parameter and the matter speed of light. Starting form \eqref{muccone}, we assume $\dot{c_g^2} = 0$ because we wish to find a relationship only between $\omega$ and $c_m^2$, since we already have the one for $c_g^2$ in subsection \eqref{mucchinella_drusina}. Therefore, using the $\dot{\omega}$ equation jointly with the $\dot{T_\omega}$ one and the definition of $\pi_\phi$ we finally obtain 
\begin{equation}
    \dot{\omega} = \alpha_3 \frac{\phi}{\dot{\phi}^2}\frac{\partial \rho}{\partial c_m^2}\dot{c}_m^2
\end{equation}
Unlike the previous case where $c_g^2 = c_g^2 (T_\omega)$, the matter speed of light increases proportionally with the Brans-Dicke parameter. This will be further investigated in \cite{bassani2023}.

\subsection{Full $\phi$ with $c_m^2 = c_m^2 (T_\phi, T_\omega)$}

In the spirit of considering all the possible extensions of $T_\omega$, we illustrate the scenario where $\bm{\alpha} = (\phi, \omega) $ and $\bm{\beta} = c_m^2$. For completeness, we also include a varying $c_g^2$ depending, just like $c_m^2$, on both $T_\phi$ and $T_\omega$ as we will see. As in previous sections, the unimodular term for the Ricci time is redundant as already provided by the canonical pair, so the only unimodular term is for $T_\omega$. The action reads
\begin{equation}
     S = V_c\int{dt \: \phi \: \dot{b}a^2 + \dot{\phi}\pi_\phi + \alpha_3 \dot{m}_i \psi_i + \phi Na(b^2 + kc_g^2) -\alpha_3Na^3 \rho - \frac{1}{4}\frac{\phi}{\omega}\frac{\pi_\phi^2}{a^3}N + \dot{\omega} T_\omega}
\end{equation}\\
As always in the scenarios were we have some dependency on $T_\phi$, we need to be more careful about which conservation equation to use. In fact, the general form \eqref{general_full_phi} should not be used, as the extra dependency of $c_m^2$ on $T_\phi$ has not been taken into consideration. Therefore, we start with a more general form, where the $\dot{\phi}$ terms have not been substituted for\\
\begin{equation}
     \dot{\rho} +3\frac{\dot{a}}{a}(\rho + p) = \frac{\dot{\phi}}{\phi}\rho -\frac{\dot{\phi}}{\phi}\frac{\rho + 3p}{2} -\frac{\dot{\phi} \pi_\phi^2}{2 \alpha_3 \omega a^6} -\frac{1}{2}\frac{\pi_\phi \dot{\pi_\phi} \phi}{\alpha_3 \omega a^6} +\frac{1}{4}\frac{\phi \pi_\phi^2}{\alpha_3 \omega^2 a^6}\dot{\omega} + \frac{k \phi}{\alpha_3 a^3}\dot{c_g^2}
\end{equation}\\
We may now proceed substituting the equations for $\dot{\phi}$ and $\dot{\pi_\phi}$ obtained from the Hamiltonian as
\begin{equation}
    \dot{\phi} = \{\phi, H\} = \frac{\partial H}{\partial \pi_\phi} = \frac{1}{2}\frac{\phi \pi_\phi N}{\omega a^3} +\alpha_3 Na^3 \frac{\partial \rho}{\partial c_m^2}\frac{\partial c_m^2}{\partial T_\phi} -\phi Na k \frac{\partial c_g^2}{\partial T_\phi} \label{how_sal}
\end{equation}
\begin{equation}
    \dot{\pi_\phi} = \{\pi_\phi, H\} = -\frac{\partial H}{\partial \phi} = -N \biggl (\frac{-\rho +3p}{2}\biggl)\frac{\alpha_3 a^3}{\phi} -\frac{1}{2}\frac{\pi_\phi^2 N}{\omega a^3}
\end{equation}
where we notice that the $\dot{\pi_\phi}$ equation is the same one we derived in subsection \eqref{master}. Furthermore, since $c_g^2$ and $c_m^2$ are also deepening on $T_\omega$, we have the additional two Hamilton's equations
\begin{equation}
    \dot{\omega} = \{\omega, H\} = \frac{\partial H}{\partial T_\omega} = -\phi Na k \frac{\partial c_g^2}{\partial T_\omega} + \alpha_3 Na^3 \frac{\partial \rho}{\partial c_m^2}\frac{\partial c_m^2}{\partial T_\omega} 
\end{equation}
\begin{equation}
    \dot{T_\omega} = \{T_\omega, H\} = -\frac{\partial H}{\partial \omega} = \frac{1}{4}\frac{\phi \pi_\phi^2 N}{\omega^2 a^3}
\end{equation}
Substituting these in equation \eqref{how_sal} we obtain 
\begin{align}
    \dot{\rho} +3\frac{\dot{a}}{a}(\rho + p) = \frac{\rho}{2}\frac{\pi_\phi N}{\omega a^3} + \frac{\alpha_3 \rho Na^3}{\phi} \frac{\partial \rho }{\partial c_m^2}\frac{\partial c_m^2}{\partial T_\phi} -\rho Na k \frac{\partial c_g^2}{\partial T_\phi} \\ \nonumber\\
    -\frac{(\rho + 3p)}{4}\frac{\pi_\phi N}{\omega a^3} - \frac{(\rho +3p) \alpha_3 Na^3}{2 \phi} \frac{\partial \rho }{\partial c_m^2}\frac{\partial c_m^2}{\partial T_\phi} + \frac{(\rho +3p)}{2} Nak \frac{\partial c_g^2}{\partial T_\phi} \\ \nonumber\\
    -\frac{1}{4}\frac{\phi \pi_\phi^3 N}{\alpha_3 \omega^2 a^9} -\frac{1}{2}\frac{\pi_\phi^2 N}{\omega a^3}\frac{\partial \rho }{\partial c_m^2}\frac{\partial c_m^2}{\partial T_\phi} +\frac{1}{2}\frac{\phi \pi_\phi^2 Nk}{\alpha_3 \omega a^5}\frac{\partial c_g^2}{\partial T_\phi} +\frac{(-\rho +3p)}{4}\frac{\pi_\phi N}{\omega a^3} \\ \nonumber \\
    +\frac{1}{4}\frac{\phi \pi_\phi^3 N}{\alpha_3 \omega^2 a^9} -\frac{1}{4}\frac{\phi^2 \pi_\phi^2 N k}{\alpha_3 \omega^2 a^5}\frac{\partial c_g^2}{\partial T_\omega} +\frac{1}{4}\frac{\phi \pi_\phi^2 N}{\omega^2 a^3} \frac{\partial \rho }{\partial c_m^2}\frac{\partial c_m^2}{\partial T_\omega}  \\ \nonumber \\
    +\frac{1}{4}\frac{\phi^2 \pi_\phi^2 kN}{\alpha_3 \omega^2 a^5}\frac{\partial c_g^2}{\partial T_\omega} -\biggl(\frac{-\rho + 3p}{2}\biggl)kNa \frac{\partial c_g^2}{\partial T_\phi} - \frac{1}{2}\frac{\phi \pi_\phi^2 k N}{\alpha_3 \omega a^5}\frac{\partial c_g^2}{\partial T_\phi}
\end{align}
which heavily simplifies to 
\begin{equation}
    \dot{\rho} +3\frac{\dot{a}}{a}(\rho + p) = \frac{\partial \rho}{\partial c_m^2} \biggl[\frac{\alpha_3 Na^3}{\phi}\frac{(\rho -3p)}{2}\frac{\partial c_m^2}{\partial T_\phi} -\frac{1}{2}\frac{\pi_\phi^2 N}{\omega a^3}\frac{\partial c_m^2}{\partial T_\phi} +\frac{1}{4}\frac{\phi \pi_\phi^2 N}{\omega^2 a^3}\frac{\partial c_m^2}{\partial T_\omega}\biggl]
\end{equation}
This shows, once again, energy violation in a scenario where a matter parameter depends on one or more gravitational clocks. Interestingly, the first source term is equivalent to the one we will find for $c_m^2 = c_m^2 (T_\phi, T_N)$, as well as the second one, arising from the canonical momentum $\pi_\phi$. However, the third term is new and it provides a source in $c_m^2$ given by the Brans-Dicke time $T_\omega$.

\subsection{Fixed $\phi_0$ with $c_m^2 = c_m^2 (T_\phi, T_N)$}

We would like to derive a scenario for the $\alpha_3$-like case. This arises when we fix  $\phi_0$ only in the canonical pair. As we have seen in subsection \eqref{de_cou_pled}, doing so leads to different equations of motion, but gives an overall conservation equation equal to the full $\phi$ (which we analyse in the next subsection). Therefore, we would like to consider the case where $\bm{\alpha} = (\phi, \alpha_3)$ and $\bm{\beta} = c_m^2$. Once again, for completeness, also $c_g^2$ is assumed to vary, such that $\bm{\beta} = c_g^2$ as well. The action we begin with is 
\begin{equation}
     S = V_c \int{dt \: \phi_0 \: \dot{b}a^2 + \dot{\phi}\pi_\phi + \alpha_3 \dot{m}_i \psi_i + \phi Na(b^2 + k) -\alpha_3Na^3 \rho - \frac{1}{4}\frac{\phi}{\omega}\frac{\pi_\phi^2}{a^3}N}
\end{equation}
where the second canonical pair act as a unimodular term for the Ricci clock. In this case, since the $\phi$ in the canonical pair is fixed, it is not necessary to express the Hamiltonian in terms of $A^2$, so we can directly consider the general  conservation equation \eqref{drusillo_oneee}
\begin{equation}
    \dot{\rho} + 3\frac{\dot{a}}{a}(\rho +p) = -\frac{\dot{\alpha_3}}{\alpha_3}\rho +\frac{\dot{\phi}}{\phi}\rho -\frac{1}{2}\frac{\phi \pi_\phi \dot{\pi_\phi}}{\alpha_3 \omega a^6} + \frac{1}{4}\frac{\phi \pi_\phi^2}{\alpha_3 \omega^2 a^6}\dot{\omega} + \frac{k \phi}{\alpha_3 a^2}\frac{dc_g^2}{dt}
\end{equation}
To find equations for the terms with $\dot{\alpha_3}$, $\dot{\phi}$ and $\dot{\pi_\phi}$ we use, once again, the Hamiltonian, giving 
\begin{equation}
     \dot{\phi} = \{\phi, H\} = \frac{\partial H}{\partial \pi_\phi} = \frac{1}{2}\frac{\phi \pi_\phi N}{\omega a^3} +\alpha_3 Na^3 \frac{\partial \rho}{\partial c_m^2}\frac{\partial c_m^2}{\partial T_\phi} -\phi Na k \frac{\partial c_g^2}{\partial T_\phi}
\end{equation}
\begin{equation}
    \dot{\pi_\phi} = \{\pi_\phi, H\} = -\frac{\partial H}{\partial \phi} = \frac{\alpha_3 \rho}{\phi}Na^3 \equiv \dot{T_\phi}
\end{equation}
where the clock equation was already obtained in subsection \eqref{drusillo_oneee} as the equation of motion of the conjugate momentum of $\phi$. On the other hand, given the $\alpha_3$ dependence, we have 
\begin{equation}
    \dot{\alpha_3} = \{\alpha_3, H \} = \frac{\partial H}{\partial T_N} =  - Na \phi k \frac{\partial c_g^2}{\partial T_N} + Na^3 \alpha_3 \frac{\partial \rho}{\partial c_m^2} \frac{\partial c_m^2}{\partial T_N}
\end{equation}
\begin{equation}
   \dot{T_\N} = \{T_N, H \} = -\frac{\partial H}{\partial \alpha_3} = - \rho Na^3    
\end{equation}
Therefore, substituting these expressions in the conservation equation and using that 
\begin{equation}
    \frac{d c_g^2}{dt} = \frac{\partial c_g^2}{\partial T_\phi}\dot{T_\phi} + \frac{\partial c_g^2}{\partial T_N}\dot{T_N}
\end{equation}
we arrive at 
\begin{equation}
    \dot{\rho} +3\frac{\dot{a}}{a}(\rho + p) = \rho Na^3 \frac{\partial \rho}{\partial c_m^2} \biggl[\frac{\alpha_3}{\phi}\frac{\partial c_m^2}{\partial T_\phi} - \frac{\partial c_m^2}{\partial T_N} \biggl]
\end{equation}\\
were we see that both the Ricci and the Newton times create a source term for the conservation equation. Interestingly, this result closely relates to the one found in \cite{magueijo2023evolving}, the main difference being the Ricci time source term. In fact, due to the fixed scalar field in the canonical term, the equation of motion for $T_R$ is different, resulting in a theory with only energy density.

\subsection{Full $\phi$ with $c_m^2 = c_m^2 (T_\phi, T_N)$}

We now consider the scenario where $c_m^2$ depends on the Ricci and the Newton times, following the same spirit of \cite{magueijo2023evolving}. As before, we include also a varying $c_g^2$ for completeness. The action is
\begin{equation}
    S = V_c \int{dt \: \phi \: \dot{b}a^2 + \dot{\phi}\pi_\phi + \alpha_3 \dot{m}_i \psi_i + \phi Na(b^2 + k) -\alpha_3Na^3 \rho - \frac{1}{4}\frac{\phi}{\omega}\frac{\pi_\phi^2}{a^3}N}
\end{equation}
giving the usual Hamiltonian as
\begin{equation}
    H = -\phi Na (b^2 + kc_g^2) + \alpha_3 Na^3 \rho + \frac{1}{4} \frac{\phi}{\omega} \frac{\pi_\phi^2}{a^3}N
\end{equation}
which may be expressed in terms of $A^2 = \phi a^2$ as
\begin{equation}
    H = -\phi N \frac{A}{\phi^\frac{1}{2}}(b^2 +kc_g^2) + \frac{A^3}{\phi^\frac{3}{2}} \alpha_3 N \rho + \frac{1}{4}\frac{\phi}{\omega} \pi_\phi^2 N \frac{\phi^\frac{3}{2}}{A^3}
\end{equation}
This form of the Hamiltonian will prove crucial in deriving the $\dot{\pi_\phi}$ equation, as the correct canonical variable is given by $\phi a^2$ when we include a Ricci time dependence. The general conservation equation, without using the equation of motion for $\dot{\phi}$ derived by the definition of $\pi_\phi$ is therefore
\begin{equation}
    \dot{\rho} +3\frac{\dot{a}}{a}(\rho + p) = -\frac{\dot{\alpha_3}}{\alpha_3}\rho + \frac{\dot{\phi}}{\phi}\rho -\frac{\dot{\phi}}{\phi}\frac{\rho + 3p}{2} -\frac{\dot{\phi} \pi_\phi^2}{2 \alpha_3 \omega a^6} -\frac{1}{2}\frac{\pi_\phi \dot{\pi_\phi} \phi}{\alpha_3 \omega a^6} + \frac{k \phi}{\alpha_3 a^3}\dot{c_g^2} \label{same_again}
\end{equation}
where we are including the $\dot{\alpha_3}$ term because $c_m^2$ and $c_g^2$ depend on the Newton time, while the $\dot{\omega}$ term has been removed as, in this scenario, $\omega$ is constant. We can now derive the equations of motion for $\phi$ and $\pi_\phi$, obtaining
\\
\begin{equation}
     \dot{\phi} = \{\phi, H\} = \frac{\partial H}{\partial \pi_\phi} = \frac{1}{2}\frac{\phi \pi_\phi N}{\omega a^3} +\alpha_3 Na^3 \frac{\partial \rho}{\partial c_m^2}\frac{\partial c_m^2}{\partial T_\phi} -\phi Na k \frac{\partial c_g^2}{\partial T_\phi}
\end{equation}
\\
\begin{equation}
    \dot{\pi_\phi} = \{\pi_\phi, H\} = -\frac{\partial H}{\partial \phi} = -N \biggl (\frac{-\rho +3p}{2}\biggl)\frac{\alpha_3 a^3}{\phi} -\frac{1}{2}\frac{\pi_\phi^2 N}{\omega a^3} \equiv \dot{T_\phi}
\end{equation}
\\
On the other hand, we have the equations of motion for the canonical pair $(\alpha_3, T_N)$
\\
\begin{equation}
    \dot{\alpha_3} = \{\alpha_3, H \} = \frac{\partial H}{\partial T_N} =  - Na \phi k \frac{\partial c_g^2}{\partial T_N} + Na^3 \alpha_3 \frac{\partial \rho}{\partial c_m^2} \frac{\partial c_m^2}{\partial T_N}
\end{equation}
\begin{equation}
    \dot{T_\N} = \{T_N, H \} = -\frac{\partial H}{\partial \alpha_3} = - \rho Na^3
\end{equation}
Therefore, combining them inside equation \eqref{same_again} and carrying out the necessary simplifications we arrive at
\\
\\
\begin{equation}
    \dot{\rho} +3\frac{\dot{a}}{a}(\rho + p) = Na^3 \frac{\partial \rho}{\partial c_m^2} \biggl[\frac{\alpha_3}{\phi} \frac{\partial c_m^2}{\partial T_\phi} \frac{\rho-3p}{2} - \frac{\partial c_m^2}{\partial T_N}\rho\biggl]
    -\frac{1}{2}\frac{N \pi_\phi^2}{\omega a^3}\frac{\partial \rho}{\partial c_m^2}\frac{\partial c_m^2}{\partial T_\phi} \label{mumu}
\end{equation}
\\
\\
This result, as the one obtained in the previous section, also closely relates to \cite{magueijo2023evolving}. In fact, the first term on the RHS of equation \eqref{mumu} is exactly the same with the scalar field giving the Ricci time. However, the second term is new and it is the result of the scalar field momentum giving an extra source term for $c_m^2$.
\\
\\
Having reached the end of our research, we provide a short overview on the general pattern followed by varying constants. To get net energy violation two possible cases can be considered: a gravitational parameter depending on a matter clock or a matter clock depending on a gravitational parameter. In the first case, we might have $\bm{\beta} = (c_g^2, \omega, \phi)$, all depending on a matter clock given by the time canonical to an $\bm{\alpha}$ such as $\alpha_3$ or $\rho_\Lambda$. These cases lead to net energy violation as the variation of the parameter can be exchanged with the constant of motion of the clock, since it is given by a matter parameter. The other scenario occurs when a matter parameter depends on a gravitational clock. This is the case, for example, when $\bm{\beta} = c_m^2$ and $\bm{\alpha} = \omega$. Once again, the varying matter parameter can exchange energy with the gravitational clock, thus giving a net energy violation. All the other combinations, as we have shown above, do not lead to any energy violation, with all the constant's variation being absorbed in the clock. As expected, with the exception of extra terms given by the canonical momentum of $\phi$, our results reproduce the pattern identified in \cite{magueijo2023evolving}.

\clearpage{\pagestyle{empty}\cleardoublepage}

\chapter{Conclusion and Discussion}

We have reached the conclusion of our initial work on varying constants in cosmological Brans-Dicke theories. Our starting point has been a conceptual analysis of the ideas surrounding physical constants and the possibility they might be varying with time. We have reviewed the main theoretical intuitions regarding this topic, ranging from Dirac's Large Number Hypothesis to Mach's Principle, presenting their consequences. Furthermore, we have explored the experimental evidences supporting these ideas, opening the way to more extensive applications of varying constants in different fields of physics, from the Standard Model to Cosmology. These all have been necessary precursors towards the main objective of this research: the Brans-Dicke theory of gravity. In fact, motivated by the natural implementation of Mach's principle and of a varying $G$ in the Brans-Dicke action, we have used it as the platform to extend varying constants theories. Doing so, we have produced several results regarding energy conservation and the Cosmological Constant problem.
\\
\\
In Chapter 2, we have presented the main open problems is the Standard Big Bang Cosmology, providing the current solution of inflation. These cosmological puzzles then lead us to consider an alternative solution, based on the idea of a time varying speed of light. We have reviewed the main results of the VSL model in cosmology, as the conceptual basis for our next steps. Generalising this idea to more constants of Nature, it became necessary an appropriate mathematical formalism. This was promptly supplied by unimodular gravity and minisuperspace. The first one allowed to obtain equations of motion for the constants directly from the action principle, having demoted them to constants of motion on-shell. Furthermore, unimodular gravity offered solid definitions of relational physical times, excellent as evolution parameters for the constants we are studying. On the other hand, reducing the Einstein-Cartan action to minisuperpace enabled us to fully exploit the Hamiltonian formalism of General Relativity. Therefore, we developed the minisuperspace reduction of the Einstein-Cartan and Einstein-Hilbert actions for a FLRW metric, showing how it naturally produces an Hamiltonian structure we can use to derive the equations of motion for the constants and their times.
\\
\\
\\
Continuing, we have applied these formalisms to obtain energy conservation results in cosmology. Doing so, we were able to characterise different scenarios where some constants provide clocks while other vary with respect to these clocks. Interestingly, when considering overall energy violation, a pattern arose, linking gravitational parameters to matter clocks and vice-versa. All the other combinations of constants have been found to be sterile, accordingly to \cite{magueijo2023evolving}, resulting in the standard conservation equation. The landscape of varying constant resulted to be a vast and manifold one, leading us to consider multiple combinations in our extensions.
\\
\\
Moreover, in Chapter 3, we developed the core of our original work: energy conservation due to varying constants in Brans-Dicke cosmologies. Following the procedure outlined in Chapter 2, we presented the energy conservations and violations resulting from time dependencies like $T_\Lambda$, $T_\phi$ and $T_N$ with a dynamical scalar field $\phi$. We also considered the dependence of several constants on the Brans-Dicke time $T_\omega$, enlarging the set of constants that can be varying. Particularly, this addition will turn useful in future work on Black Holes applications by Magueijo \cite{magueijo2023BH}. Applying the idea of varying constant to Brans-Dicke theory, we have obtained several energy conservation and violation scenarios where the $\omega$ parameter plays the important role of energy source. Specifically, considering its dependency on the cosmological constant time $T_\Lambda$, we have obtained a relationship between $\rho_\Lambda$ and $\omega$ which might lead to future developments in the context of the vacuum energy sequester \cite{bassani2023}. Furthermore, similar relationships have been obtained for $c_g^2$ and $\alpha_3$, also leading to future extensions connected to VSL and the early Universe \cite{bassani2023}. Finally, the inclusion of the Brans-Dicke parameter $\omega$ has potentially enabled us to connect our theoretical results in the field of energy conservation with cosmological observations of the Brans-Dicke theory.
\\
\\
To conclude, we would like to briefly mention the future developments and applications of these results, some of which will be included in a paper to come \cite{bassani2023}. Firstly, as Chapter 3 has taught us, the possible combinations of constants and times are plenty, to the point that only imagination is really the limit when coupling multiple different actions to gravity. Therefore, in our future work, we will be selecting the most interesting, relevant and insightful scenarios and focus on them. As mentioned earlier, these might be the ones were $\omega$ is linked to $c_g^2$ and $\rho_\Lambda$. Specifically this ladder case brings us to our next extension of this theory: the vacuum energy sequester. In fact, a scenario where $\omega = \omega (T_\Lambda)$ might provide an explanation to the Cosmological Constant problem, where the vacuum energy is absorbed by the variation of $\omega$. This is particularly interesting given the observational constraints put on $\omega$ \cite{bertotti2003test}. Secondly, when considering the Brans-Dicke kinetic term, we set the $c$ factor appearing with it to zero. However, this could not be the case and therefore we will investigate its inclusion, especially in the $ \omega = \omega(T_\Lambda)$ scenario. Thirdly, the scenarios where $\phi$ provides the Ricci time will be further extended, including the case when both the unimodular action and the canonical pair provide equivalent times definitions. Other applications of our theory could be linked to the Standard Model, where we might have coupling and particle's parameters depending on $\omega$, leading to unknown connections between cosmology and particle physics.
\\
\\
Lastly, among the many extensions of this work, the most fascinating one is a theory explaining the origin of varying constants beyond the String Theory formalism. Throughout this dissertation, we have postulated that the constants of Nature are varying with respect to relational times. To do so, we have used the unimodular formalism, which provided a solid and rigorous mathematical platform to implement this idea. However, this formalism does not explain why the constants of Nature could be varying, omitting the mechanism behind their dynamics. One simple answer to this would be, as we have done in this work, to assume that constants do vary, without asking the deeper question of why. After all, often Physics explains how events happen and not why. But what if it could be possible to formulate a theory predicting and describing the evolution of some constants? It would certainly be exciting to discover if these theory's predictions agree with the experimental evidences, leading to a radically different view of the Reality we live in. This speculation, like many other ideas, is left to the future, when maybe our posterity will finally consider trivial what we, for so long, chased into darkness.

\clearpage

\thispagestyle{empty}
\null\vfill

\settowidth\longest{\huge\itshape just as his inclination leads him;}
\begin{center}
    \parbox{\longest}{%
    \raggedright{\itshape{``Veniet tempus quo posteri nostri tam aperta nos nescisse mirentur''}\par\bigskip
     }  
    }
\end{center}
\begin{center}
    \textit{\textit{Lucius Annaeus Seneca}}
\end{center}

\vfill\vfill

%%%%%%%%%%%%%%%%%%%%%%%%%%%%%%%%%%%%
%%%%%%%%%%%%%%%%%%%%%%%%%%%%%%%%%%%%
%% bibliography
\newpage
\addcontentsline{toc}{chapter}{Bibliography}
\bibliographystyle{unsrt}
\bibliography{bibliography.bib}

\nocite{carroll2001cosmological}
\nocite{albrecht1999time}
\nocite{magueijo2023evolving}
\nocite{wood1984spelunking}
\nocite{turok2002critical}
\nocite{guth1981inflationary}
\nocite{abbott1988mystery}
\nocite{d2022introducing}
\nocite{henneaux1989cosmological}
\nocite{vilenkin1994approaches}
\nocite{vishwakarma2015machian}
\nocite{moffat1993superluminary}
\nocite{bufalo2015unimodular}
\nocite{weinberg1989cosmological}
\nocite{martin2012everything}
\nocite{bertotti2003test}
\nocite{brans1961mach}
\nocite{dirac1938new}
\nocite{bohmer2016introduction}
\nocite{lichtenegger2008mach}
\nocite{marzke1964gravitation}
\nocite{bombelli1991time}
\nocite{einstein1923electrodynamics}
\nocite{finch2003mathematical}
\nocite{halliday2013fundamentals}
\nocite{griffiths2018introduction}
\nocite{hacking1975identity}
\nocite{einstein1908relativitatsprinzip}
\nocite{magueijo2022connection}
\nocite{magueijo2021cosmological}
\nocite{uzan2011varying}
\nocite{acheson2020wonder}
\nocite{weinberg1983overview}
\nocite{robinson2011symmetry}
\nocite{levshakov2009spatial}
\nocite{levy1979importance}
\nocite{uzan2008natural}
\nocite{ellis2005c}
\nocite{sciama1957monthly}
\nocite{dicke1959new}
\nocite{chiba2011constancy}
\nocite{riess1998observational}
\nocite{hellings1989experimental}
\nocite{damour1988limits}
\nocite{damour1991orbital}
\nocite{teller1948change}
\nocite{degl1995time}
\nocite{webb1999search}
\nocite{blatt2008new}
\nocite{kanekar2004conjugate}
\nocite{martins2004wmap}
\nocite{shlyakhter1976direct}
\nocite{fujii2000nuclear}
\nocite{griffiths2005introduction}
\nocite{bekenstein1982fine}
\nocite{damour2009equivalence}
\nocite{damour2012theoretical}
\nocite{berti2015testing}
\nocite{einstein1915field}
\nocite{french2003vibrations}
\nocite{weinberg1972gravitation}
\nocite{mach1893science}
\nocite{singleton2016global}
\nocite{raine1975mach}
\nocite{ozsvath2001approaches}
\nocite{heckman1962relativistic}
\nocite{thorne2000gravitation}
\nocite{fixsen1994cosmic}
\nocite{srednicki2007quantum}
\nocite{borde1996singularities}
\nocite{padilla2015lectures}
\nocite{trautman2006einstein}
\nocite{gronwald1996gauge}
\nocite{magueijo2020equivalence}
\nocite{borowiec2022scalar}
\nocite{york1986boundary}
\nocite{lowenstein2012essentials}
\nocite{iyer1997lagrangian}
\nocite{gielen2020singularity}
\nocite{brown1993action}
\nocite{frion2019affine}
\nocite{einstein1952gravitational}
\nocite{fortin2022relational}
\nocite{rovelli1996relational}
\nocite{arthur2021leibniz}
\nocite{papagiannopoulos2017dynamical}
\nocite{almeida2021quantum}
\nocite{martel1998likely}
\nocite{bassani2023}
\nocite{magueijo2023BH}
\nocite{einstein1905electrodynamics}
\nocite{gorbunov2011introduction}
\nocite{{bruno}}
\nocite{vilenkin1986boundary}
\nocite{abbott2016observation}

\end{document}